\documentclass[twocolumn,trackchanges]{aastex631}
\usepackage{graphicx}	
\usepackage{amsmath}	
\usepackage{amssymb}	
\usepackage{rotating}
\usepackage{float}
\usepackage{hyperref} 
\usepackage[para,splitrule]{footmisc} 

\newcommand{\hi}{H{\sc i}}
\newcommand{\hii}{H{\sc ii}}

\newcommand{\ha}{H$\alpha$}
\newcommand{\hb}{H$\beta$}
\newcommand{\Msun}{M$_{\odot}$}
\newcommand{\Mstar}{$M_{\ast}$}
\begin{document}

\title[Dwarf  Galaxy Integral-field Survey]{Dwarf Galaxy Integral-field Survey (DGIS): survey overview and the result of global mass-metallicity relation}

\author[0000-0001-7634-0034]{Xin Li}
\affiliation{School of Astronomy and Space Science,
Nanjing University,
Nanjing, Jiangsu 210093, China}
\affiliation{Key Laboratory of Modern Astronomy and Astrophysics (Nanjing University),
Ministry of Education,
Nanjing, Jiangsu 210093, China}
\author[0000-0002-8614-6275]{Yong Shi}
\altaffiliation{E-mail: yong@nju.edu.cn}
\affiliation{School of Astronomy and Space Science,
Nanjing University,
Nanjing, Jiangsu 210093, China}
\affiliation{Key Laboratory of Modern Astronomy and Astrophysics (Nanjing University),
Ministry of Education,
Nanjing, Jiangsu 210093, China}
\author[0000-0002-1620-0897]{Fuyan Bian}
\altaffiliation{E-mail: fbian@eso.org}
\affiliation{European Southern Observatory,
Alonso de Cordova 3107,
Casilla 19001, Vitacura, Santiago 19, Chile}
\author[0000-0002-2937-6699]{Xiaoling Yu}
\altaffiliation{E-mail: xiaoling@mail.qjnu.edu.cn}
\affiliation{College of Physics and Electronic Engineering,
Qujing Normal University,
Qujing, Yunnan 655011, China}
\author[0000-0001-9018-9465]{Zhiyuan Zheng}
\affiliation{School of Astronomy and Space Science,
Nanjing University,
Nanjing, Jiangsu 210093, China}
\affiliation{Key Laboratory of Modern Astronomy and Astrophysics (Nanjing University),
Ministry of Education,
Nanjing, Jiangsu 210093, China}
\author[0000-0001-8112-7844]{Songlin Li}
\affiliation{Research School of Astronomy and Astrophysics,
Australian National University,
Canberra, ACT 2611, Australia}
\author[0000-0003-3226-031X]{Yanmei Chen}
\affiliation{School of Astronomy and Space Science,
Nanjing University,
Nanjing, Jiangsu 210093, China}
\affiliation{Key Laboratory of Modern Astronomy and Astrophysics (Nanjing University),
Ministry of Education,
Nanjing, Jiangsu 210093, China}
\author[0000-0002-3890-3729]{Qiusheng Gu}
\affiliation{School of Astronomy and Space Science,
Nanjing University,
Nanjing, Jiangsu 210093, China}
\affiliation{Key Laboratory of Modern Astronomy and Astrophysics (Nanjing University),
Ministry of Education,
Nanjing, Jiangsu 210093, China}
\author[0000-0003-4874-0369]{Junfeng Wang}
\affiliation{Department of Astronomy and Institute of Theoretical Physics and Astrophysics,
Xiamen University,
Xiamen, 361005, China}
\author[0000-0001-8317-2788]{Shude Mao}
\affiliation{Department of Astronomy,
Tsinghua University,
Beijing, Beijing 100084, China}
\affiliation{Department of Astronomy, 
Westlake University, 
Hangzhou 310030, Zhejiang Province, China}
\author[0000-0002-0584-8145]{Xiangdong Li}
\affiliation{School of Astronomy and Space Science,
Nanjing University,
Nanjing, Jiangsu 210093, China}
\affiliation{Key Laboratory of Modern Astronomy and Astrophysics (Nanjing University),
Ministry of Education,
Nanjing, Jiangsu 210093, China}
\author[0000-0003-1632-2541]{Hongxin Zhang}
\affiliation{Key Laboratory for Research in Galaxies and Cosmology, Department of Astronomy,
University of Science and Technology of China,
Hefei, Anhui 230026, China}
\affiliation{School of Astronomy and Space Science,
University of Science and Technology of China,
Hefei, Anhui 230026, China}
\author[0000-0002-2583-2669]{Kai Zhu}
\affiliation{Department of Astronomy,
Tsinghua University,
Beijing, Beijing 100084, China}
\author[0000-0002-7299-2876]{Zhiyu Zhang}
\affiliation{School of Astronomy and Space Science,
Nanjing University,
Nanjing, Jiangsu 210093, China}
\affiliation{Key Laboratory of Modern Astronomy and Astrophysics (Nanjing University),
Ministry of Education,
Nanjing, Jiangsu 210093, China}

\begin{abstract}
Low-mass galaxies are the building blocks of massive galaxies in the framework of hierarchical structure formation. To enable detailed studies of galactic ecosystems in dwarf galaxies by spatially resolving different galactic components, we have carried out the Dwarf Galaxy Integral-field Survey (DGIS). This survey aims to acquire observations with spatial resolutions as high as 10 to 100 pc while maintaining reasonably high signal-to-noise ratios with VLT/MUSE and ANU-2.3m/WiFeS. The whole sample will be composed of 65 dwarf galaxies with $M_{\rm \ast}$ $<$ 10$^{9}$ $\rm M_{\odot}$, selected from the Spitzer Local Volume Legacy Survey. The overall scientific goals include studying baryonic cycles in dwarf galaxies, searching for off-nuclear (intermediate)-massive black holes, and quantifying the inner density profiles of dark matter. In this work, we describe the sample selection, data reduction, and high-level data products. By integrating the spectra over the field of view for each galaxy, we obtained the integrated gas-phase metallicity and discussed its dependence on stellar mass and SFR. We find that the overall relation between metallicity and stellar mass of our DGIS nearly follows the extrapolation from the higher mass end. Its dispersion does not decrease by invoking the dependence on SFR.

\end{abstract}


\section{Introduction} \label{sec:intro}
Low-mass galaxies are the building blocks of massive galaxies within the framework of hierarchical structure formation theory \citep{Wise12, Stierwalt17} and exhibit a wide diversity of physical properties \citep{Tolstoy09, Henkel22}. Compared to spiral galaxies, dwarf galaxies have lower stellar masses, lower star formation rates (SFR), episodic star formation histories (SFH) \citep{Tolstoy09}, thicker disks or rounder morphologies \citep{Roychowdhury10, Poulain21}, and contain less dust and molecular gas \citep{Rosenberg08, Shi16}. Additionally, star-forming dwarf galaxies have lower gas-phase metallicities and higher ionization states, which are similar to those observed in high-redshift galaxies \citep{Skarleth21, Espinosa-Ponce22, Nakajima22, Henkel22, Schaerer22}.

Dwarf galaxies offer crucial laboratories to test theories of galaxy formation and evolution in many aspects. They deviate from the classical Kennicutt–Schmidt (KS) law \citep{Kennicutt98}, leading to the promotion of the search for a more universal star formation law \citep{Shi18, Du23}; some of them harbor intermediate-mass black holes (IMBH) \citep{Reines13, Mezcua18}, providing insights into the formation mechanisms of supermassive black hole seeds \cite[][and references therein]{Greene20}; because of low gravitational potential, feedback from star formation and active galactic nuclei (AGN) can have more pronounced impact on the ISM and circumgalactic media (CGM) \citep[e.g.,][]{Mashchenko08,Sharda23,Zheng23}. The above global properties and related studies have been primarily based on numerous imaging \citep[e.g.,][]{York2000, James04, Martin05, Skrutskie06, Keller07, Dale09, Kaiser10, Ferrarese12,Bennet17, Geha17, Dey19, Habas20, Ferrarese20, Kojima20, Poulain21, Drlica-Wagner21, Davis21, Carlsten22} and single-fiber/long-slit observations \citep[e.g.,][]{York2000, Liske15, Geha17, Kojima20}. 

As a result, these studies have primarily focused on the global properties of dwarf galaxies, with limited attention given to their local properties. Spatially resolved spectroscopic studies can, for example, help us explore the link between the different galactic components, the growth of black holes, the effects of AGN feedback and the modes of stellar feedback, environmental impacts including gas accretion and gas stripping, and local relations like the mass-metallicity relation (MZR) and its dependence on SFR, among others \citep{Sanchez20}. 

The development of Integral Field Spectroscopy (IFS) has significantly enhanced our understanding of the internal physical mechanisms of galaxies \citep[see the review,][]{Sanchez20}. Among the large legacy programs, \cite{Cano-Diaz22} provided the largest homogeneous dwarf galaxy sample (136) from the SDSS-IV Mapping Nearby Galaxy at APO \citep[MaNGA,][]{Bundy15}, with stellar mass ranging from 10$^{7.5}$ M$_{\odot}$ to 10$^9$ M$_{\odot}$, distance from 0.9 to 143 Mpc, and spatial resolution of hundreds to thousands of parsecs; the Sydney-AAO Multi-object Integral-field spectrograph \citep[SAMI,][]{Croom12} covers a wide range of stellar mass between $10^7 \sim 10^{12}$\,\Msun, with a seeing of about 2.1{\arcsec}; the Calar Alto Legacy Integral Field Area survey \citep[CALIFA,][]{Sanchez12} includes galaxies with $M_{\rm B}$ down to $-18$ mag in the nearby universe, with a Point Spread Function (PSF) of around 2.5{\arcsec}; The AMUSING++ survey \citep{Lopez-Coba20} is mainly from the All-weather MUse Supernova Integral-field of Nearby Galaxies \citep[AMUSING,][]{Galbany16} survey, and also includes several smaller MUSE-based IFU surveys. It also contains low-mass galaxies with stellar masses down to 10$^8\,M_{\odot}$ and a PSF ranging from 0.6{\arcsec} to 1.4{\arcsec}.  Recently, more IFS observations have been carried out for dwarf galaxies with deeper exposure and higher spatial resolution. For instance, \cite{Cairos10, Cairos15} used the Potsdam Multi-Aperture Spectrophotometer (PMAS) and the Visible Multi-Object Spectrograph (VIMOS) to observe Blue Compact Galaxies (BCG), which are a type of low-metallicity, low-luminosity galaxies and have violent star formation; their final sample comprises roughly 40 objects \citep{Cairos17}. The \hi\, Kilofibre-Optical-AAT-Lenslet-Array (KOALA) IFS Dwarf Galaxy Survey (\hi\-KIDS) conducted optical integral observations of about 100 nearby gas-rich dwarf galaxies that already had interferometric \hi\, data \citep{Lopez-Sanchez20}. The DWarf galaxies Archival Local survey for Interstellar medium investigation \citep[DWALINE,][]{Marasco23} survey used the European Southern Observatory (ESO) Very Large Telescope (VLT) for 19 nearby starburst dwarf galaxies. \cite{Heesters23} used the VLT to observe 56 dwarf galaxies beyond the Local Volume as a follow-up observation to the MATLAS \citep[Mass Assembly of early Type gaLAxies with their fine Structures,][]{Habas20} photometric surveys. Additionally, there have been many IFS observations targeting specific objects \citep[e.g.,][]{Monreal-Ibero12, Kumari17, Kumari18, Kashiwagi21, Roier22, Valle-Espinosa23}. 

\begin{figure*}
\begin{center}
  \includegraphics[width=\textwidth]{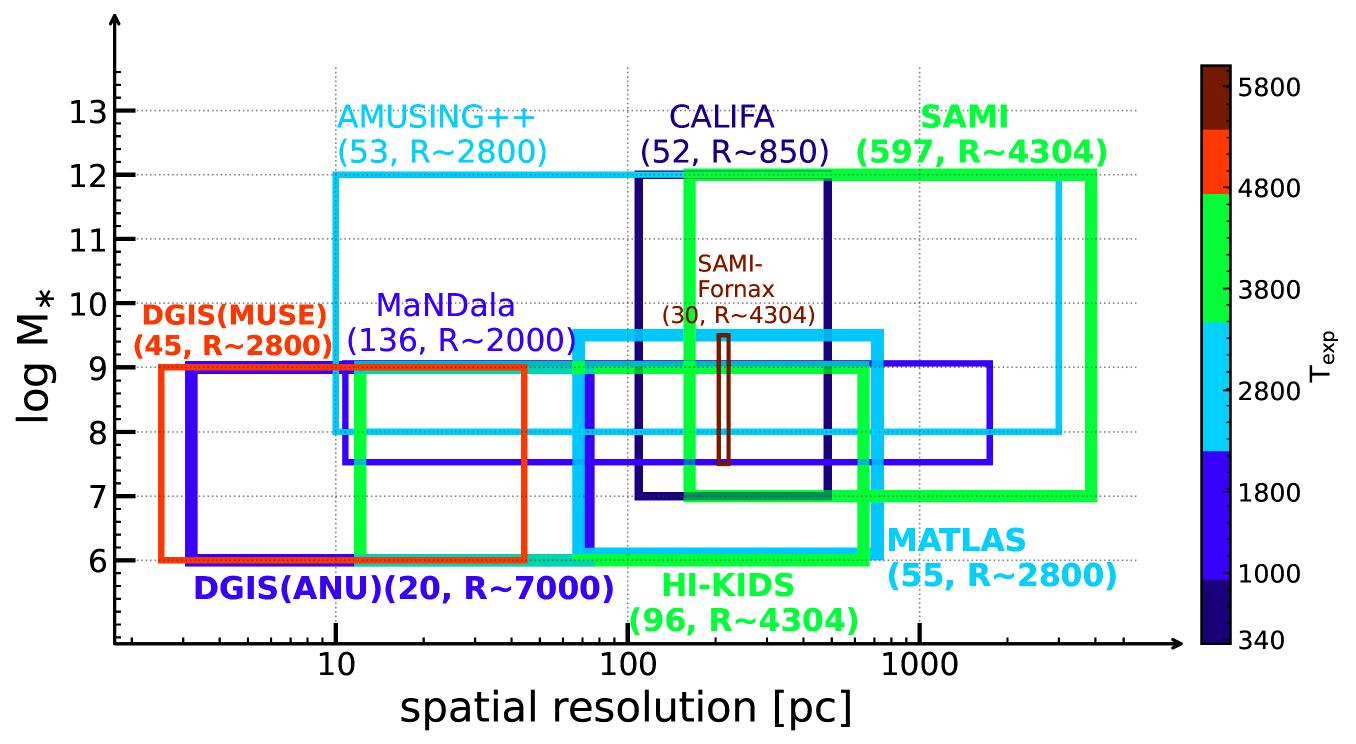}
\end{center}
\caption{Coverage of spatial resolutions and stellar masses for different IFS surveys.  The color bar is the averaged exposure time normalized to the 8-m telescope ($\rm T_{EXP} = t_{exp}/(\frac{8^2}{A^2})$, where A is the diameter of telescopes).  The names of surveys, the corresponding number of galaxies with stellar mass less than $10^9$ \Msun, and the spectral resolution around 6563 \AA\ are labeled near the rectangles and colored the same colors. DGIS survey in this work is divided to two sample based on different IFS spectrographs. SAMI (Sydney-AAO Multi-object Integral-field spectrograph, \cite{Croom12}); CALIFA (Calar Alto Legacy Integral Field Area, \cite{Sanchez12}); AMUSING++ \citep{Lopez-Coba20}; SAMI-Fornax \citep{Scott20}; MaNDala (MaNGA Dwarf galaxy, \cite{Cano-Diaz22}); MATLAS (Mass Assembly of early Type gaLAxies with 124 their fine Structures, \cite{Heesters23}), HI-KIDS(\hi\ Kilofibre-Optical-AAT- 113 Lenslet-Array (KOALA) IFS Dwarf Galaxy Survey, \cite{Lopez-Sanchez20}). }
\label{fig:survey}
\end{figure*}

However, these existing studies generally lack exposure depth and physical spatial resolution, which hampers detailed studies of local galaxy properties in a statistically meaningful manner. For this purpose, we have carried out the Dwarf Galaxies Integral-field Survey (DGIS)\footnote{\url{https://www.dgisteam.com/index.html}} that contains all dwarf galaxies (${\rm 10^{6}\,M_{\odot}}<M_{\ast}<{\rm 10^{9}\,M_{\odot}}$) in the southern hemisphere of the {\it Spitzer} Local Volume Legacy Survey \citep[LVL,][]{Dale09}. The {\it Spitzer} LVL project targeted all known galaxies in the Local Group and far-UV-flux-limited galaxies within the Local Volumes \citep{Dale09}. Statistical tests and comparisons with blind all-sky HI surveys indicate a sample completeness of $>$ 95\% \citep{Lee09a}. Two IFS instruments are applied to obtain their spatially resolved spectra. They are the Multi Unit Spectroscopic Explorer (MUSE) of the VLT, which has a large field-of-view (FOV) of 1 square arcminutes and a high spatial resolution of around 0.8{\arcsec} \citep{Bacon10}, and the Wide Field Spectrograph (WiFeS) on the Australian 2.3 m telescope (ANU-2.3m), which covers a full optical wavelength range from 3290\,\AA\ to 9120\,\AA\ with a high spectral resolution of 7000 \citep{Dopita07,Dopita10}. We achieved deep exposures for each source to achieve a high signal-to-noise ratio (S/N). As shown in Fig.~\ref{fig:survey}, we compare DGIS with existing IFS surveys in terms of stellar mass coverage, spatial resolution, spectral resolution at around 6563\,\AA, exposure depth, and the number of galaxies with stellar masses less than $10^{9}$\,\Msun, including well-known surveys like MaNGA, SAMI, and CALIFA. As shown in the figure, DGIS significantly advances dwarf galaxy surveys with its high spatial resolution and deep exposure, while maintaining a competitive sample size.

The structure of this paper is as follows: in \S\ref{sec:Sample_Description}, we introduce the sample selection and provide an overview of DGIS; in \S\ref{sec:goals}, we present the overall scientific goals of the DGIS program; in \S\ref{sec:observation}, we show the settings of observation; in \S\ref{sec:data_reduction}, we introduce the datacube reduction process; in \S\ref{sec:DataProduct}, we describe the generation of high-level data products; in \S\ref{sec:global_properties}, we list the collection and measurement of global properties; in \S\ref{sec:relation}, we make the integrated MZR as well as the metallicity fundamental relation (FMR) based on stacked spectra; in \S\ref{sec:summary}, we give the summary and conclusions. In this work, we adopt a standard $\Lambda$CDM cosmology, assuming the $\Omega_{\Lambda} = 0.7$, $\Omega_{M} = 0.3$, and $H_{0}=73\, {\rm km\,s^{-1}\,Mpc^{-1}}$, and solar metallicity from \cite{Asplund09}.

\section{Sample Description} \label{sec:Sample_Description}
\begin{figure*}
   \begin{center}
       \includegraphics[width=0.8\textwidth]{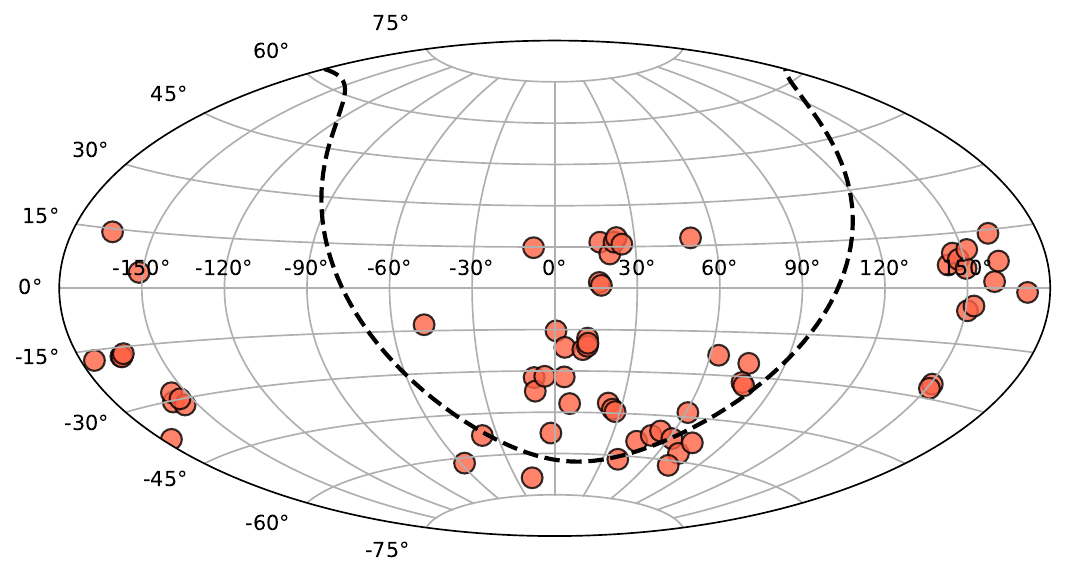}
   \end{center} 
\caption{Aitoff projection of galaxies in the DGIS survey in the equatorial coordinate. The dashed line is the Galactic plane.}
\label{fig:celestial}
\end{figure*}

The sample in DGIS is representative by including all objects in the {\it Spitzer} LVL program with declination (DEC) $<$ 20$^{\circ}$, stellar mass $< 10^9$\,\Msun, and distance $<$\,11 Mpc, based on the properties provided in Table~1 of \cite{Dale09}. The final sample is composed of 65 objects, with the Small Magellanic Cloud (SMC) excluded. Their celestial distribution is shown in Fig.~\ref{fig:celestial}. 

\begin{figure*}
\begin{center}
\includegraphics[width=0.31\textwidth]{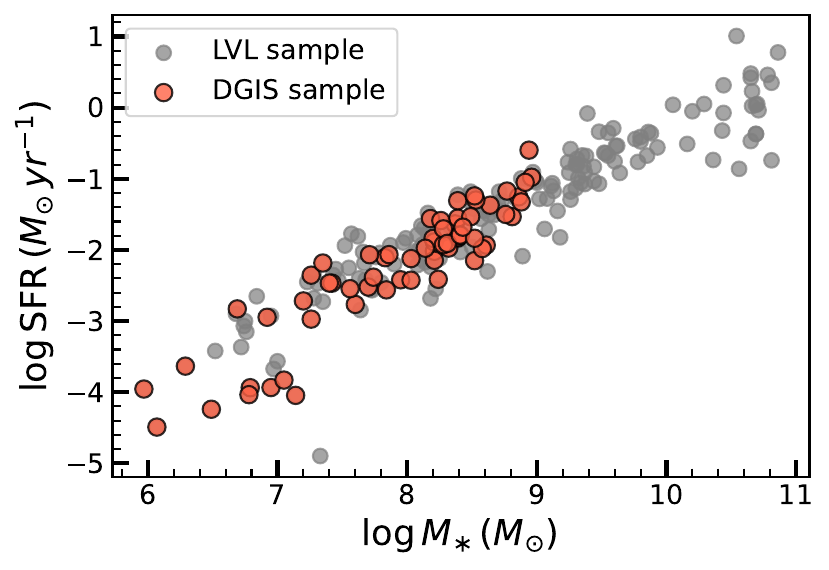}
\includegraphics[width=0.31\textwidth]{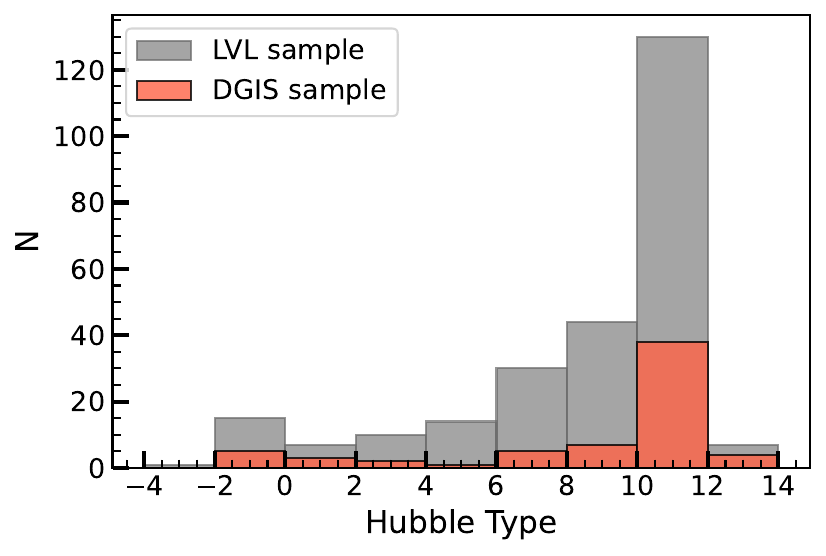}
\includegraphics[width=0.31\textwidth]{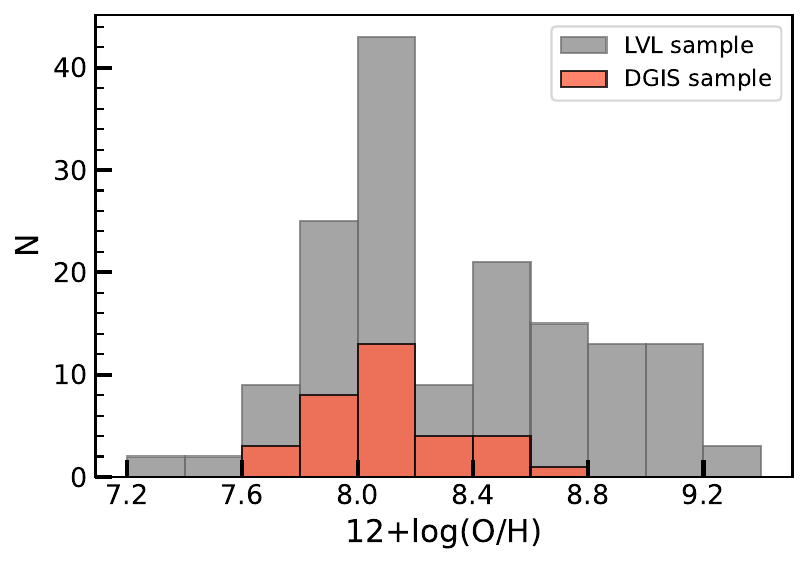}
\end{center}   
\caption{\textbf{left:} \Mstar\ versus SFR for the Local Volume Legacy survey (LVL, grey filled dots) and DGIS (tomato); \textbf{middle:} histograms of Hubble Type for LVL (faint grey) and DGIS (tomato); \textbf{right:} number distribution of metallicity (12+log(O/H)) for LVL (faint grey) and DGIS (tomato). The data with grey and tomato colors are all adopted from \cite{Dale09}, except for the the metallicity are collected from \cite{Cook14b}.}
\label{fig:sample}
\end{figure*}

Their basic properties are summarized and presented in Table~\ref{Tab:sample}, sorted by their right ascension (RA). For a detailed description, see \S\ref{sec:global_properties}. The sample covers a distance ranging from 0.44 to 11.4 Mpc, except for two objects located largely beyond 11 Mpc, according to the updated distance data from NED. As shown in the Fig.~\ref{fig:sample}, the stellar masses range from 10$^{6}$ \Msun\ to 10$^{9}$\,\Msun, with 60\% of the sample falling within the 10$^{8} \sim 10^{9}$\,\Msun\ range. The sample includes all Hubble types, with a majority of them being dwarf irregulars. 

\begin{table*}
\setlength{\tabcolsep}{3pt} 
\tiny
\begin{flushleft}
    
\begin{tabular}{cccccccccccccc}
\hline
\hline
I.D. & Target & RA & DEC & LogM$_{\ast}$ & D & redshift & T & b/a & PA & i  & Re & n & LogSFR(UV) \\
    &      & (J2000) & (J2000) & (\Msun) & (Mpc) &      &   &    & (deg) & (deg) & ({\arcsec}) &  & (\Msun\ yr$^{-1}$) \\
(1) &  (2) &    (3)  &  (4)    &   (5)  & (6) & (7)   & (8) & (9 )& (10) & (11) & (12) & (13) & (14) \\
\hline
1 & WLM & 00:01:58.16 & -15:27:39.3 & 7.42 & 0.92 & -0.0004 & 10 & 0.51 & 0 & 62 & 207.16 $\pm$ 55.22 & 1.55 $\pm$ 0.30 & -2.47 $\pm$ 0.05 \\
2 & NGC0059 & 00:15:25.13 & -21:26:39.8 & 8.52 & 5.30 & 0.0013 & -3 & 0.70 & 302 & 47 & 28.71 $\pm$ 0.58 & 1.63 $\pm$ 0.05 & -2.15 $\pm$ 0.06 \\
3 & ESO410-G005 & 00:15:31.56 & -32:10:47.8 & 7.14 & 1.90 & 0.0005 & -1 & 0.74 & 308 & 46 & 38.76 $\pm$ 1.84 & 0.86 $\pm$ 0.06 & -4.04 $\pm$ 0.11 \\
4 & ESO294-G010 & 00:26:33.37 & -41:51:19.1 & 6.79 & 1.90 & 0.0004 & -3 & 0.62 & 0 & 54 & 36.61 $\pm$ 3.83 & 1.78 $\pm$ 0.16 & -3.94 $\pm$ 0.12 \\
5 & IC1574 & 00:43:03.82 & -22:14:48.8 & 7.70 & 4.92 & 0.0012 & 10 & 0.61 & 0 & 53 & 41.09 $\pm$ 3.05 & 1.14 $\pm$ 0.13 & -2.52 $\pm$ 0.05 \\
6 & ESO540-G030 & 00:49:20.96 & -18:04:31.5 & 6.95 & 3.40 & 0.0007 & -1 & 0.88 & 0 & 30 & 84.00 $\pm$ 37.58 & 1.00 $\pm$ 0.38 & -3.94 $\pm$ 0.15 \\
7 & UGCA015 & 00:49:49.20 & -21:00:54.0 & 6.92 & 3.34 & 0.0010 & 10 & 0.52 & 28 & 65 & 25.27 $\pm$ 1.06 & 0.47 $\pm$ 0.10 & -2.95 $\pm$ 0.06 \\
8 & ESO540-G032 & 00:50:24.32 & -19:54:24.2 & 6.78 & 3.40 & 0.0008 & -3 & 0.91 & 0 & 25 & 36.40 $\pm$ 4.70 & 0.77 $\pm$ 0.14 & -4.04 $\pm$ 0.10 \\
9 & UGC00668 & 01:04:47.79 & +02:07:04.0 & 7.26 & 0.65 & -0.0008 & 10 & 0.80 & 60 & 38 & 155.84 $\pm$ 8.97 & 0.47 $\pm$ 0.13 & -2.35 $\pm$ 0.05 \\
10 & UGC00685 & 01:07:22.44 & +16:41:04.4 & 7.95 & 4.70 & 0.0005 & 9 & 0.82 & 122 & 35 & 36.52 $\pm$ 2.07 & 1.30 $\pm$ 0.12 & -2.42 $\pm$ 0.05 \\
11 & UGC00695 & 01:07:46.44 & +01:03:49.2 & 8.21 & 10.68 & 0.0022 & 6 & 0.84 & 0 & 32 & 29.01 $\pm$ 3.28 & 1.42 $\pm$ 0.21 & -2.02 $\pm$ 0.05 \\
12 & UGC00891 & 01:21:18.90 & +12:24:43.0 & 8.39 & 11.10 & 0.0021 & 9 & 0.61 & 42 & 52 & 25.68 $\pm$ 1.31 & 1.26 $\pm$ 0.14 & -1.83 $\pm$ 0.05 \\
13 & UGC01056 & 01:28:47.26 & +16:41:17.2 & 8.32 & 10.57 & 0.0020 & 10 & 0.94 & 0 & 20 & 20.63 $\pm$ 1.51 & 1.34 $\pm$ 0.19 & -1.97 $\pm$ 0.05 \\
14 & UGC01104 & 01:32:42.53 & +18:19:01.6 & 8.21 & 7.50 & 0.0023 & 9 & 0.62 & 0 & 52 & 17.89 $\pm$ 0.80 & 1.45 $\pm$ 0.12 & -1.89 $\pm$ 0.05 \\
15 & NGC0625 & 01:35:04.63 & -41:26:10.3 & 8.86 & 4.07 & 0.0014 & 9 & 0.51 & 90 & 59 & 76.66 $\pm$ 21.52 & 1.81 $\pm$ 0.51 & -1.25 $\pm$ 0.05 \\
16 & UGC01176 & 01:40:09.90 & +15:54:17.0 & 8.37 & 9.00 & 0.0021 & 10 & 0.83 & 25 & 33 & 70.46 $\pm$ 10.93 & 0.98 $\pm$ 0.18 & -1.63 $\pm$ 0.12 \\
17 & ESO245-G005 & 01:45:03.74 & -43:35:52.9 & 8.18 & 4.43 & 0.0013 & 10 & 0.71 & 318 & 45 & 83.20 $\pm$ 4.75 & 0.85 $\pm$ 0.10 & -1.56 $\pm$ 0.05 \\
18 & ESO245-G007 & 01:51:06.34 & -44:26:40.9 & 6.07 & 0.44 & -0.0001 & 10 & 0.83 & 0 & 44 & 78.44 $\pm$ 5.91 & 0.50 $\pm$ 0.11 & -4.49 $\pm$ 0.20 \\
19 & ESO115-G021 & 02:37:47.28 & -61:20:12.1 & 8.20 & 4.99 & 0.0017 & 8 & 0.28 & 221 & 74 & 43.67 $\pm$ 0.89 & 0.69 $\pm$ 0.05 & -1.84 $\pm$ 0.05 \\
20 & ESO154-G023 & 02:56:50.38 & -54:34:17.1 & 8.64 & 5.76 & 0.0019 & 8 & 0.51 & 39 & 59 & 103.33 $\pm$ 16.01 & 1.34 $\pm$ 0.24 & -1.37 $\pm$ 0.05 \\
21 & NGC1311 & 03:20:06.96 & -52:11:07.9 & 8.39 & 5.45 & 0.0019 & 9 & 0.47 & 36 & 62 & 32.83 $\pm$ 1.22 & 1.26 $\pm$ 0.10 & -1.82 $\pm$ 0.05 \\
22 & UGC02716 & 03:24:07.20 & +17:45:12.0 & 8.21 & 6.38 & 0.0013 & 8 & 0.71 & 90 & 45 & 42.36 $\pm$ 3.48 & 1.47 $\pm$ 0.13 & -2.15 $\pm$ 0.06 \\
23 & IC1959 & 03:33:12.59 & -50:24:51.3 & 8.39 & 6.06 & 0.0021 & 9 & 0.45 & 330 & 63 & 24.95 $\pm$ 0.94 & 1.24 $\pm$ 0.10 & -1.55 $\pm$ 0.05 \\
24 & NGC1510 & 04:03:32.64 & -43:24:00.4 & 8.77 & 10.08 & 0.0030 & -2 & 0.97 & 0 & 15 & 18.28 $\pm$ 1.28 & 2.13 $\pm$ 0.14 & -1.17 $\pm$ 0.05 \\
25 & NGC1522 & 04:06:07.92 & -52:40:06.3 & 8.53 & 9.54 & 0.0030 & 11 & 0.66 & 37 & 49 & 18.09 $\pm$ 1.69 & 1.85 $\pm$ 0.21 & -1.30 $\pm$ 0.05 \\
26 & ESO483-G013 & 04:12:41.12 & -23:09:32.0 & 8.81 & 10.68 & 0.0027 & -3 & 0.72 & 322 & 46 & 26.32 $\pm$ 2.30 & 2.03 $\pm$ 0.18 & -1.53 $\pm$ 0.06 \\
27 & ESO158-G003 & 04:46:17.28 & -57:20:37.6 & 8.76 & 10.20 & 0.0040 & 9 & 0.86 & 0 & 30 & 30.01 $\pm$ 1.72 & 0.98 $\pm$ 0.15 & -1.50 $\pm$ 0.05 \\
28 & ESO119-G016 & 04:51:29.20 & -61:39:03.4 & 8.40 & 10.08 & 0.0032 & 10 & 0.52 & 26 & 58 & 53.61 $\pm$ 10.88 & 1.24 $\pm$ 0.28 & -1.79 $\pm$ 0.07 \\
29 & NGC1705 & 04:54:13.50 & -53:21:39.8 & 8.39 & 5.10 & 0.0021 & 11 & 0.72 & 220 & 44 & 19.15 $\pm$ 0.38 & 2.08 $\pm$ 0.04 & -1.31 $\pm$ 0.05 \\
30 & ESO486-G021 & 05:03:19.69 & -25:25:22.5 & 8.26 & 9.11 & 0.0029 & 2 & 0.86 & 90 & 30 & 14.42 $\pm$ 1.10 & 1.32 $\pm$ 0.20 & -1.59 $\pm$ 0.05 \\
31 & NGC1800 & 05:06:25.72 & -31:57:15.2 & 8.88 & 8.44 & 0.0027 & 9 & 0.71 & 107 & 44 & 22.45 $\pm$ 1.00 & 1.63 $\pm$ 0.11 & -1.32 $\pm$ 0.05 \\
32 & UGCA106 & 05:11:59.32 & -32:58:21.4 & 8.96 & 10.01 & 0.0031 & 9 & 0.76 & 14 & 40 & 64.97 $\pm$ 2.02 & 0.86 $\pm$ 0.07 & -0.98 $\pm$ 0.05 \\
33 & CGCG035-007 & 09:34:44.72 & +06:25:31.8 & 7.60 & 4.92 & 0.0018 & 5 & 0.76 & 63 & 41 & 17.38 $\pm$ 1.41 & 1.69 $\pm$ 0.19 & -2.77 $\pm$ 0.05 \\
34 & IC0559 & 09:44:43.89 & +09:36:54.0 & 7.84 & 10.00 & 0.0018 & 5 & 0.92 & 63 & 23 & 25.37 $\pm$ 1.64 & 1.49 $\pm$ 0.14 & -2.56 $\pm$ 0.05 \\
35 & UGC05288 & 09:51:17.00 & +07:49:39.0 & 8.14 & 11.40 & 0.0019 & 8 & 0.86 & 331 & 30 & 26.71 $\pm$ 1.71 & 1.56 $\pm$ 0.15 & -1.98 $\pm$ 0.05 \\
36 & UGC05373 & 10:00:00.10 & +05:19:56.0 & 7.56 & 1.44 & 0.0010 & 10 & 0.80 & 90 & 37 & 93.69 $\pm$ 8.07 & 0.86 $\pm$ 0.12 & -2.54 $\pm$ 0.06 \\
37 & UGCA193 & 10:02:36.00 & -06:00:49.0 & 8.28 & 9.70 & 0.0022 & 7 & 0.24 & 14 & 77 & 27.18 $\pm$ 1.85 & 1.44 $\pm$ 0.15 & -1.93 $\pm$ 0.05 \\
38 & NGC3109 & 10:03:06.88 & -26:09:34.5 & 8.28 & 1.34 & 0.0013 & 9 & 0.22 & 91 & 78 & 132.72 $\pm$ 16.78 & 0.91 $\pm$ 0.20 & -1.70 $\pm$ 0.19 \\
39 & AM1001-270 & 10:04:04.10 & -27:19:51.6 & 6.29 & 1.30 & 0.0012 & 10 & 0.57 & 319 & 78 & 58.73 $\pm$ 24.66 & 0.95 $\pm$ 0.40 & -3.63 $\pm$ 0.50 \\
40 & UGC05456 & 10:07:19.64 & +10:21:42.5 & 7.83 & 10.50 & 0.0018 & 5 & 0.75 & 322 & 41 & 24.11 $\pm$ 1.04 & 1.09 $\pm$ 0.12 & -2.11 $\pm$ 0.05 \\
41 & SextansA & 10:11:00.80 & -04:41:34.0 & 7.35 & 1.46 & 0.0011 & 10 & 0.77 & 35 & 40 & 81.52 $\pm$ 7.15 & 0.42 $\pm$ 0.21 & -2.18 $\pm$ 0.05 \\
42 & UGC05797 & 10:39:25.18 & +01:43:06.8 & 8.03 & 10.40 & 0.0024 & 10 & 0.98 & 0 & 11 & 32.70 $\pm$ 4.02 & 1.70 $\pm$ 0.22 & -2.43 $\pm$ 0.05 \\
43 & UGC05889 & 10:47:22.30 & +14:04:10.0 & 8.61 & 7.73 & 0.0019 & 9 & 0.94 & 0 & 19 & 42.19 $\pm$ 1.75 & 1.06 $\pm$ 0.10 & -1.93 $\pm$ 0.05 \\
44 & UGC05923 & 10:49:07.57 & +06:55:02.2 & 8.24 & 7.33 & 0.0024 & 0 & 0.62 & 353 & 52 & 8.80 $\pm$ 0.24 & 1.43 $\pm$ 0.07 & -2.41 $\pm$ 0.05 \\
45 & UGC06457 & 11:27:12.25 & -00:59:40.7 & 8.31 & 10.53 & 0.0032 & 10 & 0.74 & 19 & 42 & 29.94 $\pm$ 2.07 & 1.60 $\pm$ 0.12 & -1.91 $\pm$ 0.06 \\
46 & ESO321-G014 & 12:13:49.62 & -38:13:52.9 & 7.26 & 3.20 & 0.0020 & 10 & 0.61 & 22 & 54 & 34.58 $\pm$ 2.33 & 1.02 $\pm$ 0.13 & -2.97 $\pm$ 0.15 \\
47 & ISZ399 & 12:19:59.51 & -17:23:31.0 & 8.81 & 15.94 & 0.0045 & 11 & 0.77 & 314 & 39 & 9.26 $\pm$ 0.72 & 2.16 $\pm$ 0.18 & -- \\
48 & UGC08091 & 12:58:40.44 & +14:13:03.0 & 6.69 & 2.13 & 0.0007 & 10 & 0.74 & 32 & 48 & 62.47 $\pm$ 17.65 & 2.17 $\pm$ 0.29 & -2.83 $\pm$ 0.05 \\
49 & UGCA319 & 13:02:14.39 & -17:14:15.1 & 8.03 & 5.70 & 0.0025 & 9 & 0.69 & 24 & 46 & 26.79 $\pm$ 0.85 & 0.96 $\pm$ 0.06 & -2.12 $\pm$ 0.05 \\
50 & UGCA320 & 13:03:16.74 & -17:25:22.9 & 8.52 & 6.03 & 0.0025 & 9 & 0.33 & 114 & 71 & 50.34 $\pm$ 3.67 & 1.21 $\pm$ 0.19 & -1.24 $\pm$ 0.09 \\
51 & MCG-03-34-002 & 13:07:56.65 & -16:41:20.9 & 8.41 & 7.90 & 0.0031 & 4 & 0.62 & 320 & 52 & 18.12 $\pm$ 1.12 & 1.89 $\pm$ 0.12 & -1.79 $\pm$ 0.05 \\
52 & IC4247 & 13:26:44.43 & -30:21:44.7 & 7.74 & 4.97 & 0.0014 & 2 & 0.60 & 333 & 53 & 19.60 $\pm$ 0.45 & 1.07 $\pm$ 0.06 & -2.38 $\pm$ 0.05 \\
53 & ESO444-G084 & 13:37:19.99 & -28:02:42.0 & 7.40 & 4.61 & 0.0020 & 10 & 0.70 & 310 & 47 & 31.77 $\pm$ 2.04 & 0.85 $\pm$ 0.10 & -2.47 $\pm$ 0.07 \\
54 & NGC5253 & 13:39:55.96 & -31:38:24.4 & 8.94 & 3.15 & 0.0013 & 11 & 0.76 & 44 & 40 & 34.98 $\pm$ 0.72 & 2.28 $\pm$ 0.04 & -0.60 $\pm$ 0.05 \\
55 & NGC5264 & 13:41:36.68 & -29:54:47.1 & 8.58 & 4.53 & 0.0016 & 9 & 0.84 & 66 & 32 & 52.43 $\pm$ 2.08 & 1.08 $\pm$ 0.09 & -1.98 $\pm$ 0.05 \\
56 & KKH086 & 13:54:33.55 & +04:14:34.8 & 6.49 & 2.60 & 0.0010 & 10 & 0.63 & 0 & 61 & 29.52 $\pm$ 4.37 & 1.01 $\pm$ 0.24 & -4.24 $\pm$ 0.16 \\
57 & IC4951 & 20:09:31.77 & -61:51:01.7 & 8.49 & 9.57 & 0.0026 & 8 & 0.41 & 355 & 66 & 31.24 $\pm$ 3.15 & 1.64 $\pm$ 0.19 & -1.53 $\pm$ 0.05 \\
58 & DDO210 & 20:46:51.81 & -12:50:52.5 & 5.97 & 0.94 & -0.0005 & 10 & 0.48 & 103 & 90 & 80.00 $\pm$ 42.20 & 1.64 $\pm$ 0.46 & -3.96 $\pm$ 0.05 \\
59 & NGC7064 & 21:29:02.98 & -52:46:03.4 & 8.91 & 9.69 & 0.0027 & 5 & 0.32 & 90 & 71 & 31.46 $\pm$ 3.48 & 1.31 $\pm$ 0.23 & -1.05 $\pm$ 0.05 \\
60 & IC5256 & 22:49:45.81 & -68:41:26.4 & 8.52 & 49.88 & 0.0127 & 8 & 0.61 & 22 & 52 & 10.46 $\pm$ 0.67 & 1.59 $\pm$ 0.16 & -1.84 $\pm$ 0.05 \\
61 & UGCA438 & 23:26:27.52 & -32:23:19.5 & 7.20 & 2.22 & 0.0002 & 10 & 0.86 & 0 & 31 & 33.50 $\pm$ 4.01 & 0.98 $\pm$ 0.32 & -2.72 $\pm$ 0.05 \\
62 & ESO347-G017 & 23:26:56.21 & -37:20:48.9 & 8.43 & 9.63 & 0.0023 & 9 & 0.61 & 90 & 52 & 32.86 $\pm$ 2.12 & 1.31 $\pm$ 0.15 & -1.68 $\pm$ 0.05 \\
63 & UGC12613 & 23:28:36.25 & +14:44:34.5 & 7.05 & 0.98 & -0.0006 & 10 & 0.58 & 113 & 58 & 88.06 $\pm$ 2.59 & 0.31 $\pm$ 0.07 & -3.83 $\pm$ 0.06 \\
64 & UGCA442 & 23:43:45.55 & -31:57:24.4 & 7.86 & 4.27 & 0.0009 & 9 & 0.48 & 43 & 62 & 54.98 $\pm$ 7.01 & 1.26 $\pm$ 0.18 & -2.07 $\pm$ 0.05 \\
65 & ESO149-G003 & 23:52:02.80 & -52:34:39.8 & 7.71 & 7.01 & 0.0020 & 10 & 0.40 & 332 & 68 & 50.99 $\pm$ 14.77 & 1.84 $\pm$ 0.40 & -2.07 $\pm$ 0.05 \\
\hline
\end{tabular}
\end{flushleft}
\caption{(1) The index of sample, ranged by RA; (2) the names of targets; (3) $\sim$ (4) right ascension and declination of galaxies from Dale et al. (2009); (5) stellar mass; (6) distance; (7) spectroscopic redshift; (8) $\sim$ (10) hubble type, major-to-minor axis ratio and position angle taken from Dale et al. (2009); (11) inclination angles; (12) $\sim$ (13) half-light radius and S\'ersic index; (14) \texttt{GALEX} UV-based SFR.}
\label{Tab:sample}
\end{table*}

\section{Overall Scientific Goals}\label{sec:goals}

The DGIS aims to study dwarf galaxies in detail through its high spatial resolution and high S/N data. Its overall science goals are composed of three parts:

{\bf (1) Baryonic cycles in dwarf galaxies: } Cycles between different baryonic components are the driving mechanisms for galaxy formation and evolution: gas collapses into stars, causing star formation processes; stars pollute the ISM by producing metals; the mix of metals and dust with gas regulates cooling efficiency; the CGM freshes the gas reservoir and dilutes the metallicity in galaxies; outflows/radiation from stars and central massive black holes influence the ISM and CGM in various aspects; radial migration and interaction with satellites reshape the morphologies and kinematics of gas and stars \citep[see reviews,][]{Tumlinson17,Peroux20}. These processes have been extensively studied in massive galaxies, demonstrating that a galaxy is a complicated ecosystem that can only be fully understood by studying all these processes.\citep[e.g.,][]{Tacconi13,Ford16,Kreckel19,Ginolfi20,Donahue22, Tortora22}. DGIS will advance studies of dwarf galaxies by enabling measurements of various galactic components, as well as their spatial distribution and kinematics, by combining with its rich ancillary data. Combined with IFU data for massive galaxies available in the literature, DGIS will enable a more complete understanding of the physical processes driving galaxy formation and evolution across a wide stellar mass scale from 10$^{6}$ to 10$^{11}$ M$_{\odot}$.

{\bf (2) Intermediate massive black holes in dwarf galaxies:} Massive black holes born in the early Universe left relics in dwarf galaxies today.  Instead of sinking to the galactic center, they could exist across the galaxy given their long dynamical timescale.  The population of these IMBHs and their physical properties offer crucial insights into the formation mechanisms of SMBH seeds and their growth \citep{Latif16,Reines22,Matteo23}. Hunting for these IMBHs has been extremely difficult because (i) their bolometric luminosity is of the same order of magnitude as that of massive evolved stars; (ii) light pollution from ambient stellar emission significantly dilutes the radiation from IMBHs; (iii) searching over the galaxy body is more observationally challenging than those with single-slit/fiber observations \citep{Reines22}. Currently, through various methods, roughly 1\% of dwarf galaxies within the stellar mass range of DGIS are found to host active BHs, with almost all of them located at galactic centers \citep{Greene07, Ho08, Shi08, Reines13}. DGIS has been designed to maximize the success of IMBH hunting in dwarfs by covering a significant fraction of the galaxy body (roughly to $R_{e}$), providing high spatial resolution combined with high S/N to eliminate stellar dilution significantly, and offering multiple diagnostics (BPT diagram, He II, broad emission line) to identify active BH accretion.

{\bf (3) The inner density profile of dark matter: } While the $\Lambda$CDM has achieved tremendous success in matching the observations at the large scale of the Universe, there are distinct differences at small scales. The inner profile of dark matter is one of them, with core profiles preferred in observations versus cuspy profiles predicted in simulations. Because dwarf galaxies are dominated by dark matter even at their centers, they have been ideal laboratories for understanding the above core/cuspy problem \cite{Adams14, Oh15, Karukes17}.  To have a reliable measurement of the inner profile, one needs to probe the kinematics within the central 100 pc, which requires a spatial resolution as small as tens of parsec. As shown in the collections of rotation curves of dwarf galaxies by \cite{Karukes17}, only a few objects have kinematic measurements within the central 100 pc. Our proposed observations will remedy this situation by increasing the sample size by a few factors.

\section{Observations} \label{sec:observation}

Observations of the DGIS sample started in September 2019. They were carried out with two facilities: 70\% with VLT/MUSE and the remaining with WiFeS on the ANU-2.3m telescope (PI: Y. Shi, Co-PI: F. Bian). The configurations are listed in Table~\ref{Tab:teles}.

\begin{table*}
\begin{center}
\caption{The basic information of the DGIS project. }
\label{Tab:teles}
\begin{tabular}{lll}
\hline
Properties          &    WiFeS              & MUSE \\
\hline
Number of Galaxies    & 15$^{\ast}$ (20 in total)    & 40$^{\ast}$ (45 in total) \\
Wavelength Range      & 3500 - 9000 \AA$^{\star}$    & 4650 - 9300 \AA\ \\
Wavelength Resolution & 7000                & 2000-3000  \\
On-source Time & $\sim$ 3-13 hrs$^{\dagger}$    & $\sim$ 0.6-4 hrs$^{\ddagger}$ \\
FOV                   & 25{\arcsec}$\times$38{\arcsec}      & 60{\arcsec}$\times$60{\arcsec} \\
Median Seeing         & 1.5{\arcsec}                & 0.8{\arcsec}           \\
\hline
\end{tabular}\\
$^{\ast}$The number of objects whose observations have been carried out.\\
$^{\star}$The edge wavelengths are thrown away during data processing.\\
$^{\dagger}$The total exposure time on target including both blue- and red-end exposures.\\
$^{\ddagger}$The total exposure time on target including multiple pointings.
\end{center}
\end{table*}

\subsection{WiFeS Observations} \label{subsec:WiFeS_obs}

WiFeS is an integral field, double-beam, concentric, image-slicing spectrograph mounted on the 2.3 m telescope at Siding Spring Observatory (SSO). The high-resolution grisms, including U7000, B7000, R7000, and I7000, are employed to cover the full optical wavelength range from 329 to 912 nm, with a resolution of about 7000. It provides 25 slitlets, each 38{\arcsec} long and 1{\arcsec} wide, offering a FOV of 25{\arcsec} $\times$ 38{\arcsec} and a pixel scale of 1{\arcsec} $\times$ 0.5{\arcsec} \citep[see][for details]{Dopita07,Dopita10}.

Each source is observed with a single pointing. Each exposure lasts 20 to 30 minutes, followed by an off-target sky exposure of 5 to 10 minutes. A standard star is observed every 1 to 2 hours. Only exposures taken under clear weather conditions are included in the final data products. The final integrated on-source exposure time per spatial pixel per wavelength grid is between 2 and 7 hrs, as listed in Table~\ref{Tab:spectra}. Among them, the U7000 and R7000 are observed simultaneously with the configuration of RT480, and the B7000 and I7000 are observed simultaneously with the RT615. Spatial binning of $1 \times 2$ is applied to a final pixel scale of 1{\arcsec} $\times$ 1{\arcsec}.

WiFeS observations are 80\% complete. The basic information of the observations are listed in Table~\ref{Tab:teles}.

\subsection{MUSE Observations} \label{subsec:MUSE_obs}
MUSE is a second-generation instrument of the VLT. It has 24 integral field units (IFU). In Wide Field Mode (WFM)-NOAO-E mode, it covers the optical wavelength range from 465 to 930 nm with a spectral resolution of 2800 at 6563 \AA, a FOV of 60{\arcsec}$\times$60{\arcsec}, and a pixel size of 0.2{\arcsec} $\times$ 0.2{\arcsec} \citep[see][for details]{Bacon10}. The final datacube has a binned spaxel size of 0.4{\arcsec} $\times$ 0.4{\arcsec} and an average seeing of around 0.8{\arcsec}. 

Most sources have a single pointing, except for WLM with 4 pointings, UGC00668 with 9 pointings, and UGC12613 with 3 pointings. The total on-source exposure time per spaxel for most sources ranges from 2200 to 4800 seconds, as listed in Table~\ref{Tab:spectra}. This time is divided into several exposures, each lasting between 5 and 20 minutes, followed by an off-target sky exposure of around 1 to 2 minutes.

So far, MUSE observations are 88\% complete, with 40/45 galaxies observed. The basic information about these observations are also listed in Table~\ref{Tab:teles}.

\section{Data Reduction} \label{sec:data_reduction}
\subsection{WiFeS data reduction} \label{subsec:WiFeS_reduction}
The observed single frame datacubes were generated and pre-processed using the \texttt{PYTHON} package \texttt{pyWiFeS} \citep{Childress14}. This process includes cosmic ray removal, wavelength calibration, telluric correction, atmospheric extinction correction, and flux calibration using standard stars. The reduced individual exposure frames were further post-processed to produce the final mosaic data cubes, as described below.

(1) The barycentric correction was done for the wavelength solution of each exposure frame through \texttt{astropy.AP}. We then visually checked the relative wavelength calibration among frames through strong skylines and galaxy emission lines.

(2) We checked whether the WiFeS pipeline produced the correct variance array by checking the sky exposures, where the pipeline-produced errors should be equal to the standard deviations of their fluxes over the FOV at each wavelength. The ratio of the two was found to be around unity with a scatter of about 20\%.
 
(3) For each target frame, the sky frame observed right before or after the target was used for sky subtraction. At each wavelength, the mean sky was estimated from the sky frame through the \texttt{PYTHON} code \texttt{mmm.mmm} and then was subtracted from the target frame. The error of the mean sky was then added quadratically to the error array of the target. We further masked the wavelength regions of strong skylines by identifying wavelength grids that were 10-$\sigma$ above the sky continuum in the sky datacubes. In practice, the mask was done by multiplying the variance at these wavelengths with -1.
 
(4) For a given grism, the sky-subtracted target frames obtained at different epochs were first reprojected to a common wavelength grid in a flux-conserving manner. The mean of these frames was then adopted as the final combined spectrum for this grism.

(5) To combine different grisms, a wavelength grid that starts at 3500 \AA\, and ends at 9000 \AA\ with a constant interval in logarithm of 1/7000/2/ln(10) was created. All grisms were then projected onto this  wavelength grid with flux conservation. U7000 and I7000 were normalized to B7000 and R7000, respectively.  These two were then further normalized to each other. Note that we will further calibrate the flux based on the broad-band images so that the exact grism for the flux normalization does not matter.

(6) For the final combined datacube, an initial world coordinate system (WCS) was added using the input position and PA from the observation. The datacube was then rotated to have north up and east to the left.

(7) A synthetic broad-band image in the $r$ band was created from the datacube and then compared to the  observed broad-band image from SkyMapper \citep{Onken24} to further update the WCS. The final accuracy of WCS is $\lesssim$ 1 arcsecond, limited by the FOV or the lack of point sources in the field.

(8) For the final flux calibration, we produced synthetic $v$, $g$, $r$, and $i$ images from the datacube and measured their synthetic photometry within large apertures. We then compared this synthetic photometry with measurements from SkyMapper broad-band images taken in the same bands and with the same apertures. The filter curves for the four broad bands adopted cover the spectral range of our WiFeS spectra. We fitted a double power-law flux calibration curve to the relative flux ratios in bands detected in the synthetic and real broad-band images. 

(9) Finally, the datacubes were corrected for the Milky Way (MW) foreground extinction by applying the extinction curve from \cite{Cardelli89} with $R_{\rm v}=3.1$. The E(B-V) were from \cite{Dale09}, which they adopted from \cite{Schlegel98}.

\subsection{MUSE data reduction} \label{subsec:MUSE_reduction}
(1) The MUSE pipeline \citep{Weilbacher20} automatically reduces the raw data under the workflow engine of the $\texttt{esoreflex}$ environment \citep{Freudling13}. Briefly, the pipeline corrects for bias, dark current, flat-field, and sky-flat. It further calibrates wavelengths by arc lamps, flux by standard stars, geometric and astrometric coordinates, as well as atmospheric and telluric extinction. For each frame, a sky background is observed for an off-target blank sky.

(2) Residual sky removal for individual frames: Our MUSE observations were carried out under all weather conditions. To improve the flux calibration, we carried out further post-processing on the individual frames. CGCG035-007 is selected as an example, which is illustrated in the Appendix.~\ref{sec:appendix_calib}. 

Firstly, to account for possible variations in sky brightness between targets and off-target sky pointings, we used broad-band images to guide additional sky subtraction in the target frames. The broad-band images were collected from the data archives of DESI \citep{Dey19}, Pan-STARRS \citep{Chambers16}, or SkyMapper \citep{Onken24}. Given that DESI images have deep exposure, good S/N, and cover most of our targets, flux calibration was primarily based on DESI, then Pan-STARRS and SkyMapper. The used reference broad-band images are listed in Table~\ref{Tab:calib}. Based on the white images of each exposure produced by the MUSE pipeline, the WCS of all frames were matched to the $r$-band images by aligning (like-) point sources, then they were reprojected to a common frame by \texttt{PYTHON} package of \texttt{reproject.reproject\_interp}. 

Based on IRAC 3.6 $\mu$m images, broad $r$-band images, and MUSE white images, we selected several clean backgrounds and subtracted their median spectra from the entire datacube. This process helps reduce residual sky after the off-target sky subtraction performed by the pipeline. The standard deviations of the subtracted median sky spectra were added to the errors of datacubes. If no clean background was available within the FOV, this step was skipped, as labeled in Table~\ref{Tab:calib}. For some galaxies where diffuse optical light fills the entire FOV, we still performed background subtraction, following the principle of choosing the lesser of two evils.

(3) Calibration of the spectral shape of individual frames: For each target frame, we selected an aperture covering the main body of the target within the MUSE FOV and used the median spectra within this aperture as the integrated spectrum. We found that each frame still showed discrepancies in absolute flux and, particularly towards the blue end, variations in spectral shape, as shown in Fig.~\ref{fig:subsky}. To correct the former, we convolved the integrated spectrum of each frame with the filter curves from the DESI, Pan-STARRS, or SkyMapper surveys to obtain synthetic photometry. This synthetic photometry was then compared to the observed photometry from broad-band images to correct the absolute flux.

To calibrate the spectral shape, we first discarded frames with either negative fluxes over a significant portion of the wavelength range or highly complex spectral shapes. For the remaining frames, we selected the frame with a smooth median spectrum and a synthetic broad-band color similar to the color obtained from broadband imaging data as the reference. Each spectrum was smoothed by calculating the median over wavelength bins of 150 \AA. The ratio of each spectrum to the reference was fitted with a 3rd-order polynomial function. The fitting wavelength range stopped at 8800 \AA\ to avoid strong skylines but still covered the Ca{\sc ii} triplet. We visually inspected each fit, and when the fitted polynomial showed distorted tails, a simple linear function was used instead, as detailed in Table~\ref{Tab:calib}. Each pixel of every frame was then scaled by the corresponding fitted ratios, and these ratios were also applied to the errors to maintain the S/N. Finally, we combined all frames to produce the final datacube, weighting by exposure time.

(4) Final Flux Calibration: We produced synthetic photometry from the final datacube and compared it to the broad-band images within a large aperture that covered the main body of the galaxy. The ratio between the two was fitted with a 3rd-order polynomial function or, if the polynomial resulted in distorted tails, a linear function. This fitted ratio was then applied to all pixels to achieve the final flux calibration.

(5) The datacubes were 2$\times$2-pixel rebinned, resulting in a final spatial sampling of 0.4{\arcsec} $\times$ 0.4{\arcsec}. This process promotes the S/N of each single spectrum. Besides, the wavelengths were also resampled to a constant interval in logarithm of 1/2500/2/ln(10).

(6) The WCS was corrected by comparing the position of the brightest point source in the MUSE white image with that in the Pan-STARRS, DESI, or SkyMapper images, again.

(7) Finally, the MW foreground extinction was corrected with  \cite{Cardelli89} extinction curve
and adopted E(B-V) from \cite{Dale09}.

\subsection{Spectral Analysis of Global Spectra}\label{subsec:Spec_analy}
In this work, we stacked all spectra with a continuum S/N higher than 0.5 within 1 $R_{\rm e}$ of each galaxy. The determination of the $R_{\rm e}$ ellipse is described in the next section. If the FOV was smaller than 1 $R_{\rm e}$, we mosaicked all pixels that met our S/N criteria. For WLM, UGC00668, and UGC12613, which had multiple FOVs, we stacked all their single-pointing datacubes into one. For these global spectra:

(1) We used \texttt{pPXF} \citep[\texttt{Penalized PiXel-Fitting}, ][]{Cappellari17} code and adopted \texttt{HR-PYPOPSTAR} stellar population models \citep{Millan21}, together with gas kinematics and 3rd-order polynomial function (\texttt{degree=3, mdegree=-1}), to perform the spectral fitting. The \texttt{HR-PYPOPSTAR} library provides single stellar populations (SSP) with a wide range of ages (0.1 Myr $\sim$ 13.8 Gyr), four metallicity values (0.04, 0.08, 0.02, and 0.05), and four different initial mass function (IMF) parameters, but without accounting for the $\alpha$ enhancement. Their theoretical spectra cover wavelengths from 91 to 24000 \AA\ with a very high spectral resolution (50 000 at 5000 \AA\ ). To subtract the underlying continuum and improve fitting efficiency, we used stellar populations with a fixed metallicity of 0.008 Z$_{\odot}$, picked a part of stellar age, and with Chabrier IMF \citep{Chabrier03}. The fittings were performed up to 7500 \AA, and the model spectral resolutions were scaled to match the instruments. The spectral resolution of the model is sufficient to match that of WiFeS.

(2) The emitting spectra were obtained by subtracting the continuum fitted with \texttt{pPXF}. Each emission line was fitted with a single Gaussian kernel using the \texttt{scipy.optimize.curve\_fit} from \texttt{PYTHON}. To reduce contamination from nearby emission lines, the following were fitted together: [O{\sc iii}] $\lambda \lambda 4959,5007$, \ha\ and [N{\sc ii}] $\lambda \lambda 6548,6583$ doublet, and [S{\sc ii}] $\lambda \lambda 6716,6731$ doublet. To promote fitting efficiency for the weak one in doublets, the flux ratios were fixed: [O{\sc iii}] $\lambda \lambda 4959,5007$ to 1:3, [O{\sc i}] $\lambda \lambda 6300,6364$ to 3:1, and [N{\sc ii}] $\lambda \lambda 6548,6583$ to 1:3. The S/N ratio was calculated as the peak of the emission line divided by the standard deviation of its 100 \AA\ spectral windows on either side. Flux errors include both the uncertainties from \texttt{curve\_fit} and the propagation of errors from the datacube.

(3) Finally, we corrected for dust attenuation for each emission line. We adopted the averaged Small Magellanic Clouds (SMC)-bar extinction curve proposed by \cite{Gordon03}, where $R_{\rm v} = A_{\rm v}/E(B-V) \approx 2.74$, the theoretical flux ratio of \ha-to-\hb\ is 2.86 for Case B recombination, and E(B-V)$_{\rm star} = 0.4 \times$ E(B-V)$_{\rm gas}$. The SMC-bar extinction curve was chosen because galaxies with lower masses and shallower optical depths tend to have steeper attenuation curves \citep{Zhang17,Salim18}.

The stacked spectra and their errors, continuum fittings, gas emission line fittings from \texttt{pPXF}, and 3.6 $\mu$m images overlapped with observing pointings, FOVs, and $R_{e\rm }$ ellipses, are shown in Fig.~\ref{fig:spec}. For galaxies without observations, we only display their {\it IRAC} 3.6 $\mu$m images, $R_{\rm e}$ ellipses, and planned pointings and FOVs.

\section{High-level Data product} \label{sec:DataProduct}
We generated high-level data products for DGIS using the Galaxy IFU Spectroscopy Tool (\texttt{GIST}, \cite{Bittner19}) \footnote{\url{https://abittner.gitlab.io/thegistpipeline/documentation/generalRemarks/generalRemarks.html}}. The \texttt{PYTHON}-based \texttt{GIST} pipeline reads IFS datacube, performs spaxel binning using Voronoi, fits spectra with \texttt{pPXF}, fits emission/absorption lines with \texttt{GandALF} (Gas and Absorption Line Fitting software, \cite{Sarzi06}), and extracts stellar population properties and non-parametric SFH from \texttt{pPXF}. The pipeline provides spaxel- and/or bin-level data products, including stellar and gas kinematics, emission and absorption line fluxes, as well as stellar age, stellar metallicity, and SFH. Furthermore, \texttt{GIST} provides an interactive visualization tool \texttt{Mapviewer}, which facilitates the quick generation of various parameter maps and allows users to check pixel-specific spectra and their spectral fits. 

\begin{figure*}
\begin{center}
\includegraphics[width=\textwidth]{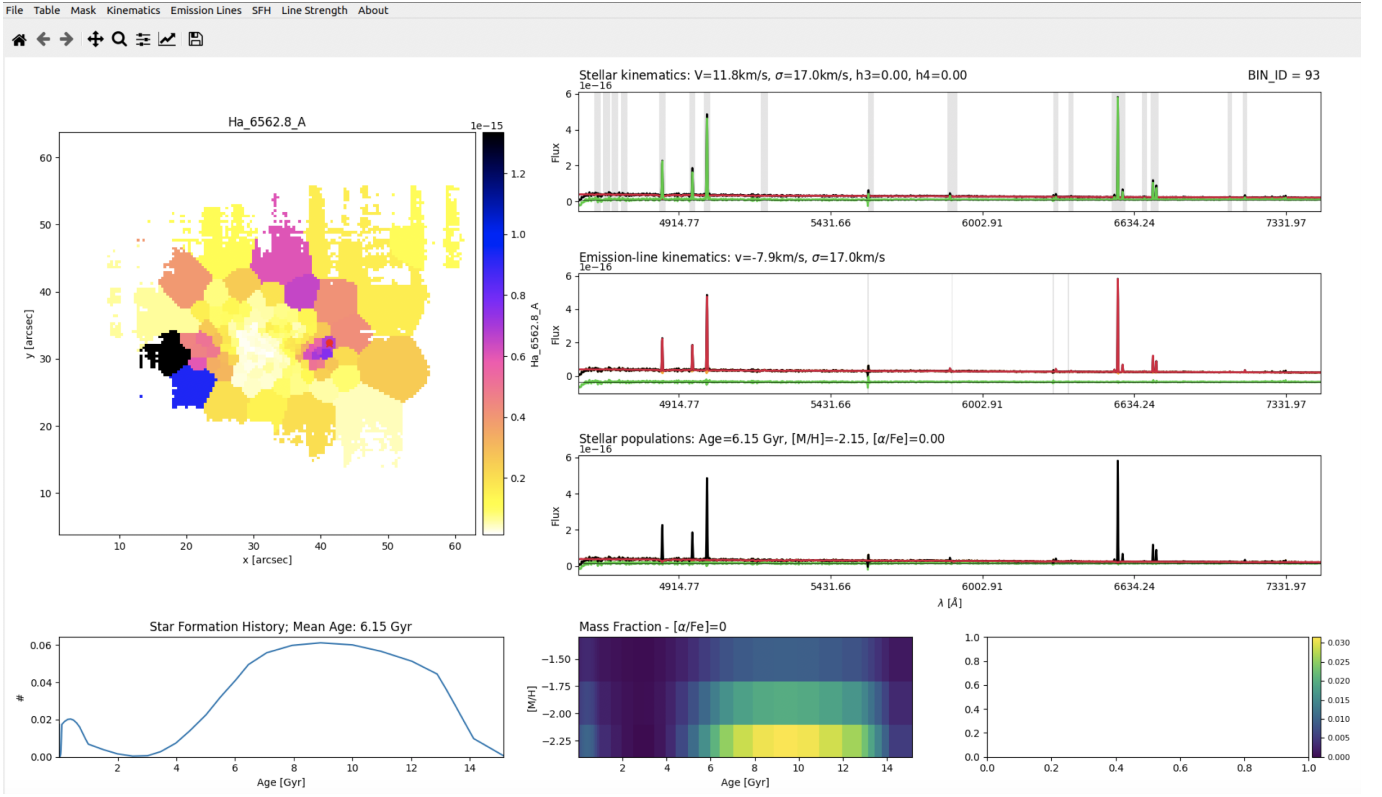}
\end{center}   
\caption{Data products of CGCG035-007 in `BIN' level, provided by \texttt{Mapviewer}.}
\label{fig:Mapviewer}
\end{figure*}

For spectral fitting, we employed the full \texttt{HR-PYPOPSTAR} stellar template with \cite{Chabrier03} IMF. Emission lines and skylines were masked during the continuum fitting process. Pixels with a continuum S/N lower than 1.5 were excluded, and an S/N of 20 was required after Voronoi binning. The pipeline was executed at both the SPAXEL and BIN levels. An example of the results displayed by \texttt{Mapviewer} is shown in Fig.~\ref{fig:Mapviewer}.

\section{Global Properties} \label{sec:global_properties}
As shown in Table~\ref{Tab:sample}, we list the global properties of our sample, ordered by their RA. The coordinates were obtained from \cite{Kennicutt08}. 

(1) {\bf Stellar masses ($M_{\ast}$)}: They are adopted from \cite{Cook14b}. These are derived from {\it Spitzer} 3.6 $\mu$m with a mass to light ratio $\Upsilon_{3.6{\mu}m}$=0.5 M$_{\odot}$/L$_{\odot,3.6{\mu}m}$, where L$_{\odot, 3.6{\mu}m} = 1.4 \times 10^{32}\, {\rm erg\,s^{-1}}$ \citep{Oh08}.

(2) {\bf Distance ($D$)}: They are taken from \cite{Kennicutt08} and updated to the values measured by the Tip of the red-giant branch (TRGB) from NED (updated to 2020). Distances measured by the Hubble flow and lacking additional redshift-independent measurements are recalculated based on \cite{Kennicutt08}, incorporating the updated Local Group velocity from NED and the cosmological parameters used in this work (H$_{\rm 0} = 73\,{\rm km/s/Mpc}$). Note that after updating distances, two objects -- ISZ 399 and IC 5256 -- are largely beyond our distance criteria of 11 Mpc. Despite this, they are still included in the analysis.

(3) {\bf Heliocentric redshifts ($z$)}: $z = {\rm c}z/{\rm c}$, where c$z$ are adopted from \cite{Dale09} and updated according to NED and \texttt{pPXF} fittings in this work. The c is the speed of light. 

(4) {\bf Hubble Type ($T$)}: They are collected from \cite{Dale09}, from the HyperLeda database.

(5) {\bf Position angle (PA) and axis ratio ($b/a$)}: They are adopted from \cite{Dale09}, whose apertures include all infrared (IR) emission lines. 

(6) {\bf Inclination angles ($i$)}: They are derived from the axis ratio \citep{Hubble1926}: $\cos{i} = \sqrt{\frac{(b/a)^2-q_{\rm 0}^2}{1-q_{\rm 0}^2}}$, where $q_{\rm 0}$ is the intrinsic axis ratio, and $i=90^{\circ}$ if $b/a < q_{\rm 0}$ \citep{Giovanelli94}. Based on \cite{Karachentsev17}, for $T\,\geqslant \,0$, we adopt $q_{\rm 0} = 5.128-1.114\log M_{\ast}+0.0612 (\log M_{\ast})^2$; for dwarf spherical galaxies with  $T\,<\,0$, we adopt $\log q_{\rm 0} = -0.43 - 0.053\times T$ from \cite{Paturel97}.

(7) {\bf S\'ersic index ($n$) and effective radius ($R_{\rm e}$)}: They are based on {\it Spitzer} 3.6 $\mu$m broad-band images. The post-processed 3.6 $\mu$m images from \cite{Dale09} have been masked for bad pixels and had the background sky subtracted \footnote{\url{https://irsa.ipac.caltech.edu/data/SPITZER/LVL/}}. According to \cite{Dale09}, the MW foreground extinction was corrected through the \cite{Cairos09} extinction curve, with $R_{\rm v} = 3.1$ and $E(B-V)$ from Table~1 of \cite{Dale09}.  Point sources were identified using the \texttt{PYTHON} package \texttt{photutils}, and foreground stars with a flux ratio of $f_{\rm 3.6 \mu m}/f_{\rm 4.8 \mu m} > 8$ were masked, as well as background extended objects with ellipticity larger than 0.5. Multiple co-centered elliptical rings were applied to the targets, with the position angle, axis ratios, and outermost radii defined by \cite{Dale09}. Each ring was corrected for the point-spread function (PSF) from Eq.~2 of \cite{Dale09}. The flux within each elliptical ring was measured to obtain flux gradients. Finally, the flux gradients were fitted with a single S\'ersic profile:$I(R) = I_{\rm e}\exp \left\{-b_{\rm n}[(\frac{R}{R_{\rm e}})^{1/n}-1]\right\}$, where $n$ is the S\'ersic index, $b_{\rm n} \approx 2n-1/3$, and $R_{\rm e}$ is the effective radius containing half the light of the whole galaxy.

(8) {\bf SFR(FUV)}: The {\it GALEX} FUV integrated photometry are taken from Table 3 of \cite{Lee11}, where apertures defined by \cite{Dale09} were used and the data were corrected for MW foreground extinction. FUV attenuation $A_{\rm FUV}$ was adopted from \cite{Cook14b} so the dust-corrected flux density to be calculated as $f_v({\rm FUV})= 10^{-0.4(m_{\rm AB}-A_{\rm FUV}+48.6)}$. \cite{Murphy11} proposed equations that transfer FUV luminosity ($L_{\rm FUV}$) to SFR(FUV) using Kroupa initial mass functions (IMF, \cite{Kroupa03}):
\begin{equation}
    {\rm SFR(FUV)} = 4.42 \times 10^{-44} [\frac{L_{\rm FUV}}{\rm erg\,s^{-1}}],
\label{eq:SFR(FUV)}
\end{equation}
where $L_{\rm FUV} = \nu L_{\rm \nu,FUV}$ at 1539 \AA. The errors of SFR(FUV) only include photometric uncertainties. 
\newline
\newline
Bellowing global spectroscopic properties are based on the associated stacked spectra and spectral fittings. We summarized these properties in Table~\ref{Tab:spectra}. 

(9) {\bf Point-Spread Function (PSF)}: The full width at half maximum (FWHM) of the MUSE $V$-band PSF are given by fitting a circular Moffat profile \citep{Moffat69} to a relatively isolated star in the FOV, according to \cite{Fusco20}. For galaxies without isolated stars, the median \texttt{DIMM SEEING} from all exposures in the observing night logs are used to represent the PSF, as shown in Fig.~\ref{fig:psf}. Since the \texttt{DIMM SEEING} are measured at the zenith, thus the $V$-band stellar FWHM were normalized to airmass=1, by dividing \texttt{AIRMASS}$^{3/5}$, according to the ESO user manual. The linear fitting is performed by the \texttt{python} package of \texttt{LtsFit} \citep{Cappellari13}. The fitted values were multiplied \texttt{AIRMASS}$^{3/5}$ to obtain their observing PSF, and the errors include the standard deviation of \texttt{DIMM SEEING}, the scatter of the fittings, and the \texttt{AIRMASS}. The sample observed by MUSE has a median FWHM of around 0.8{\arcsec}. For WiFeS observations, where the FOV is not large enough to cover a star for PSF measurement, the average seeing is adopted as the PSF, with a median FWHM of around 1.5{\arcsec}.

(10) {\bf SFR(\ha)}: The \ha-based SFR (SFR(\ha)) are derived from \cite{Murphy11} with Kroupa IMFs:
\begin{equation}
    {\rm SFR(H\alpha)} = 5.37 \times 10^{-42} [\frac{L_{\rm H\alpha}}{\rm erg\,s^{-1}}].
\label{eq:SFR(Ha)}
\end{equation}
The $L_{\rm H\alpha}$ are corrected for SMC dust attenuation as described in \ref{subsec:Spec_analy}, and the S/N of \ha\ flux should be higher than 7. The errors of SFR(\ha) are from flux uncertainties only.

(11) {\bf The gas-phase metallicity (12+log(O/H))}: The global metallicity are given by eight different calibrations: N2S2\ha\ from \cite{Dopita16} (hereafter, D16):
\begin{equation}
 \begin{aligned}
    {\rm 12+\log(O/H) = 8.77 + y + 0.45(y+0.3)^5,} \\
    {\rm y =  \log \frac{[NII]\lambda 6583}{[SII]\lambda \lambda 6716,31}+0.264 \log \frac{[NII]\lambda 6583}{H\alpha}},
\label{eq:D16}    
\end{aligned}   
\end{equation}
where $y$ is in the range of [$-1,\,0.5$]. Since this method is built by a photoionization model, an additional error of 0.2 dex is added apart from flux errors. 

The N2 calibrators from \cite{Curti20} by solving the equations (hereafter C20):
\begin{equation}
 \begin{aligned}
    & {\rm N2 = \log \frac{[NII]\lambda 6583}{H\alpha}}, \\
    & {{\rm N2} = -0.489+1.513x-2.554x^2-5.293x^3-2.867x^4,}    
\end{aligned}   
\end{equation}
where $x={\rm 12+\log(O/H)}-8.69$. The fitted errors and the dispersion are 0.16 dex and 0.1 dex. 

Calibrations with O3N2 and N2 ratios from \cite{Marino13} (hereafter M13):
\begin{equation}
 \begin{aligned}
    & {\rm O3N2 =  \log(\frac{[OIII]\lambda 5007}{H\beta}\times \frac{H\alpha}{[NII]\lambda 6583}),} \\
    & {\rm 12+\log(O/H) = 8.533[\pm 0.012]-0.214[\pm 0.012] \times O3N2,}
\label{eq:O3N2}    
\end{aligned}   
\end{equation}

\begin{equation}
 \begin{aligned}
    & {\rm N2 = \log \frac{[NII]\lambda 6583}{H\alpha}}, \\
    & {\rm 12+\log(O/H) = 8.743[\pm 0.027]+0.462[\pm 0.024] \times N2,}    
\end{aligned}   
\end{equation}
where -1.1$<$O3N2$<$1.7 with dispersions of 0.18 dex, and -1.6$<$N2$<$-0.2 with a dispersion of 0.16 dex. 

Calibrations with O3N2 and N2 from \cite{Pettini04} (hereafter PP04):
\begin{equation}
    \rm 12+\log(O/H) = 8.73-0.32 \times O3N2,
\end{equation}
\begin{equation}
    \rm 12+\log(O/H) = 9.37+2.03 \times N2+1.26\times N2^2 +0.32\times N2^3,
\end{equation}
where -1$<$ O3N2 $<$1.9 with a dispersion of 0.14 dex, and -2.5$<$ N2 $<$-0.3 with a  dispersion of 0.18 dex. The same intrinsic errors as M13 are adopted for these two calibrations. 

Calibrations with O3N2 and N2 from \cite{Perez-Montero09} (hereafter PMC09), after correcting for nitrogen abundance, are given as:
\begin{equation}
  \begin{aligned}
   & \rm N2S2 = \log(\frac{[NII]\lambda 6564}{[SII]\lambda\lambda6717,31}),\\
   & \rm \log(N/O) = 1.26\times N2S2-0.86,    
  \end{aligned}
\end{equation}
\begin{equation}
    \rm 12+\log(O/H) = 8.33-0.31\times O3N2-0.35 \times \log(N/O),
\end{equation}
\begin{equation}
    \rm 12+\log(O/H) = 0.78 \times N2-0.56 \times \log(N/O)+8.41,
\label{eq:PMC09}
\end{equation}
where N2S2 is in the range of -1 $\sim$ 0.5 with a dispersion of 0.31 dex, -1$<$O3N2$<$1.9 with a dispersion of 0.24 dex, and -2.5$<$N2$<$-0.3 with a dispersion of 0.21 dex.

Each emission line used in Eq.~\ref{eq:D16}$\sim$\ref{eq:PMC09} is corrected for dust extinction as described in \S\ref{subsec:Spec_analy} and meets the criteria of an S/N higher than 7.  Line ratios exceeding the applicable range of the relations are also excluded. The errors include flux uncertainties, 1$\sigma$ scatters, and intrinsic errors of the relations. For our data, we note that the errors are dominated by the systematic errors.

(12) {\bf Electron density ($n_{\rm e}$)}: At an electron temperature of $T_{\rm e} = 10^4$ K, the electron density ($n_{\rm e}$) is determined by the ratio $R$=[S{\sc ii}]$\lambda$6716/[S{\sc ii}]$\lambda$6731 \citep{Proxauf14}:
\begin{equation}
\begin{aligned}
\log(n_e {\rm [cm^{-3}]}) = & 0.0543 \tan(-3.0553R+2.8506)\\
& +6.98-10.6905R+9.9186R^2 \\
& -3.5442R^3,   
\end{aligned}
\end{equation}
where $n_{\rm e}$ within [40,~1000] cm$^{-3}$. Each emission line has been corrected for dust attenuation and satisfied S/N $>$ 7.
\newline 
\newline
We also list the metallicity gradients for all galaxies with available global metallicity in Table~\ref{Tab:gradient}.

(13) {\bf Metallicity gradient ($\rm \nabla(O/H)$)}: The metallicity gradient up to 1 $\rm R_e$ is determined using two methods. The first is averaging spectra within concentric elliptical annuli along the major axis and then calculating the corresponding metallicity, labeled as ``Spectra" in Table~\ref{Tab:gradient}; the second is adopting the median values of the elliptical annuli within the metallicity map, labeled as ``Spaxel". The center, R$_{\rm e}$, position angle, and axis ratio of the elliptical annulus are from Table~\ref{Tab:sample}. The step of each annulus is the FWHM of PSF, and the distance to the center is normalized by $\rm R_e$. For the ``Spaxel" method, the median value and its errors are determined through a bootstrapping process. Pixels classified as Seyfert or LINER based on the [SII]-BPT diagram, those with \ha\ equivalent widths less than 6 \AA, and metallicity diagnostics beyond their applicable boundaries are excluded. For the ``Spectra" method, pixels with an \ha\ emission line S/N lower than three are masked before averaging the spectra, after which the same criteria are applied to the mean spectra. Additionally, for both methods, bins containing more than 20 pixels are selected, and gradients are calculated only if more than five bins are available. The metallicity gradient is defined as the slope of the linear fittings, with the gradient and its associated error determined using bootstrapped fittings. Table~\ref{Tab:gradient} presents the ``Spectra" and ``Spaxel" gradients for the `D16$\_$N2S2\ha', `PP04$\_$O3N2', and `PP04$\_$N2' calibrations, along with the corresponding fitting radius ranges. The table also includes the median gradient for each method.

\begin{table*}[htbp]
\setlength{\tabcolsep}{3pt} 
\noindent
\tiny
\begin{flushleft}
\rotatebox{90}{
\begin{minipage}{\textheight}
\resizebox{\textwidth}{!}{
\begin{tabular}{ccccccccccccccc}
\hline
\hline
I.D. & Target & TEL & ExpTime & FWHM &   LogSFR(\ha)   & D16$\_$N2S2\ha\ & C20  & M13$\_$O3N2 & M13$\_$N2 & PP04$\_$O3N2 & PP04$\_$N2 & PMC09$\_$O3N2 & PMC09$\_$N2& $n_{\rm e}$ \\
    &        &     &   (s)   & ({\arcsec})  &(\Msun yr$^{-1}$)&      (dex)      & (dex)&     (dex)   &    (dex)  &    (dex)     &    (dex)   &     (dex)    &    (dex)  & (cm$^{-3}$) \\
(1) &  (2)   & (3) &   (4)   & (5)  &   (6)           &      (7)        & (8)  &      (9)    &     (10)  &     (11)     &     (12)   &      (13)    &     (14)  &    (15)   \\
\hline
1 & WLM  & MUSE & 4800+3*2320 & 0.79 $\pm$ 0.81$^{\dagger}$ & -4.00 & -- & -- & -- & -- & -- & -- & -- & -- & <40  \\
2 & NGC0059  & MUSE & 3360 & 0.43 $\pm$ 0.47$^{\dagger}$ & -2.02 & 7.94 & 8.28 & 7.90$^{\ast}$ & 8.16 & 8.14 & 8.19 & 8.27 & 8.39 & <40  \\
3 & ESO410-G005  & WiFeS & 14400(BI)+7200(UR) & 1.40 $\pm$ 0.34 & -5.23 & -- & -- & -- & -- & -- & -- & -- & -- & --  \\
4 & ESO294-G010  & MUSE & 4800 & 0.86 $\pm$ 0.14 & -4.71 & -- & -- & -- & -- & -- & -- & -- & -- & --  \\
5 & IC1574  & MUSE & 4480 & 0.43 $\pm$ 0.32$^{\dagger}$ & -3.38 & -- & -- & -- & -- & -- & -- & -- & -- & --  \\
7 & UGCA015  & MUSE & 4560 & 0.76 $\pm$ 0.43 & -- & -- & -- & -- & -- & -- & -- & -- & -- & --  \\
8 & ESO540-G032  & MUSE & 4560 & 0.43 $\pm$ 0.35 & -- & -- & -- & -- & -- & -- & -- & -- & -- & --  \\
9 & UGC00668  & MUSE & 8370 & 0.72 $\pm$ 0.46$^{\dagger}$ & -4.74 & -- & -- & -- & -- & -- & -- & -- & -- & 68.55  \\
10 & UGC00685  & MUSE & 7920 & 1.41 $\pm$ 0.25 & -2.41 & 7.70 & 8.27 & 8.07 & 8.16 & 8.26 & 8.18 & 8.47 & 8.53 & <40  \\
11 & UGC00695  & MUSE & 3*310+6*990 & 1.46 $\pm$ 0.14 & -2.04 & 7.53$^{\ast}$ & 8.26 & 8.17 & 8.15 & 8.33 & 8.16 & 8.58 & 8.59 & <40  \\
12 & UGC00891  & MUSE & 2240 & 0.69 $\pm$ 0.17 & -2.55 & 7.74 & 8.37 & 8.19 & 8.24 & 8.35 & 8.28 & 8.56 & 8.68 & <40  \\
13 & UGC01056  & MUSE & 15120 & 0.78 $\pm$ 0.77$^{\dagger}$ & -1.91 & 7.73 & 8.30 & 8.03 & 8.18 & 8.23 & 8.20 & 8.44 & 8.55 & <40  \\
15 & NGC0625  & MUSE & 2240 & 0.45 $\pm$ 0.14 & -1.31 & 8.07 & 8.32 & 7.95$^{\ast}$ & 8.19 & 8.18 & 8.22 & 8.27 & 8.36 & 41.59  \\
17 & ESO245-G005  & MUSE & 4800 & 0.84 $\pm$ 0.41$^{\dagger}$ & -3.58 & -- & -- & -- & -- & -- & -- & -- & -- & <40  \\
19 & ESO115-G021  & MUSE & 4800 & 0.70 $\pm$ 0.14 & -2.74 & 7.80 & 8.31 & 8.11 & 8.19 & 8.29 & 8.22 & 8.47 & 8.54 & <40  \\
21 & NGC1311  & MUSE & 4400 & 0.48 $\pm$ 0.32$^{\dagger}$ & -2.20 & 7.81 & 8.34 & 8.10 & 8.21 & 8.28 & 8.24 & 8.47 & 8.58 & <40  \\
22 & UGC02716  & MUSE & 4480 & 0.77 $\pm$ 0.35$^{\dagger}$ & -2.44 & 7.71 & 8.25 & 7.99 & 8.14 & 8.20 & 8.16 & 8.40 & 8.48 & <40  \\
23 & IC1959  & MUSE & 2240 & 0.73 $\pm$ 0.37$^{\dagger}$ & -2.05 & 7.93 & 8.34 & 8.05 & 8.21 & 8.25 & 8.25 & 8.39 & 8.51 & <40  \\
24 & NGC1510  & WiFeS & 9000(BI)+23500(UR) & 1.65 $\pm$ 0.22 & -1.22 & 8.32 & 8.45 & 8.03 & 8.31 & 8.23 & 8.36 & 8.23 & 8.40 & <40  \\
25 & NGC1522  & MUSE & 4400 & 0.66 $\pm$ 0.14 & -1.67 & 8.01 & 8.36 & 8.01 & 8.23 & 8.22 & 8.26 & 8.34 & 8.48 & <40  \\
26 & ESO483-G013  & MUSE & 2240 & 0.77 $\pm$ 0.28$^{\dagger}$ & -1.39 & 8.09 & 8.39 & 8.05 & 8.26 & 8.25 & 8.30 & 8.34 & 8.47 & <40  \\
27 & ESO158-G003  & MUSE & 2240 & 1.04 $\pm$ 0.24 & -1.92 & 8.18 & 8.55 & 8.33 & 8.40 & 8.45 & 8.48 & 8.53 & 8.71 & <40  \\
28 & ESO119-G016  & MUSE & 2240 & 0.63 $\pm$ 0.14 & -2.38 & 7.91 & 8.43 & -- & 8.29 & -- & 8.34 & -- & 8.68 & 41.69  \\
30 & ESO486-G021  & MUSE & 8400 & 1.37 $\pm$ 0.19 & -1.26 & 7.94 & 8.40 & 7.75$^{\ast}$ & 8.26 & 8.04$^{\ast}$ & 8.31 & 8.20$^{\ast}$ & 8.61 & <40  \\
31 & NGC1800  & MUSE & 3160 & 0.98 $\pm$ 0.34$^{\dagger}$ & -0.52 & 8.20 & 8.45 & 7.72$^{\ast}$ & 8.31 & 8.01$^{\ast}$ & 8.37 & 8.08$^{\ast}$ & 8.51 & <40  \\
33 & CGCG035-007  & MUSE & 4400 & 0.57 $\pm$ 0.23 & -3.02 & 7.79 & 8.37 & 8.18 & 8.24 & 8.34 & 8.28 & 8.53 & 8.65 & <40  \\
35 & UGC05288  & MUSE & 8960 & 1.11 $\pm$ 0.62$^{\dagger}$ & -1.58 & 7.88 & 8.37 & 8.09 & 8.24 & 8.27 & 8.28 & 8.44 & 8.59 & <40  \\
36 & UGC05373  & MUSE & 4800 & 0.70 $\pm$ 0.37$^{\dagger}$ & -3.80 & -- & -- & -- & -- & -- & -- & -- & -- & <40  \\
37 & UGCA193  & MUSE & 8920 & 1.73 $\pm$ 0.31 & -2.25 & 7.87 & 8.39 & 8.20 & 8.26 & 8.35 & 8.30 & 8.53 & 8.64 & <40  \\
39 & AM1001-270  & MUSE & 2280 & 1.99 $\pm$ 1.64 & -- & -- & -- & -- & -- & -- & -- & -- & -- & --  \\
44 & UGC05923  & MUSE & 4560 & 0.88 $\pm$ 0.46$^{\dagger}$ & -2.24 & 8.06 & 8.45 & 8.20 & 8.31 & 8.36 & 8.37 & 8.47 & 8.62 & <40  \\
45 & UGC06457  & MUSE & 2280 & -- & -- & -- & -- & -- & -- & -- & -- & -- & -- & --  \\
46 & ESO321-G014  & MUSE & 5200 & 1.37 $\pm$ 0.06 & -3.55 & -- & -- & -- & -- & -- & -- & -- & -- & 68.87  \\
47 & ISZ399  & WiFeS & 16200(BI)+14076(UR) & 1.50 $\pm$ 0.17 & -0.81 & 8.53 & 8.60 & 8.43 & 8.45 & 8.51 & 8.54 & 8.45 & 8.53 & 57.28  \\
48 & UGC08091  & MUSE & 2280 & 0.45 $\pm$ 0.17 & -2.94 & 7.74 & 8.13 & 7.96$^{\ast}$ & 8.05 & 8.18 & 8.05 & 8.35 & 8.28 & <40  \\
50 & UGCA320  & MUSE & 4480 & 1.02 $\pm$ 0.20 & -1.71 & 7.61$^{\ast}$ & 8.12 & 7.90$^{\ast}$ & 8.05 & 8.14 & 8.05 & 8.35 & 8.34 & <40  \\
51 & MCG-03-34-002  & MUSE & 2280 & 0.81 $\pm$ 0.13 & -2.39 & 7.89 & 8.38 & 8.31 & 8.24 & 8.43 & 8.29 & 8.59 & 8.60 & <40  \\
52 & IC4247  & MUSE & 2280 & 1.06 $\pm$ 0.98 & -3.02 & 7.72 & 8.25 & 8.10 & 8.14 & 8.28 & 8.16 & 8.48 & 8.48 & <40  \\
53 & ESO444-G084  & MUSE & 4480 & 0.32 $\pm$ 0.09 & -3.00 & -- & -- & -- & -- & -- & -- & -- & -- & 365.59  \\
54 & NGC5253  & WiFeS & 14400(BI)+9000(UR) & 1.50 $\pm$ 0.15 & -1.28 & 8.19 & 8.44 & 8.14 & 8.30 & 8.31 & 8.36 & 8.37 & 8.50 & 66.22  \\
55 & NGC5264  & WiFeS & 10800(BI)+14400(UR) & 1.60 $\pm$ 0.16 & -3.02 & 8.15 & 8.52 & 8.34 & 8.37 & 8.45 & 8.44 & 8.54 & 8.67 & <40  \\
56 & KKH086  & MUSE & 5280 & 0.81 $\pm$ 0.58$^{\dagger}$ & -- & -- & -- & -- & -- & -- & -- & -- & -- & --  \\
57 & IC4951  & WiFeS & 27000(BI)+21240(UR) & 1.25 $\pm$ 0.39 & -1.66 & -- & -- & -- & -- & -- & -- & -- & -- & <40  \\
58 & DDO210  & WiFeS & 19800(BI)+27936(UR) & 1.75 $\pm$ 0.32 & -- & -- & -- & -- & -- & -- & -- & -- & -- & --  \\
59 & NGC7064  & WiFeS & 21600(BI)+21960(UR) & 1.25 $\pm$ 0.46 & -1.94 & -- & -- & -- & -- & -- & -- & -- & -- & <40  \\
60 & IC5256  & WiFeS & 14400(BI)+23400(UR) & 1.70 $\pm$ 0.22 & -0.02 & 8.47 & 8.60 & 8.38 & 8.45 & 8.48 & 8.53 & 8.45 & 8.58 & <40  \\
61 & UGCA438  & MUSE & 2240 & 1.16 $\pm$ 0.08 & -- & -- & -- & -- & -- & -- & -- & -- & -- & --  \\
63 & UGC12613  & MUSE & 3*1035 & 0.75 $\pm$ 0.41$^{\dagger}$ & -- & -- & -- & -- & -- & -- & -- & -- & -- & --  \\
64 & UGCA442  & MUSE & 2240 & 0.96 $\pm$ 0.25 & -2.55 & 7.74 & 8.21 & 8.03 & 8.11 & 8.23 & 8.12 & 8.42 & 8.40 & <40  \\
\hline
$\Delta$ &     &      &      &                 &        & 0.2 & 0.19 & 0.18 & 0.16 & 0.14 & 0.18 & 0.26 & 0.29 &      \\
\hline
\end{tabular}
}
\caption{Spectroscopic properties for observed galaxies. (1) $\sim$ (2) The index and name of sample, same as Table~2; (3) the used telescope; (4) on-source time. For MUSE observation, n$\times$T represent the number of FOVs and on-source time of single FOV; for WiFeS, T(BI)+T(UR) represents total exposure time on B+I grism plus exposure time on U+R grism; (5) FWHM of PSF and its errors. The marked values are derived from the median \texttt{DIMM SEEING}; (6) \ha-based SFR from global spectrum; (7) $\sim$ (14) metallicity calculated by N2S2\ha\ from \cite{Dopita16}, N2 calibration combined with O3N2 calibration from \cite{Curti20}, O3N2 and N2 calibration of \cite{Marino13}, O3N2 and N2 calibrations of \cite{Pettini04}, and O3N2 and N2 calibrations of \cite{Perez-Montero09}. Metallicity which is marked with asterisks, their emission line ratios that exceed the range of application given by the corresponding methods; (15) electron density. The errors of SFR(\ha) are negligible as the high S/N of \ha. The errors of metallicity in columns (7)$\sim$(14) are dominated by systematic errors of corresponding calibrations, as listed in the last lines.}
\label{Tab:spectra}
\end{minipage}
}
\end{flushleft}
\end{table*}

\begin{table*}[htbp]
\setlength{\tabcolsep}{3pt} 
\begin{center}
\rotatebox{90}{
\begin{minipage}{\textheight}
\resizebox{\textwidth}{!}{
\begin{tabular}{cc|cc|cc|cc|cc|cc|cc}
\hline
\hline
I.D. & Target &       \multicolumn{4}{c|}{D16$\_$N2S2\ha}         &          \multicolumn{4}{c|}{PP04$\_$O3N2}            &     \multicolumn{4}{c}{PP04$\_$N2}                \\ 
\hline
    &        & Spectra  & R/R$_{\rm e}$ &  Spaxel    & R/R$_{\rm e}$ & Spectra  & R/R$_{\rm e}$ &  Spaxel    & R/R$_{\rm e}$ & Spectra  & R/R$_{\rm e}$ &  Spaxel    & R/R$_{\rm e}$ \\
(1) & (2)    &  (3)     &     (4)     &        (5)    &    (6)     &         (7)   &    (8)     &      (9)       &   (10)     &     (11)     &    (12)    &        (13)   &     (14) \\
\hline
2 & NGC0059  & -0.16 $\pm$ 0.03$^{\ast}$ & [0.06,0.46] & -0.12 $\pm$ 0.05$^{\ast}$ & [0.12,0.42] & 0.66 $\pm$ 0.04$^{\ast}$ & [0.06,0.46] & 0.45 $\pm$ 0.15$^{\ast}$ & [0.16,0.42] & 0.68 $\pm$ 0.03$^{\ast}$ & [0.06,0.46] & 0.59 $\pm$ 0.15$^{\ast}$ & [0.12,0.42] \\
10 & UGC00685  & 0.12 $\pm$ 0.06$^{\ast}$ & [0.08,0.46] & 0.15 $\pm$ 0.03 & [0.08,0.69] & 0.21 $\pm$ 0.02 & [0.08,0.73] & 0.19 $\pm$ 0.03 & [0.08,0.69] & 0.03 $\pm$ 0.02 & [0.08,0.73] & 0.04 $\pm$ 0.03 & [0.08,0.69] \\
11 & UGC00695  & -- & -- & 0.29 $\pm$ 0.07$^{\ast}$ & [0.10,0.45] & 0.41 $\pm$ 0.03 & [0.00,0.60] & 0.32 $\pm$ 0.05$^{\ast}$ & [0.05,0.45] & 0.18 $\pm$ 0.03 & [0.00,0.65] & 0.36 $\pm$ 0.08 & [0.05,0.50] \\
12 & UGC00891  & -0.10 $\pm$ 0.02 & [0.19,0.99] & -- & -- & -0.02 $\pm$ 0.01 & [0.19,0.99] & -- & -- & -0.05 $\pm$ 0.02 & [0.19,0.99] & -- & -- \\
13 & UGC01056  & -0.14 $\pm$ 0.06$^{\ast}$ & [0.04,0.38] & 0.06 $\pm$ 0.02 & [0.08,0.94] & 0.16 $\pm$ 0.01 & [0.04,0.98] & 0.21 $\pm$ 0.02 & [0.08,0.94] & 0.11 $\pm$ 0.01 & [0.04,0.98] & 0.16 $\pm$ 0.02 & [0.08,0.98] \\
15 & NGC0625  & -0.01 $\pm$ 0.01 & [0.03,0.65] & 0.13 $\pm$ 0.01 & [0.05,0.56] & 0.07 $\pm$ 0.01 & [0.03,0.65] & -0.09 $\pm$ 0.02 & [0.05,0.56] & 0.11 $\pm$ 0.01 & [0.03,0.65] & -0.00 $\pm$ 0.02 & [0.05,0.56] \\
19 & ESO115-G021  & -0.27 $\pm$ 0.01 & [0.14,1.00] & 0.05 $\pm$ 0.08$^{\ast}$ & [0.23,0.43] & 0.25 $\pm$ 0.01 & [0.14,1.00] & 0.12 $\pm$ 0.03 & [0.26,1.01] & -0.01 $\pm$ 0.01 & [0.14,1.00] & -0.00 $\pm$ 0.03 & [0.21,1.01] \\
21 & NGC1311  & -0.10 $\pm$ 0.01 & [0.07,0.99] & -0.13 $\pm$ 0.02 & [0.18,0.82] & 0.08 $\pm$ 0.01 & [0.07,0.99] & 0.10 $\pm$ 0.03 & [0.18,0.82] & 0.01 $\pm$ 0.01 & [0.07,0.99] & 0.01 $\pm$ 0.03 & [0.16,0.85] \\
22 & UGC02716  & -0.10 $\pm$ 0.04$^{\ast}$ & [0.02,0.42] & 0.99 $\pm$ 0.07$^{\ast}$ & [0.04,0.24] & 0.72 $\pm$ 0.04$^{\ast}$ & [0.07,0.46] & 1.35 $\pm$ 0.17$^{\ast}$ & [0.07,0.24] & 0.32 $\pm$ 0.03$^{\ast}$ & [0.02,0.49] & 0.97 $\pm$ 0.18$^{\ast}$ & [0.04,0.24] \\
23 & IC1959  & -0.09 $\pm$ 0.01 & [0.06,0.99] & -0.06 $\pm$ 0.01 & [0.12,1.02] & -0.00 $\pm$ 0.01 & [0.06,0.99] & -0.07 $\pm$ 0.01 & [0.12,1.02] & -0.05 $\pm$ 0.01 & [0.06,0.99] & -0.06 $\pm$ 0.02 & [0.12,1.02] \\
24 & NGC1510  & -0.13 $\pm$ 0.07$^{\ast}$ & [0.09,0.45] & 0.01 $\pm$ 0.03 & [0.18,0.81] & 0.44 $\pm$ 0.07$^{\ast}$ & [0.09,0.45] & 0.26 $\pm$ 0.07 & [0.18,0.72] & 0.39 $\pm$ 0.07$^{\ast}$ & [0.09,0.45] & 0.22 $\pm$ 0.07 & [0.18,0.81] \\
25 & NGC1522  & -0.17 $\pm$ 0.02 & [0.07,0.73] & -0.13 $\pm$ 0.03 & [0.11,0.65] & 0.17 $\pm$ 0.02 & [0.07,0.73] & 0.02 $\pm$ 0.05 & [0.11,0.65] & 0.15 $\pm$ 0.02 & [0.07,0.73] & 0.03 $\pm$ 0.06 & [0.11,0.65] \\
26 & ESO483-G013  & -0.26 $\pm$ 0.03 & [0.03,0.53] & -0.04 $\pm$ 0.01 & [0.06,1.03] & 0.24 $\pm$ 0.03 & [0.03,0.53] & 0.23 $\pm$ 0.02 & [0.06,1.03] & 0.20 $\pm$ 0.03 & [0.03,0.53] & 0.14 $\pm$ 0.02 & [0.06,1.03] \\
27 & ESO158-G003  & -0.16 $\pm$ 0.01 & [0.03,0.97] & -0.07 $\pm$ 0.01 & [0.07,1.01] & -0.09 $\pm$ 0.01 & [0.03,0.97] & -0.11 $\pm$ 0.02 & [0.07,1.01] & -0.12 $\pm$ 0.01 & [0.03,0.97] & -0.12 $\pm$ 0.03 & [0.07,1.01] \\
28 & ESO119-G016  & -0.09 $\pm$ 0.01 & [0.02,0.94] & -- & -- & -0.31 $\pm$ 0.01 & [0.04,0.94] & -- & -- & -0.20 $\pm$ 0.01 & [0.02,0.94] & -- & -- \\
30 & ESO486-G021  & -0.12 $\pm$ 0.02 & [0.00,0.95] & -0.08 $\pm$ 0.01 & [0.10,1.05] & -0.06 $\pm$ 0.02 & [0.00,0.95] & -0.02 $\pm$ 0.02 & [0.10,1.05] & -0.05 $\pm$ 0.02 & [0.00,0.95] & -0.04 $\pm$ 0.03 & [0.10,1.05] \\
31 & NGC1800  & -0.06 $\pm$ 0.02 & [0.13,0.96] & -0.09 $\pm$ 0.00 & [0.13,1.00] & -0.13 $\pm$ 0.02 & [0.13,0.96] & -0.03 $\pm$ 0.02 & [0.13,1.00] & -0.13 $\pm$ 0.02 & [0.13,0.96] & -0.06 $\pm$ 0.02 & [0.13,1.00] \\
33 & CGCG035-007  & -0.15 $\pm$ 0.02 & [0.07,0.96] & 0.03 $\pm$ 0.02 & [0.10,0.86] & 0.07 $\pm$ 0.01 & [0.07,0.99] & 0.07 $\pm$ 0.03 & [0.10,0.86] & -0.06 $\pm$ 0.01 & [0.07,0.99] & -0.04 $\pm$ 0.04 & [0.10,0.89] \\
35 & UGC05288  & -0.17 $\pm$ 0.02 & [0.00,1.00] & -0.15 $\pm$ 0.01 & [0.08,0.79] & 0.18 $\pm$ 0.02 & [0.00,1.00] & 0.12 $\pm$ 0.02 & [0.08,0.79] & 0.06 $\pm$ 0.01 & [0.00,1.00] & 0.00 $\pm$ 0.03 & [0.08,0.79] \\
37 & UGCA193  & -0.11 $\pm$ 0.02 & [0.06,0.95] & 0.06 $\pm$ 0.01 & [0.13,1.02] & 0.14 $\pm$ 0.02 & [0.06,0.95] & 0.16 $\pm$ 0.02 & [0.13,1.02] & -0.02 $\pm$ 0.02 & [0.06,0.95] & 0.04 $\pm$ 0.03 & [0.13,1.02] \\
44 & UGC05923  & -0.16 $\pm$ 0.04 & [0.10,0.70] & 0.03 $\pm$ 0.04 & [0.20,0.80] & -0.08 $\pm$ 0.04 & [0.10,0.70] & -0.12 $\pm$ 0.07 & [0.20,0.80] & -0.06 $\pm$ 0.04 & [0.10,0.70] & -0.05 $\pm$ 0.08 & [0.20,0.80] \\
47 & ISZ399  & -0.20 $\pm$ 0.04 & [0.32,0.97] & -0.19 $\pm$ 0.01 & [0.49,1.13] & 0.08 $\pm$ 0.04 & [0.32,0.97] & 0.07 $\pm$ 0.06 & [0.49,1.13] & 0.09 $\pm$ 0.04 & [0.32,0.97] & 0.08 $\pm$ 0.07 & [0.49,1.13] \\
48 & UGC08091  & -0.15 $\pm$ 0.03 & [0.02,0.55] & 0.14 $\pm$ 0.02 & [0.04,0.61] & 0.07 $\pm$ 0.01 & [0.02,0.73] & 0.08 $\pm$ 0.02 & [0.03,0.61] & -0.03 $\pm$ 0.01 & [0.02,0.73] & 0.02 $\pm$ 0.02 & [0.03,0.61] \\
50 & UGCA320  & 0.05 $\pm$ 0.03 & [0.04,0.99] & 0.03 $\pm$ 0.04 & [0.24,0.79] & -0.10 $\pm$ 0.01 & [0.04,0.99] & -0.03 $\pm$ 0.03 & [0.20,0.83] & -0.03 $\pm$ 0.01 & [0.04,0.99] & 0.01 $\pm$ 0.04 & [0.20,0.97] \\
51 & MCG-03-34-002  & -0.20 $\pm$ 0.02 & [0.04,0.93] & 0.06 $\pm$ 0.02 & [0.09,0.93] & 0.16 $\pm$ 0.02 & [0.04,0.98] & 0.21 $\pm$ 0.02 & [0.09,0.93] & 0.09 $\pm$ 0.02 & [0.04,0.98] & 0.18 $\pm$ 0.03 & [0.09,0.93] \\
52 & IC4247  & -0.03 $\pm$ 0.06$^{\ast}$ & [0.05,0.49] & -0.03 $\pm$ 0.05 & [0.16,0.59] & -0.14 $\pm$ 0.03 & [0.05,0.65] & -0.16 $\pm$ 0.06 & [0.11,0.65] & -0.04 $\pm$ 0.03 & [0.05,0.65] & -0.03 $\pm$ 0.08 & [0.11,0.65] \\
54 & NGC5253  & -1.00 $\pm$ 0.09$^{\ast}$ & [0.09,0.34] & -0.44 $\pm$ 0.02 & [0.13,0.56] & -0.29 $\pm$ 0.09$^{\ast}$ & [0.09,0.34] & -0.07 $\pm$ 0.07 & [0.13,0.56] & -0.33 $\pm$ 0.09$^{\ast}$ & [0.09,0.34] & -0.06 $\pm$ 0.09 & [0.13,0.56] \\
55 & NGC5264  & -0.01 $\pm$ 0.05$^{\ast}$ & [0.09,0.43] & -- & -- & 0.21 $\pm$ 0.05$^{\ast}$ & [0.09,0.43] & -- & -- & -0.16 $\pm$ 0.05$^{\ast}$ & [0.09,0.43] & -- & -- \\
57 & IC4951  & -0.00 $\pm$ 0.02 & [0.20,0.88] & -- & -- & 0.11 $\pm$ 0.02 & [0.20,0.88] & -- & -- & 0.02 $\pm$ 0.02 & [0.20,0.88] & -- & -- \\
59 & NGC7064  & -0.24 $\pm$ 0.03 & [0.24,0.79] & -- & -- & 0.01 $\pm$ 0.02 & [0.24,0.99] & -- & -- & -0.10 $\pm$ 0.02 & [0.24,0.99] & -- & -- \\
60 & IC5256  & -0.04 $\pm$ 0.04 & [0.33,0.98] & -0.04 $\pm$ 0.03 & [0.49,1.14] & -0.10 $\pm$ 0.04 & [0.33,0.98] & -0.11 $\pm$ 0.11 & [0.65,1.14] & -0.07 $\pm$ 0.04 & [0.33,0.98] & -0.07 $\pm$ 0.07 & [0.49,1.14] \\
64 & UGCA442  & -0.02 $\pm$ 0.02 & [0.02,0.87] & 0.29 $\pm$ 0.02 & [0.05,0.80] & 0.21 $\pm$ 0.01 & [0.02,1.00] & -0.03 $\pm$ 0.02 & [0.05,0.79] & 0.06 $\pm$ 0.01 & [0.02,1.00] & 0.11 $\pm$ 0.03 & [0.05,0.80] \\
\hline
 & Median & -0.12 $\pm$ 0.08 &      & -0.03 $\pm$ 0.14 &      & 0.07 $\pm$ 0.15 &      & 0.04 $\pm$ 0.12 &      & -0.02 $\pm$ 0.10 &      & 0.01 $\pm$ 0.11 &      \\
\hline
\end{tabular}
}
\caption{Metallicity gradients for the ``Spectra" and ``Spaxel" methods are provided for the `D16$\_$N2S2\ha', `PP04$\_$O3N2', and `PP04$\_$N2' calibrations. The starting and ending radii of the fittings are also listed. Gradients with fitting radii less than 0.5 $R_{\rm e}$ are marked with a star to indicate that they may exhibit steeper slopes.}
\label{Tab:gradient}
\end{minipage}
}
\end{center}
\end{table*}

\section{Metallicity scaling relation} \label{sec:relation}
 \subsection{The relation between stellar mass and gas-phase metallicity}
Metallicity is the product of generations of star formations and is regulated by the physical activities in the ISM and CGM. Stellar mass is also a result of star formation, resulting in the well-known stellar mass-metallicity relation (MZR), which is simultaneously regulated by gas accretion/ejection \citep{Sanchez19,Maiolino19}. It is generally agreed that stellar mass is positively correlated with gas-phase metallicity for star-forming galaxies, although the slopes and shapes can differ depending on the metallicity tracers and sample selection \citep[e.g][]{Kewley08,Hunt16,Sanchez19,Maiolino19,Curti20}. For dwarf galaxies, it is debated whether metallicity is still positively correlated with stellar mass \citep{Sanchez19,Lee06}. 

The MZR at the low-mass end is affected by: (1) the metallicity calibrations. For the extension of MZR established by SDSS to the lower mass end, metallicity calculated using the [O{\sc iii}]$\lambda4363$-based direct-T$_{\rm e}$ method is expected to be linearly correlated with \Mstar, but the relation tends to become flat for some strong-line methods, like using O3N2 or N2 diagnostics \citep{Kewley08, Andrews13}. Similar phenomena are also found by IFU-based studies \citep{Sanchez17, Sanchez19}. \cite{Zahid12} collected a sample of dwarf irregulars \citep{Lee06}, BCD \citep{Zhao10}, and dwarf galaxies from the SDSS and DEEP2 surveys. They found that the direct T$_{\rm e}$ methods tend to select those metal-poor galaxies, while the N2 diagnostic tends to have higher metallicity with larger scatter. Besides, based on the same sample, the strong-line methods deviate from the MZR established by the direct T$_{\rm e}$ methods and show large scatters \citep{Zhao10,Berg12,Sanchez19}. (2) Sample selection: according to \cite{Hunt16}, in the local Universe, BCD \citep{Hunt10} and star-bursting dwarf galaxies \citep{Engelbracht08} are more metal-poor than dwarf irregulars, all of which have direct T$_{\rm e}$-based metallicity.

(3) Aperture effect: single-fiber or long-slit spectroscopy only covers a small fraction of a galaxy, which may not be representative of the entire galaxy. For example, for a representative sample of dwarf galaxies in the local Universe, \cite{Berg12} used MMT long-slit spectrograph (1{\arcsec} $\times$ 180{\arcsec}) to target only a few \hii\ regions within a galaxy, thereby missing a large part of star-forming regions. Our DGIS sample can largely alleviate this last caveat by sampling the majority of star-forming regions. MUSE has a large FOV that covers around 1 $R_{\rm e}$ of the targets. Besides, although the FOV of WiFeS is smaller, it still covers a large fraction of nebular emitting regions.
 
\begin{figure*}
\begin{center}
\includegraphics[width=\textwidth]{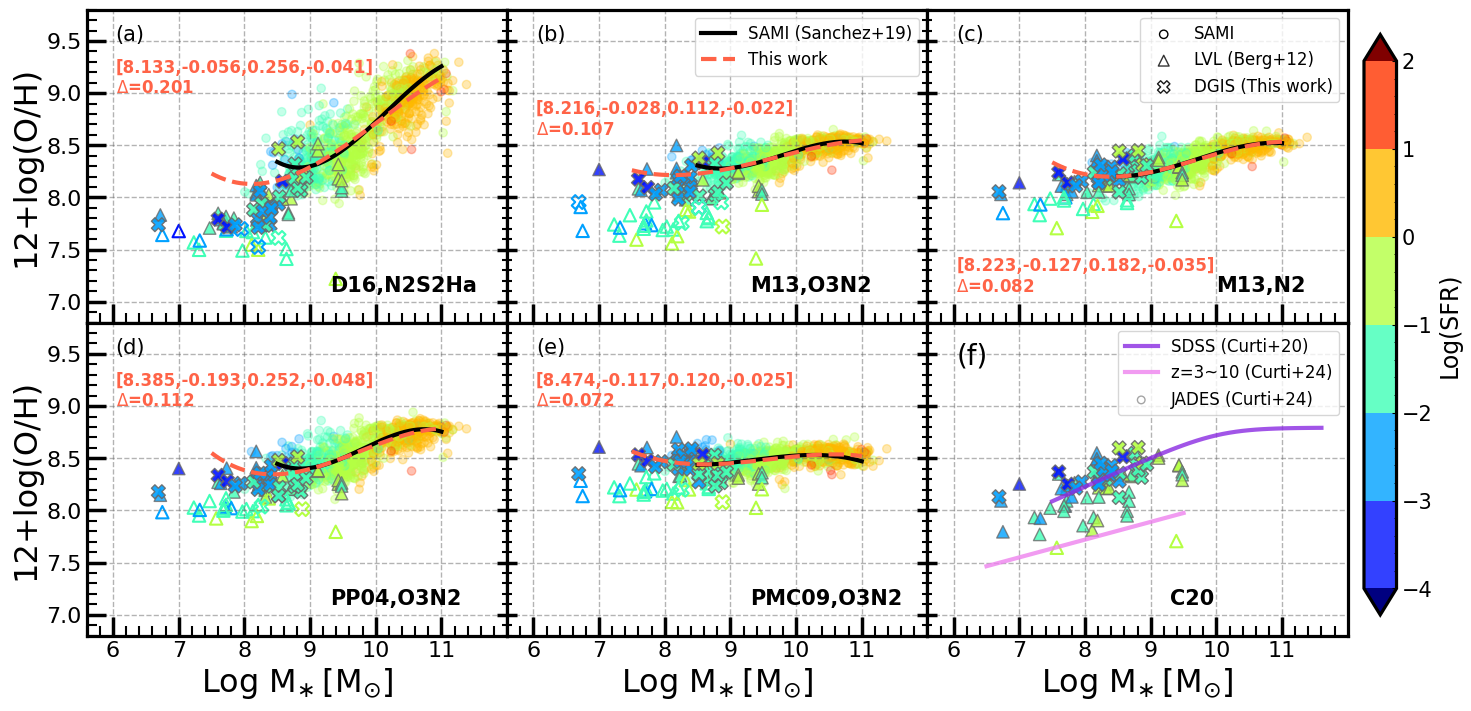}
\end{center}   
\caption{Gas-phase metallicity versus stellar mass at galactic scale, colored with star formation rate. The corresponding metallicity calibration of each panel labeled at right corners: (a) N2S2\ha\ calibrations from \cite{Dopita16} (D16); (b) and (c): O3N2 and N2 calibrations from \cite{Marino13}(M13); (d): O3N2 calibration from \cite{Pettini04} (PP04); (e): O3N2 calibration from \cite{Perez-Montero09}(PMC09); (f) Combined multiple strong-line calibrations from \cite{Curti20} (C20), this work we mainly use their N2 calibrator. One point represents one galaxy. The circular dots are SAMI galaxies collected from \cite{Sanchez19}; the triangles are dwarf galaxies in the \cite{Berg12}, and the metallicity based on their provided emission line table; the crosses are dwarf galaxies in this work. The data points without face color have a line ratio exceeding the applicable range of the corresponding methods. The black solid lines are MZR proposed by \cite{Sanchez19} for SAMI galaxies, and the red dashed lines are MZR of this work fitted for both SAMI and dwarf galaxies from DGIS and \cite{Berg12}, with fitted stellar mass ranged from $10^{8\sim11}$\Msun. The fitting results are also annotated in the corner. The last panel doesn't include SAMI but high-redshift galaxies from \cite{Curti24} (dots with dim grey edge color), and its purple and pink lines are MZRs collected from SDSS and {\it JWST}, respectively.}
\label{fig:MZR}
\end{figure*}

The global MZR of dwarf galaxies in this work is illustrated in Fig.~\ref{fig:MZR} and  colored according to the global SFR. The metallicity calibrations for each panel include N2S2\ha\ from D16, O3N2 and N2 from M13, O3N2 from PP04 and PMC09, and the strong line method calibrated by C20. For comparison with spiral galaxies, we collected data from the SAMI survey as reported in Table~2 of \cite{Sanchez19} and referenced their established polynomial MZRs for each calibration method. We also referenced the MZR relation established by SDSS with the C20 method \citep{Curti20}. To compare with dwarf galaxies, we substituted emission lines from Table~3 of \cite{Berg12} into Eq.~\ref{eq:D16}$\sim$\ref{eq:PMC09} of this work to get metallicity, thereby assessing the significance of the aperture effect. We also present the MZR relation established by \cite{Curti24} for low-mass galaxies in the early universe from {\it JWST} with C20 calibration.

As shown in Fig.~\ref{fig:MZR}, using different metallicity calibrations, the MZR at the lower-mass end has a different shape. The `D16$\_ $N2S2\ha' calibration gives a steeper slope and lower metallicity, while the `PMC09$\_ $O3N2' calibration derives very flat relations with higher metallicity than other calibrations. Using the same calibrations, our results closely align with the extrapolations of MZRs established by the SAMI survey. However, they tend to extend the decreasing trend down to $10^8$ \Msun\ instead of stopping around $10^9$ \Msun\ for `D16$\_ $N2S2\ha', `PP04$\_ $O3N2', and `M13$\_ $O3N2' calibrations, as indicated by the red dashed lines. The fitting results [$p_{\rm 0},\,p_{\rm 1},\,p_{\rm 2},\,p_{\rm 3}$] for each calibration, which combines data from SAMI, dwarf galaxies from \cite{Berg12}, and DGIS, are annotated in the upper left corner of the graph. These results utilize the same polynomial function as described by \cite{Sanchez19}: ${\rm 12+\log(O/H)}=\sum_{i=0}^4p_{\rm i}x^{i},\,x=\log(M_{\ast}/{\rm M_{\odot}})-8$, where $M_{\ast}$ ranges from $10^{8}$ to $10^{11}$ \Msun. Compared to higher-redshift low-mass galaxies observed by {\it JWST}, it is not surprising that the metallicity of our local dwarf galaxies is around 0.5 dex higher, since galaxies at higher redshift have a more pristine environment. The comparison of our results with \cite{Berg12} shows that using a similar sample and the same metallicity calculations, the metallicity obtained from IFS is around 0.1 dex higher than that derived from long-slit spectra, indicating a marginal impact of aperture effect at the low-mass end. However, the slight offset could still be related to the replenished fresh gas within star-forming regions targeted by long-slit fiber.

\subsection{Metallicity Fundamental Relation}
The relation between the stellar mass-SFR-metallicity is described by metallicity fundamental relation (FMR), indicating the importance of the secondary dependence of SFR. We define:
\begin{equation}
    \mu_{\rm \alpha} \equiv \log(M_{\ast}/{\rm M_{\odot}})-\alpha \log({\rm SFR}),
\label{eq:FMR}
\end{equation}
where introducing $\alpha$ produces the minimum scatter between metallicity and $\mu_{\rm \alpha}$. If $\alpha$ = 0, the relation of metallicity-$\mu_{\rm \alpha}$ becomes the MZR relations, indicating that the metallicity is only determined by stellar mass. The case with $\alpha$=1 is the relation between the metallicity and specific SFR (sSFR), suggesting that metallicity is determined by star-forming activity. 

\begin{figure*}
\begin{center}
\includegraphics[width=0.7\textwidth]{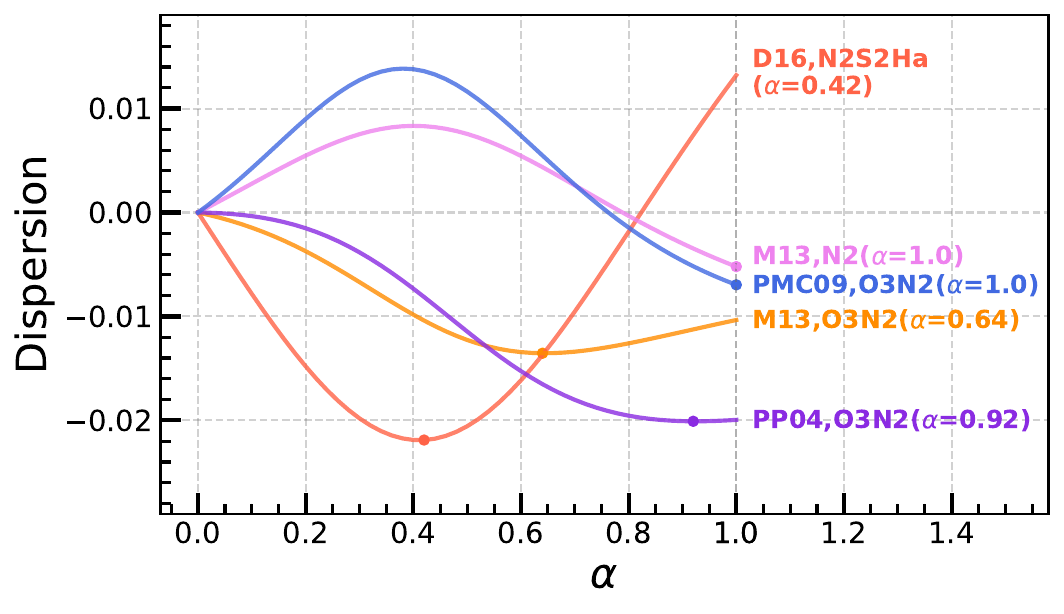}
\end{center}   
\caption{Metallicity dispersion versus adopted $\alpha$ in Eq.~\ref{eq:FMR}, and the dispersions are relative to the $\alpha=0$. One colored solid line represents one metallicity calibration and is labeled at the end. The $\alpha$ value which achieve the minimum dispersion is also labeled at the end with a dot. The samples within the plot include DGIS, \cite{Berg12}, and SAMI with $10^{8}<M_{\ast}/{\rm M_{\odot}}<10^9$\Msun\ \citep{Sanchez19}.}
\label{fig:FMR}
\end{figure*}

We collected dwarf galaxies from \cite{Sanchez19}, \cite{Berg12}, and DGIS, calculated the scatters with increasing $\alpha$, and compared these scatters relative to the MZR, where $\alpha$=0. As shown in Fig.~\ref{fig:FMR}, the scatters do not show obvious variation when changing the proportion of SFR for all calibrations. Furthermore, Fig.~\ref{fig:MZR} does not display an obvious color gradient perpendicular to the slopes. Therefore, our results indicate that SFR is not a significant factor in reducing the scatter between metallicity and stellar mass.

\section{Summary} \label{sec:summary}
In order to resolve the galactic ecosystems of dwarf galaxies at high spatial resolution with enough S/N, we proposed and conducted our DGIS project. The sample comprises 65 dwarf galaxies with stellar mass between 10$^6$ and 10$^9$ \Msun\ at distances within 11 Mpc, except for two objects whose updated distances are beyond our distance boundary. The observations are carried out using VLT/MUSE and ANU-2.3m/WiFeS and have largely been completed. Part of the reduced datacube and data products will be provided on the DGIS website (\url{https://www.dgisteam.com/index.html}), and the new observations also will be updated on the website. Here, we present the sample selection, data reduction, high-level data product generation, and global galaxy properties measurement. As the first science result, we discuss the global MZR by integrating spectra within the effective radius. We find that the overall relation between metallicity and stellar mass of our DGIS can be the extrapolation of the higher mass end and shows a weak trend with the SFR. By comparing to a similar sample but with long-slit spectra of star-forming regions in galaxies, our IFS results are about $\sim$ 0.1 dex higher for the same metallicity calibration, indicating the existence of the aperture effect in MZR studies.

\appendix

\section{MUSE PSF}
\renewcommand{\thefigure}{A\arabic{figure}}%
\setcounter{figure}{0}
\begin{figure}
   \begin{center}
       \includegraphics[width=0.5\textwidth]{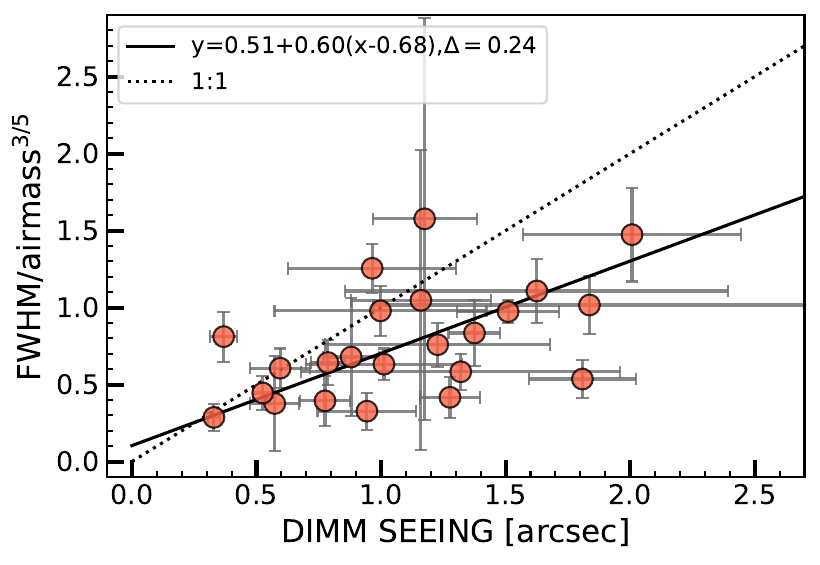}
   \end{center} 
\caption{Correlation between the point-star measured PSF and median \texttt{DIMM SEEING} from night logs.}
\label{fig:psf}
\end{figure}

\section{Spectral normalization and absolute flux calibration}
\label{sec:appendix_calib}
The galaxy CGCG035-G007 is chosen as an example to illustrate the process of spectral normalization and absolute flux calibration. Basic information about flux calibration is listed in the Table~\ref{Tab:calib}.

At first, the DESI $g$-, $r$-, $i$-, and $z$-band images are collected from the data archive. The MUSE datacube of each exposure produced by the pipeline is convolved with the $g$-, $r$-, $i$-, and $z$-band filter curves of DESI to produce synthetic broad-band images, as shown in the right panel of Fig.~\ref{fig:aperture}. The datacubes are matched to the same WCS as DESI by aligning the positions of (like-) point sources in the FOV.

The red rectangles are apertures for the target, enclosing the main body of the galaxy. The yellow rectangles are the background regions are used for background subtraction from the whole datacube after taking their median. These background regions also avoid stellar light in the IRAC 3.6 $\mu$m image. For the DESI images, the same regions are selected to subtract the background during photometry.

CGCG035-007 has 8 times exposures, each with 560s on-source time, as shown in the left panels of Fig.~\ref{fig:subsky}. As shown in Fig.~\ref{fig:subsky}, the first line is the photometry ratios of the MUSE synthetic broad-band images to the DESI real broad-band images for each exposure. A ratio close to 1 indicates that the flux of MUSE is comparable to DESI, and a flat ratio across wavelengths suggests the spectrum has a similar shape to the DESI SED. The middle row shows the median spectra within the target apertures for each exposure. The third line displays the spectral shapes of these individual spectra more clearly. As shown in Fig.~\ref{fig:subsky}(d), the pipeline-produced spectra have diverse flux levels and shapes; some exposures are even minus. After subtracting the median spectra of the selected background regions from the whole datacubes, as shown in the middle column, the spectra have more consistent flux levels and overall shape with each other, aligning better with the DESI SED, indicating that although the pipeline subtracted the background from the off-target sky and a fraction of background from datacube itself, there is residual background that over- or under- subtracted. During this subtraction process, the standard deviation of the background spectra as errors are added to the datacubes.

The spectra are distorted at shorter wavelengths but smoother and more uniform at longer wavelengths, because the shorter wavelengths are more sensitive to poor weather conditions, as shown in Fig.~\ref{fig:subsky}(e) and (g). The weather-affected, distorted spectra would contribute scatter to the final combined datacube and cause an unreliable spectral shape, since with the same exposure time, each exposure is equally weighted in the final datacube, as shown in the faint golden spectrum in Fig.~\ref{fig:fluxcal}(d). However, the pipeline can not handle these elaborate distortions. Therefore, normalizing the spectra would be meaningful to promote precision.

Firstly, the spectra are smoothed by adopting the median of bins at every 150 \AA, which outlines the overall shape and keeps detailed features, as shown in Fig.~\ref{fig:fluxcal}(a). Two exposures with notable distorted spectra shapes are removed (`20-12-26T04:13' and `20-12-27T04:15', shown in sky blue and forest green, respectively, with thinner lines). The exposure `20-12-27T04:30' (shown in blue, with thicker lines and bigger markers) is chosen as the reference since it has a smooth spectral shape and is more consistent with the DESI SED (Fig.~\ref{fig:subsky}(b)). The ratios between each spectrum and the referenced one are shown in Fig.~\ref{fig:fluxcal}(b). The ratio variations with wavelengths are fitted with a 3rd-order polynomial function, as indicated by the dotted lines in the corresponding colors. The fitting wavelength is up to 8800 \AA\ to include Ca{\sc ii} triplets while avoiding clustered skylines that affect the overall shape and cause large fluctuations. The fitted ratios for the median spectra are applied to each pixel in the datacube of every exposure, as well as to the errors, to keep S/N unchanged. After calibration, their spectra are smoother and more consistent, as shown in the right column of Fig.~\ref{fig:subsky} and smoothed spectra in Fig.~\ref{fig:fluxcal}(c). Combining all calibrated single-frame datacubes, its median spectrum has a more reasonable shape, as shown in Fig.~\ref{fig:fluxcal}(d), the slate blue one.

After normalization, all spectra are scaled to approach the referenced spectrum and then combined. Absolute flux calibration ensures that the flux observed by spectroscopy is comparable to the broad-band images, which have better light collection and flux calibration. Similarly, synthetic broad-band images of the MUSE combined datacube are produced and compared with the DESI images, as illustrated by the slate blue dots in Fig.~\ref{fig:abscal}(a). A 3rd-order polynomial function is used to fit the ratio, as shown by the slate dash-dotted line. Linear fitting induces fewer artificial effects but is less effective at modifying unreasonable curvature. High-order fitting is used cautiously; a lower degree is used instead once the fluctuation becomes large. The finally calibrated datacube shows a comparable SED to the DESI broad-band images, as shown by the magenta dots and dashed lines. The spectral comparison before and after absolute flux calibration is shown in Fig.~\ref{fig:abscal}(b).

\renewcommand{\thefigure}{B\arabic{figure}}%
\setcounter{figure}{0}

\begin{figure*}
\begin{center}
  \includegraphics[width=0.8\textwidth]{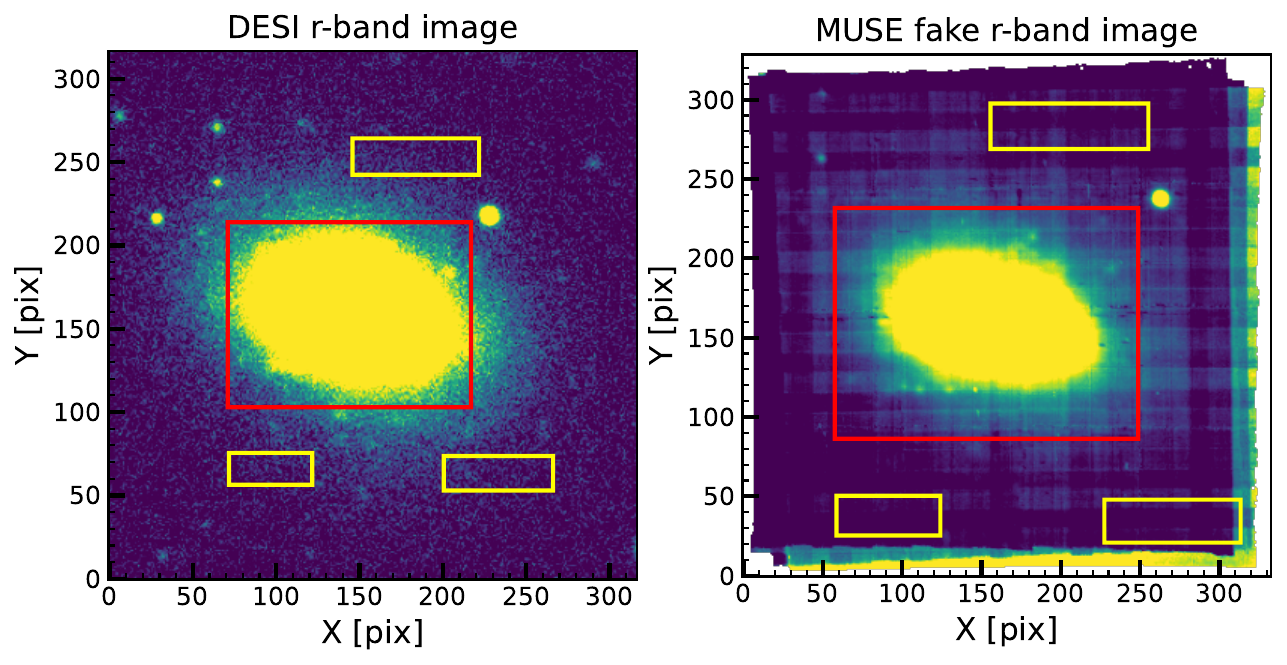}
  \caption{DESI $r$-band image (left) and MUSE moke $r$-band image (right) of CGCG035-007, overlaying apertures of the target (red) and selected backgrounds (yellow).}
  \label{fig:aperture}
\end{center}
\end{figure*}

\begin{figure*}
\begin{center}
  \includegraphics[width=0.3\textwidth]{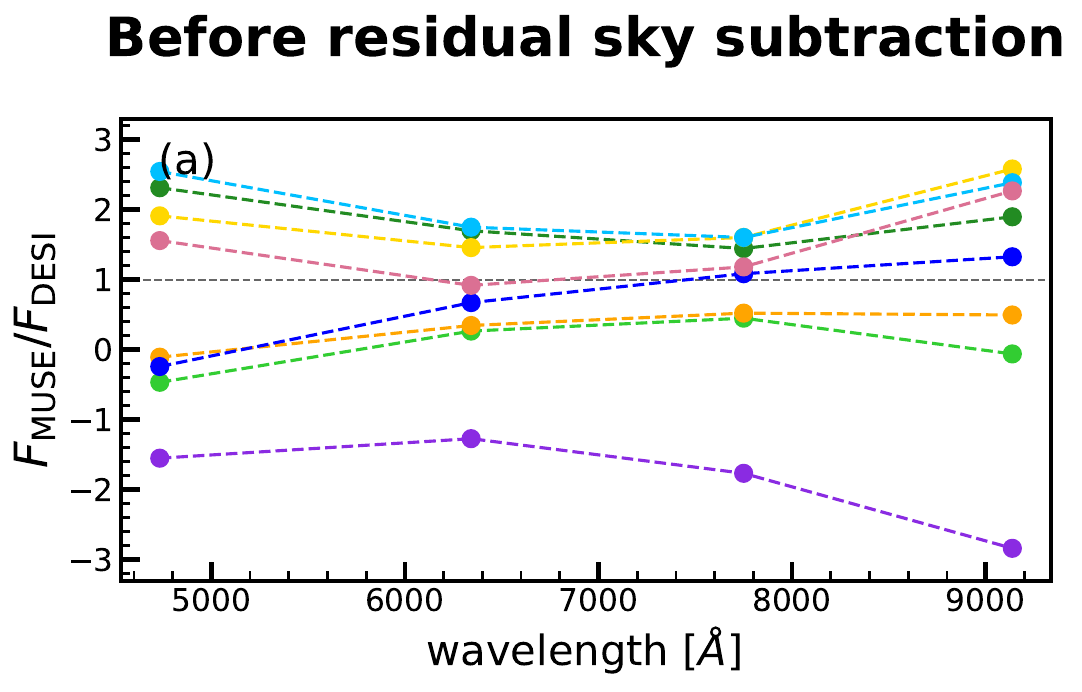}
  \includegraphics[width=0.3\textwidth]{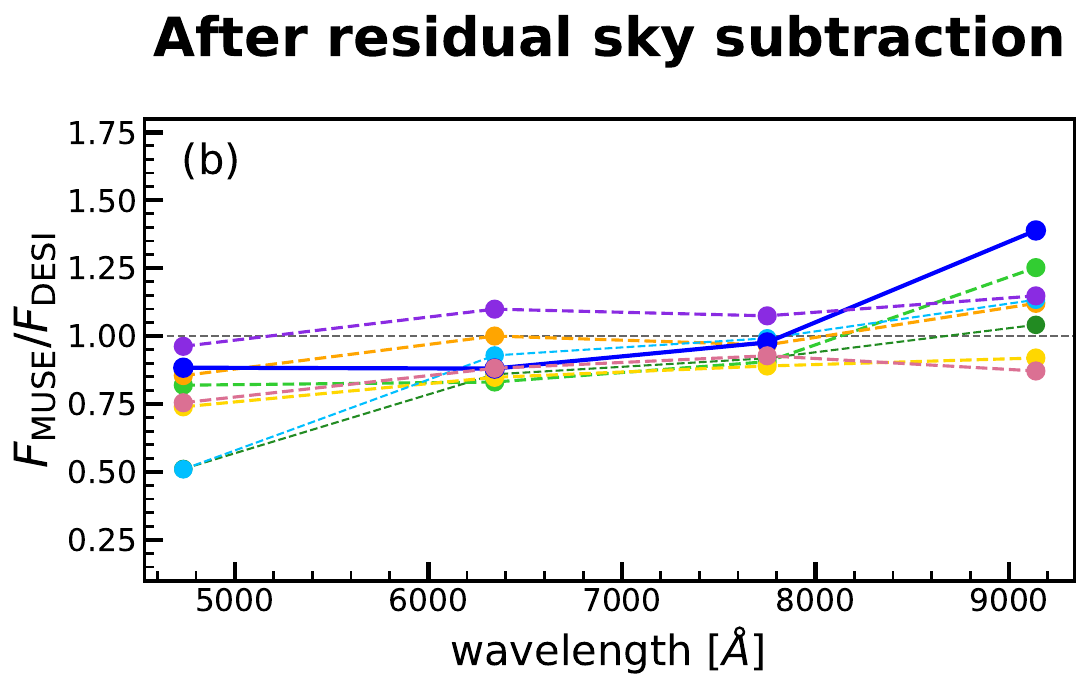}
  \includegraphics[width=0.3\textwidth]{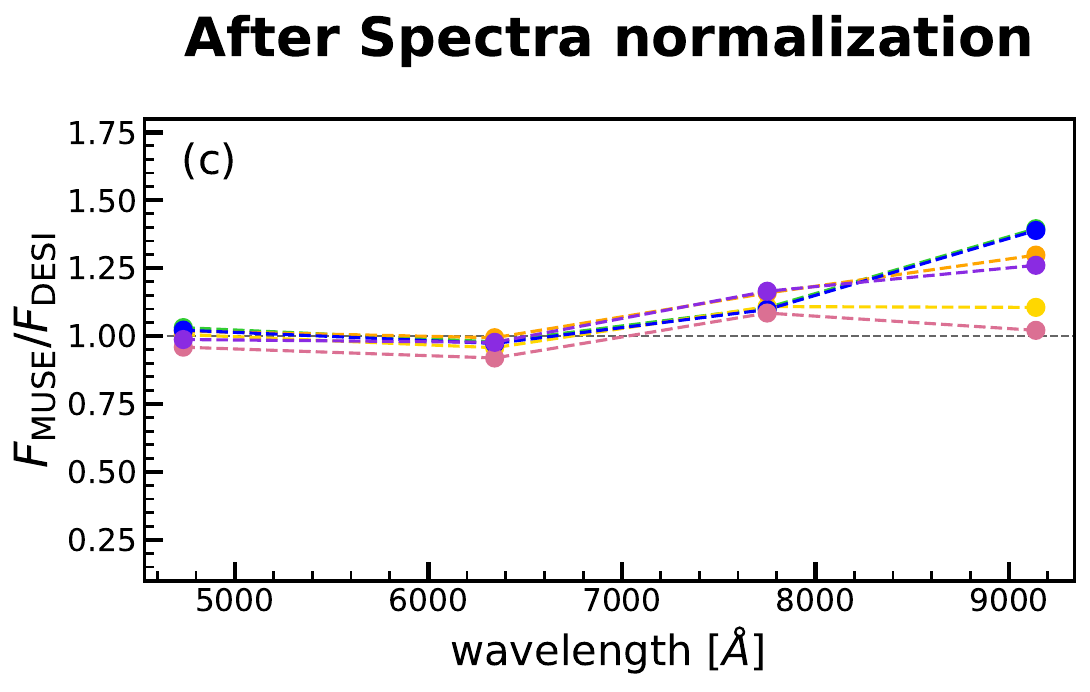}
  \includegraphics[width=0.3\textwidth]{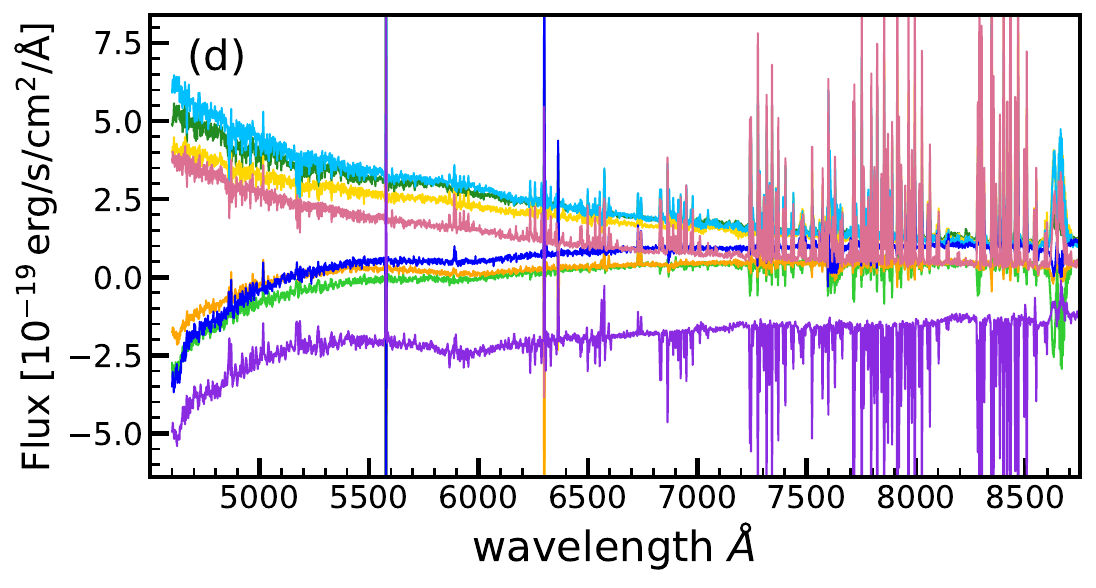}
  \includegraphics[width=0.3\textwidth]{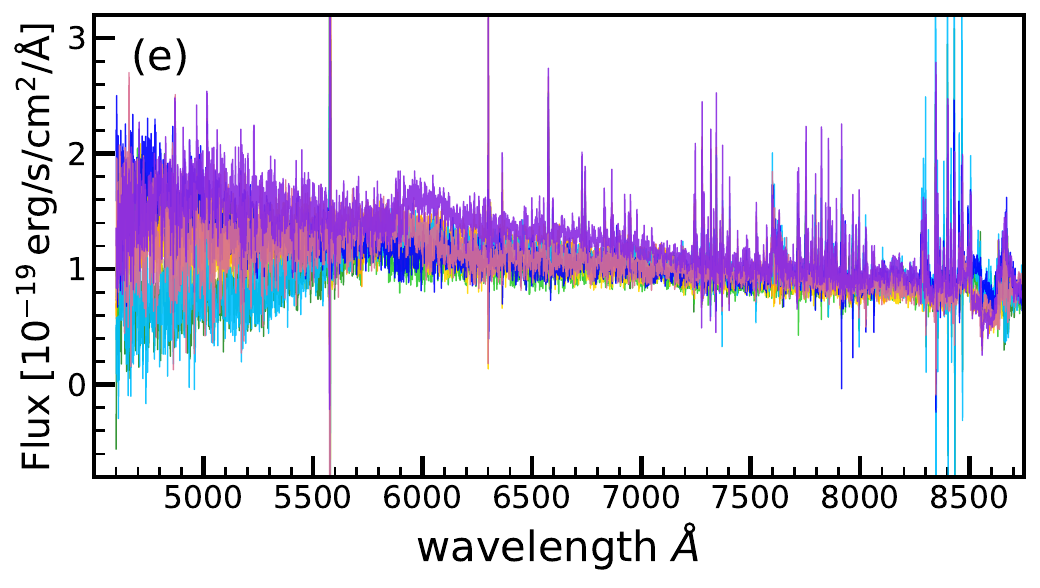}
  \includegraphics[width=0.3\textwidth]{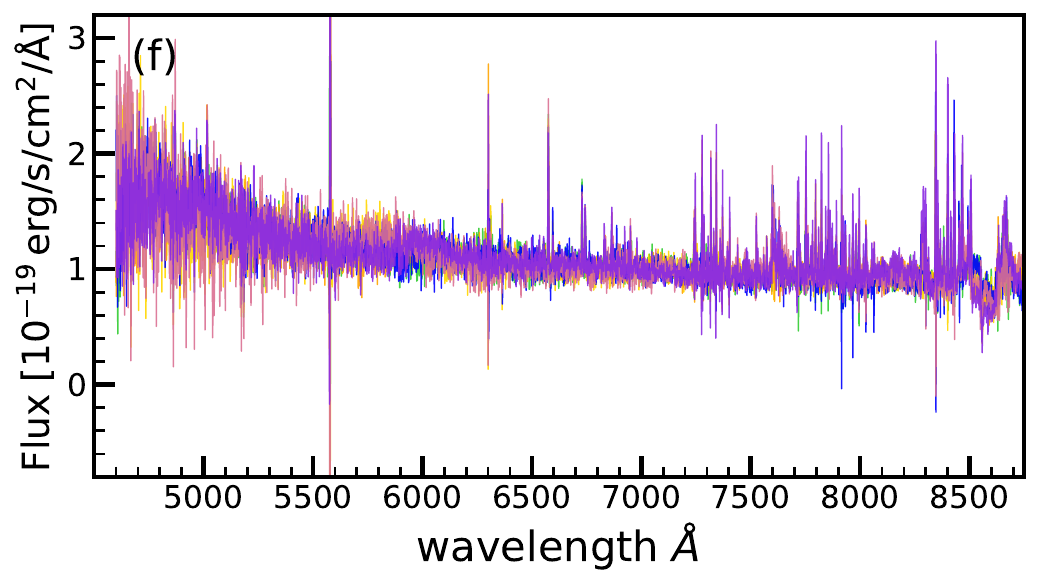}
  \includegraphics[width=0.33\textwidth]{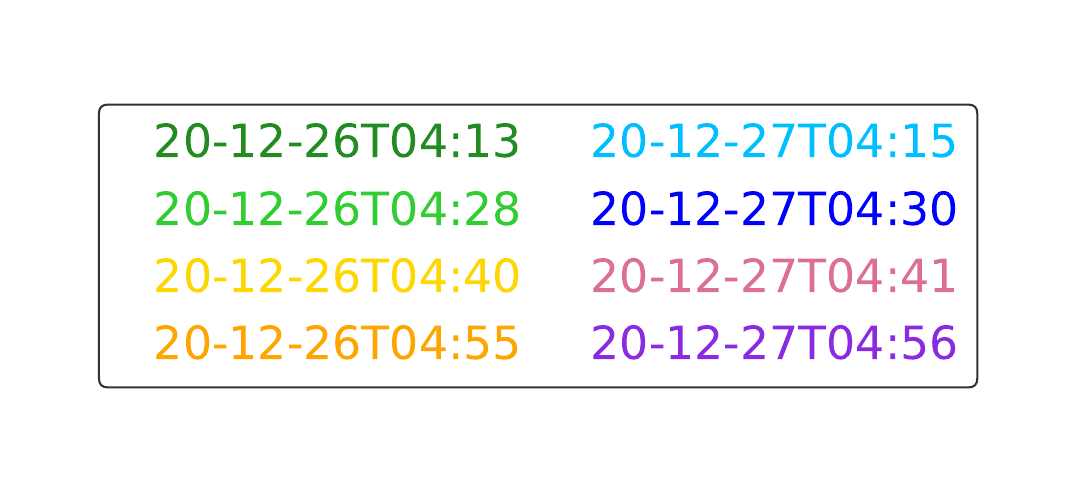}
  \includegraphics[width=0.3\textwidth]{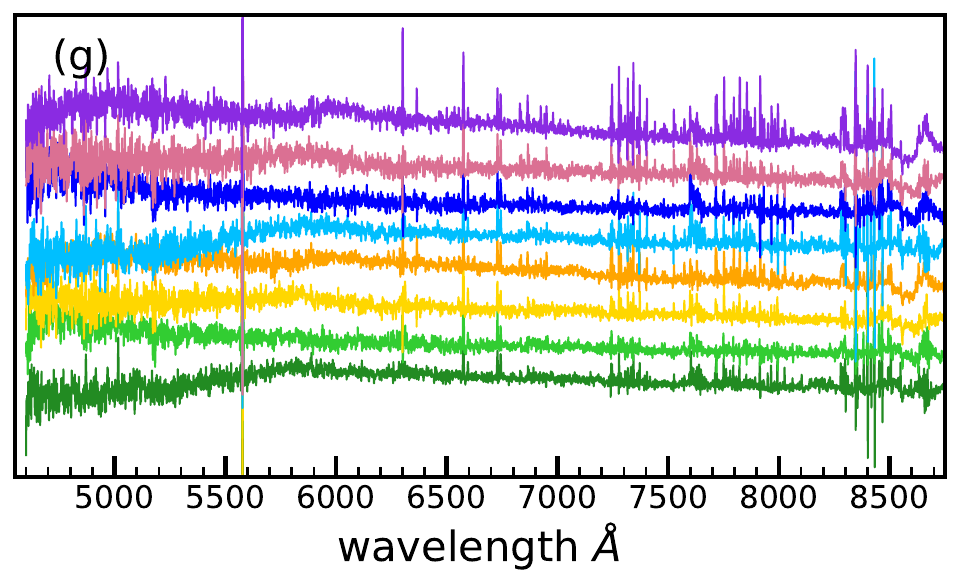}
  \includegraphics[width=0.3\textwidth]{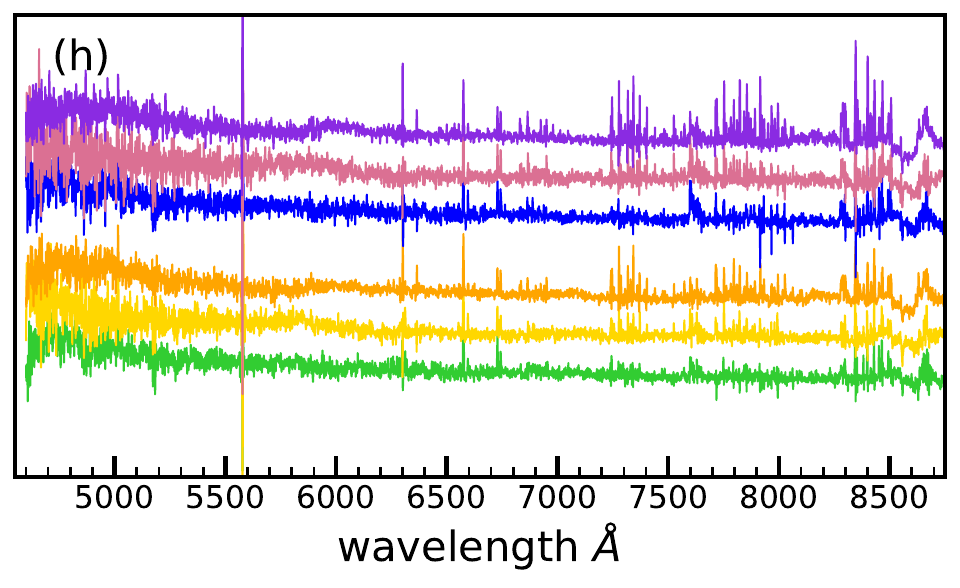}
  \caption{The single-frame datacubes of CGCG035-007 after pipeline processing (left), after subtracting residual background (middle), and after spectral normalization (right). Noted that the datacube have been subtracted the off-target sky by pipeline. (a) $\sim$ (c): MUSE fake broad-band SED of single-frame observation compared to the DESI broad-band SED;  (d) $\sim$ (f): Median spectra within apertures of single-frame observations;  (g) and (h): Same as (e) and (f) but separate them to make the shape more clearly.  The left column shows the photometric comparison and spectra for single-frame datacubes reduced by pipeline;  the middle column shows photometry and spectra of single-frame datacubes after subtracting our selected residual backgrounds;  the right column is for datacubes after performing spectral normalization. The corresponding colors of every single exposure are shown in the legend.  The reference exposure, `20-12-27T04:30', has a thicker line width with blue color.  Two exposures, `20-12-26T04:13' and `20-12-27T04:15', are thrown away due to distorted shapes, have forest green and sky blue colors, respectively, and have thinner line widths.}
  \label{fig:subsky}
\end{center}
\end{figure*}

\begin{figure*}
\begin{center}
  \includegraphics[width=0.45\textwidth]{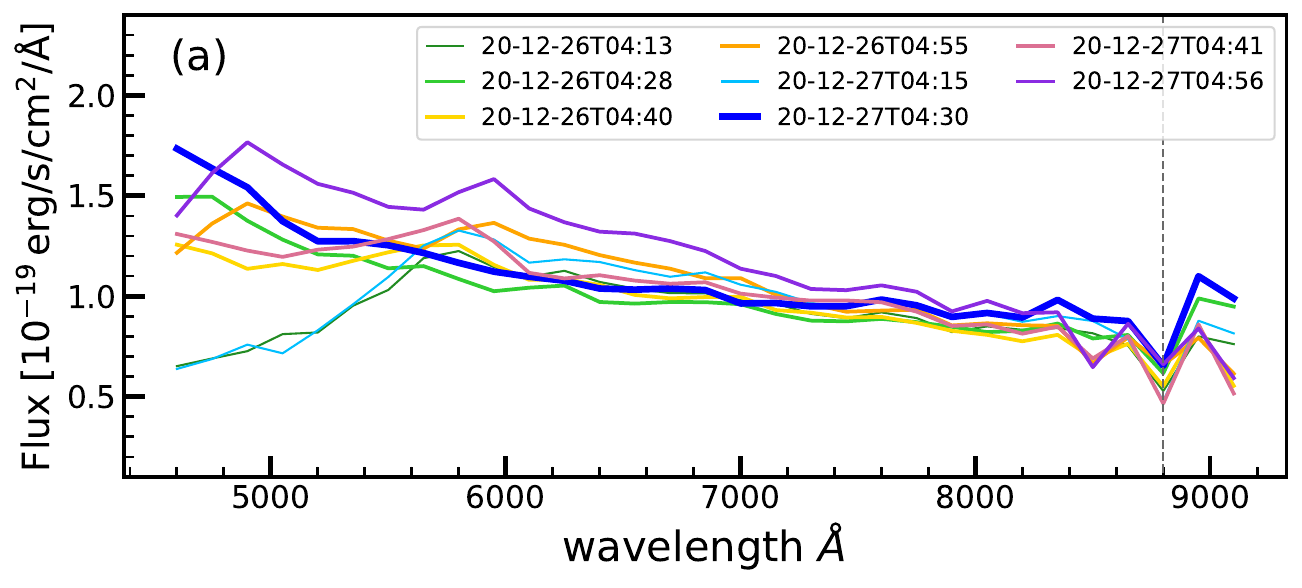}
  \includegraphics[width=0.45\textwidth]{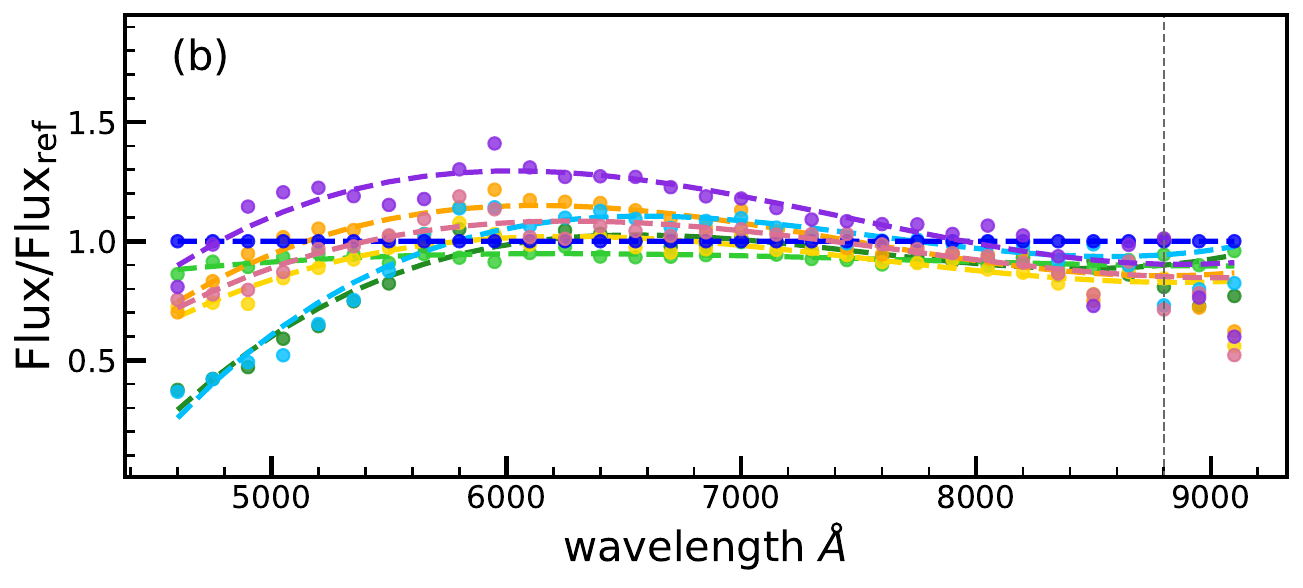}
  \includegraphics[width=0.45\textwidth]{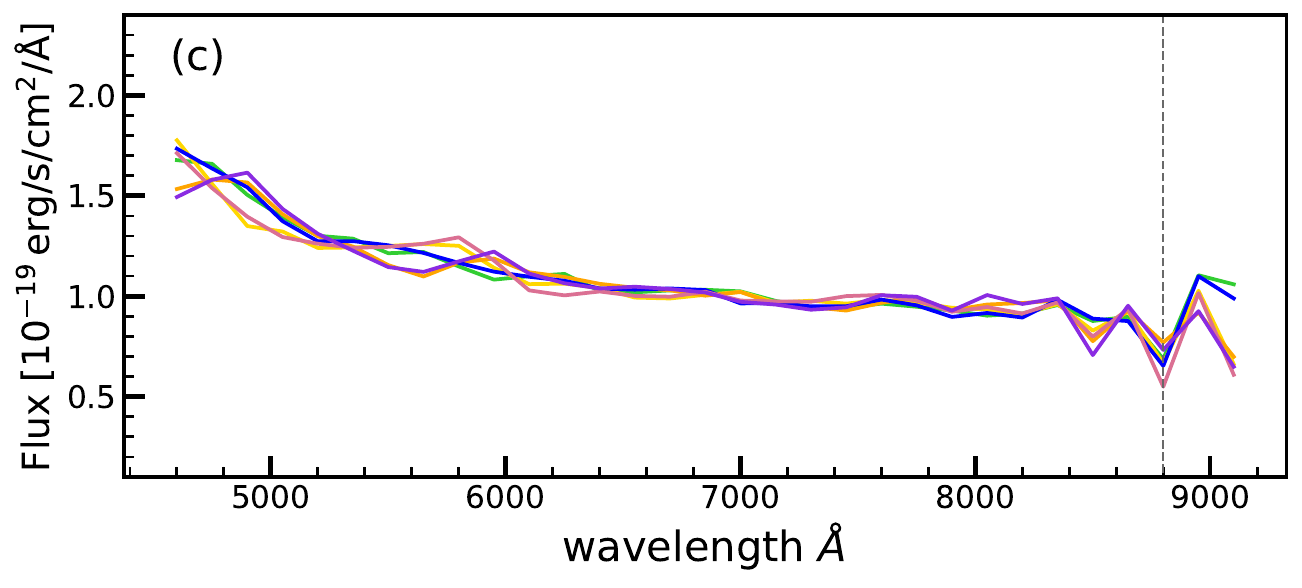}
  \includegraphics[width=0.45\textwidth]{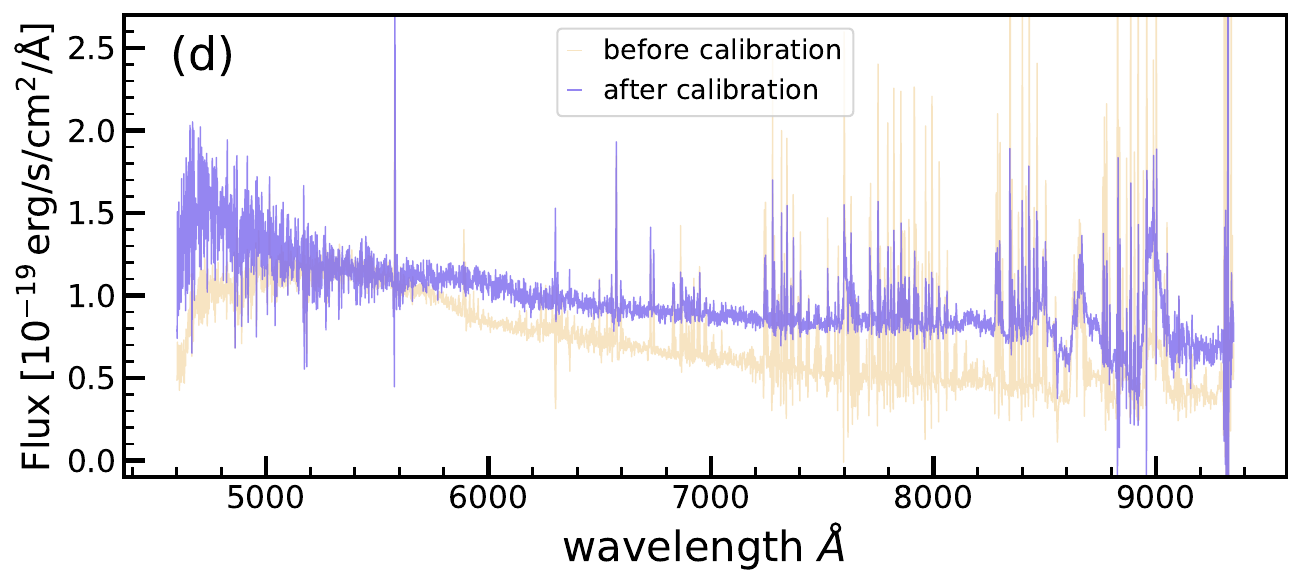}
  \caption{The procedure of spectral normalization for CGCG035-007. (a) Median spectra of single-frame spectra within the target aperture after subtracting selected backgrounds, smoothed by adopting median value at every 150 \AA.  All the spectra are normalized to the reference spectrum `20-12-27T04:30' with a blue color and thicker line.  The excluded two exposures are labeled thinner lines. (b) The dots are flux ratios between the referenced smoothed spectrum and the others, the dashed lines are the fitted 3-rd polynomial functions. The fittings are up to a wavelength of 8800 \AA, as shown in the vertical line. (c) Smoothed spectra of single-frame median spectra after normalization, excluding the two exposures with distorted shapes. (d) Median spectra of combined datacubes produced by pipeline (faint gold) and this work after subtracting residual background and normalizing spectral shape (blue-violet).}  
  \label{fig:fluxcal}
\end{center}
\end{figure*}

\begin{figure*}
\begin{center}
  \includegraphics[width=0.43\textwidth]{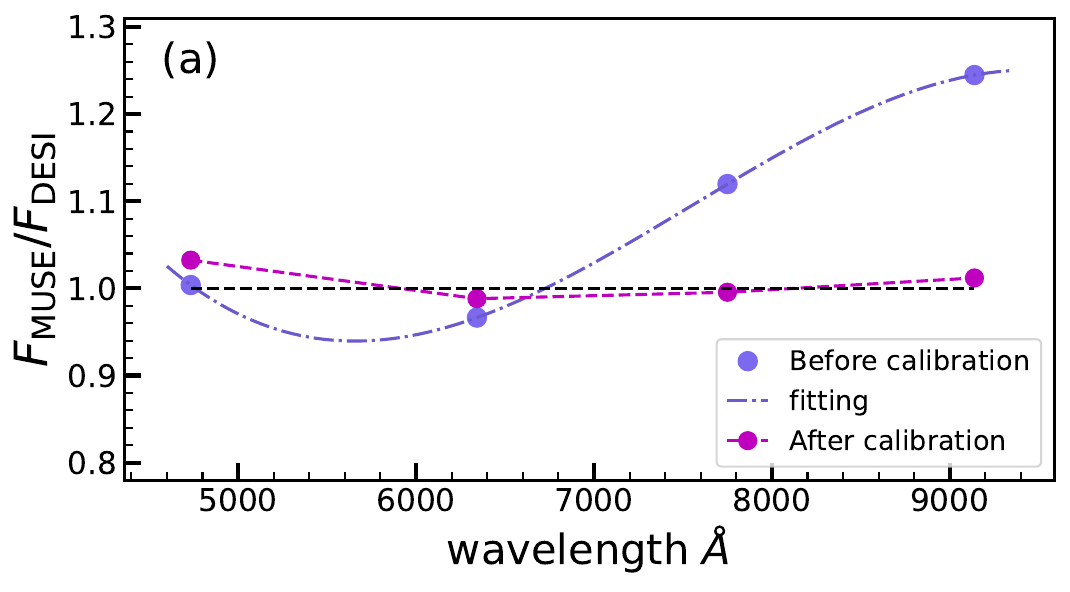}
  \includegraphics[width=0.5\textwidth]{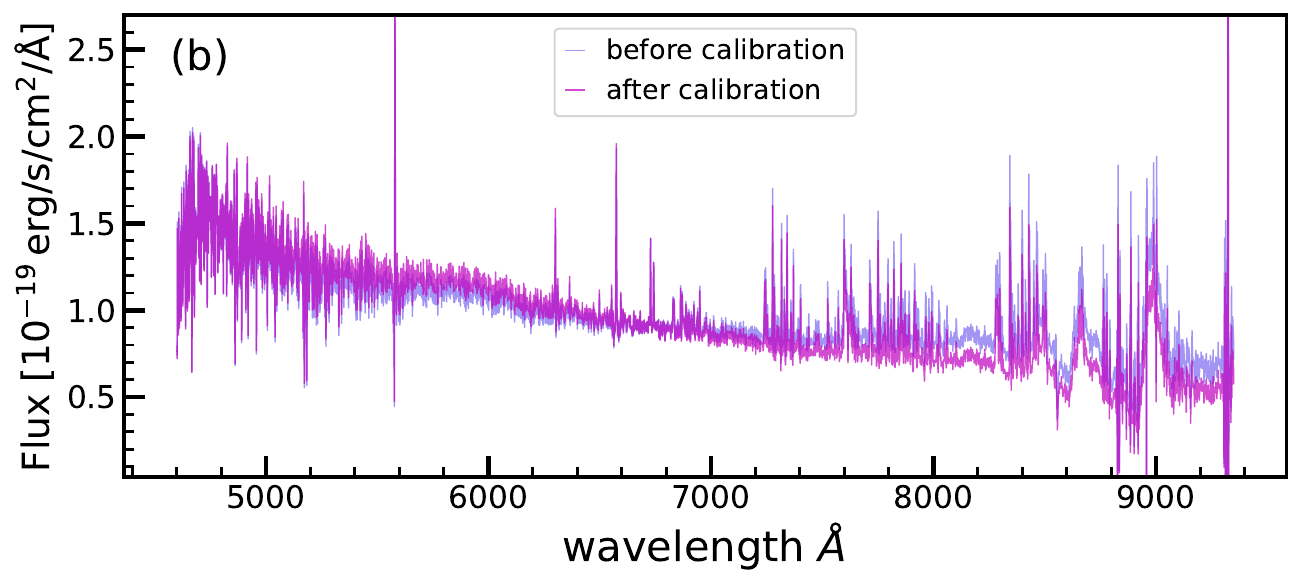}
  \caption{Absolute flux calibration for CGCG035-007. The combined datacube produced by the previous step in blue-violet color, after absolute calibration in magenta. (a) Flux densities within the target aperture of MUSE fake broad-band image compared to DESI. The blue-violet dashed line is the fitted 3-rd polynomial function. (b) Median spectra within the target aperture of combined datacubes before and after absolute flux calibration.}
  \label{fig:abscal}
\end{center}
\end{figure*}

\renewcommand{\thetable}{B\arabic{table}}%
\setcounter{table}{0}
\begin{table*}
\setlength{\tabcolsep}{3pt} 
\noindent
\tiny
\begin{flushleft}
\begin{tabular}{cccccccccccccccccc}
\hline
\hline
I.D. & Target &  RA  &   DEC   & ExpTime & Survey &  Background & Poly &  &  & I.D. & Target &  RA  &   DEC   & ExpTime & Survey & Background & Poly \\
    &     & (J2000) & (J2000) &  (s)    &        &              &      &  &  &     & (J2000) & (J2000) &  (s)    &        &              & \\
(1) & (2) &  (3)    & (4)     &  (5)    &   (6)  &      (7)     & (8)  &  &  & (1) & (2) &  (3)    & (4)     &  (5)    &   (6)  &      (7)     & (8)\\
\hline
1 & WLM & 0.49233 & -15.46092 & 3600.0 & PanStarrs & N & 3 &  &  & 22 & UGC02716 & 51.03024 & 17.75304 & 3920.0 & PanStarrs & Y & 3 \\
  &     & 0.50833 & -15.46139 & 1160.0 & PanStarrs & N & 3 &  &  & 23 & IC1959 & 53.3024 & -50.4142 & 2240.0 & DESI & Y & 3 \\
  &     & 0.50417 & -15.49333 & 1740.0 & PanStarrs & N & 3 &  &  & 25 & NGC1522 & 61.53292 & -52.66831 & 2200.0 & DESI & Y & 3 \\
  &     & 0.4875  & -15.44667 & 1740.0 & PanStarrs & N & 1 &  &  & 26 & ESO483-G013 & 63.17133 & -23.15889 & 2240.0 & DESI & Y & 3 \\ 
2 & NGC0059 & 3.85469 & -21.44436 & 3360.0 & PanStarrs & Y & 3   &  &  & 27 & ESO158-G003 & 71.56958 & -57.34306 & 1680.0 & DESI & Y & 3 \\
4 & ESO294-G010 & 6.63904 & -41.85531 & 1800.0 & DESI & N & 1 &  &  & 28 & ESO119-G016 & 72.87167 & -61.65083 & 1680.0 & DESI & Y & 1 \\
5 & IC1574 & 10.76592 & -22.24689 & 4480.0 & DESI & Y & 3 &  &  & 30 & ESO486-G021 & 75.83204 & -25.42292 & 8400.0 & DESI & Y & 3 \\
7 & UGCA015 & 12.455 & -21.015 & 2280.0 & DESI & Y & 1 &  &  &  31 & NGC1800 & 76.60717 & -31.95422 & 3160.0 & DESI & Y & 3 \\
8 & ESO540-G032 & 12.60133 & -19.90672 & 2280.0 & DESI & Y & 3 &  &  & 33 & CGCG035-007 & 143.68633 & 6.4255 & 3300.0 & DESI & Y & 3 \\
9 & UGC00668 & 16.22583 & 2.11194 & 810.0 & PanStarrs & N & 3 &  &  & 35 & UGC05288 & 147.82184 & 7.82557 & 6720.0 & DESI & Y & 1 \\
  &          & 16.22579 & 2.13331 & 990.0 & PanStarrs & N & 3 &  &  & 36 & UGC05373 & 150.00042 & 5.33222 & 2400.0 & DESI & Y & 3 \\
  &          & 16.20444 & 2.13333 & 990.0 & PanStarrs & N & 1 &  &  & 37 & UGCA193 & 150.64958 & -6.01361 & 5560.0 & DESI & Y & 3 \\
  &          & 16.24722 & 2.09077 & 990.0 & PanStarrs & N & 3 &  &  & 39 & AM1001-270 & 151.01708 & -27.331 & 2280.0 & PanStarrs & N & 1 \\
  &          & 16.22583 & 2.15472 & 990.0 & PanStarrs & N & 3 &  &  & 44 & UGC05923 & 162.28145 & 6.91739 & 1140.0 & DESI & Y & 1 \\
  &          & 16.24722 & 2.15472 & 810.0 & PanStarrs & N & 3 &  &  & 46 & ESO321-G014 & 183.45675 & -38.23136 & 5200.0 & SkyMapper & Y & 3 \\
  &          & 16.20444 & 2.11194 & 810.0 & PanStarrs & N & 3 &  &  & 48 & UGC08091 & 194.66747 & 14.21771 & 2280.0 & DESI & Y & 3 \\
  &          & 16.20444 & 2.15472 & 990.0 & PanStarrs & N & 1 &  &  & 50 & UGCA320 & 195.81973 & -17.42303 & 3920.0 & PanStarrs & Y & 1 \\
  &          & 16.24722 & 2.11194 & 990.0 & PanStarrs & N & 1 &  &  & 51 & MCG-03-34-002 & 196.98607 & -16.6891 & 2280.0 & PanStarrs & Y & 3 \\
10 & UGC00685 & 16.84276 & 16.685 & 7920.0 & DESI & Y & 3  &  &  & 52 & IC4247 & 201.68708 & -30.3625 & 2280.0 & PanStarrs & Y & 3 \\
11 & UGC00695 & 16.9434 & 1.06364 & 7832.794 & DESI & Y & 3 &  &  & 53 & ESO444-G084 & 204.33329 & -28.045 & 3360.0 & PanStarrs & Y & 3 \\
12 & UGC00891 & 20.32954 & 12.41172 & 1120.0 & DESI & Y & 3 &  &  & 56 & KKH086 & 208.63979 & 4.243 & 2400.0 & DESI & N & 3 \\
13 & UGC01056 & 22.19679 & 16.68797 & 15120.0 & DESI & Y & 3 &  &  & 61 & UGCA438 & 351.61467 & -32.38875 & 2240.0 & SkyMapper & Y & 2 \\
15 & NGC0625 & 23.7693 & -41.4362 & 2240.0 & DESI & Y & 3 &  &  & 63 & UGC12613 & 352.14667 & 14.74306 & 690.0 & DESI & N & 3 \\
17 & ESO245-G005 & 26.26558 & -43.59803 & 2400.0 & DESI & N & 1  &  &  &     &          & 352.16828 & 14.73127 & 1035.0 & DESI & N & 2 \\
19 & ESO115-G021 & 39.447 & -61.33669 & 4800.0 & DESI & Y & 3 &  &  &     &          & 352.12506 & 14.75484 & 810.0 & DESI & N & 1 \\
21 & NGC1311 & 50.02892 & -52.18547 & 3300.0 & DESI & Y & 2   &  &  & 64 & UGCA442 & 355.93979 & -31.95677 & 2240.0 & SkyMapper & Y & 3 \\
\hline
\end{tabular}
\end{flushleft}
\caption{Parameters of flux calibration for MUSE observation. (1) $\sim$ (2) The index and name of sample, same as Table~2; (3) $\sim$ (4) RA and DEC of the pointings; (5) on-source time of finally adopted exposure of the synthetic datacube; (6) the referenced broad-band image for flux calibration; (7) if subtract additional background in the FOV of datacube; (8) the degree of polynomial function at spectral normalization.}
\label{Tab:calib}
\end{table*}

\clearpage
\section{Stacked spectra}

\renewcommand{\thefigure}{C\arabic{figure}}%
\setcounter{figure}{0}
\begin{figure*}
\label{fig:spec}
  \includegraphics[width=\textwidth]{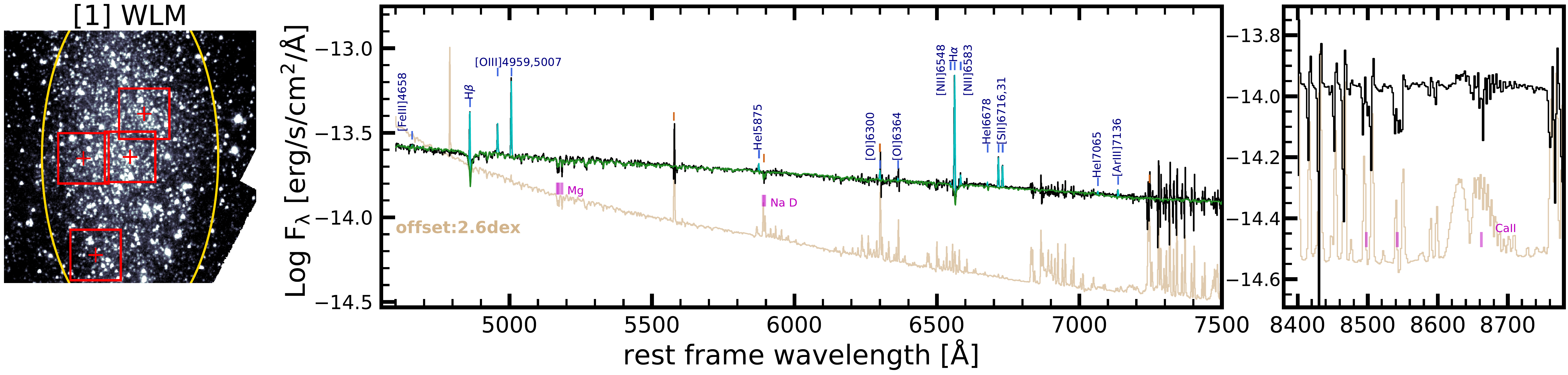}
  \includegraphics[width=\textwidth]{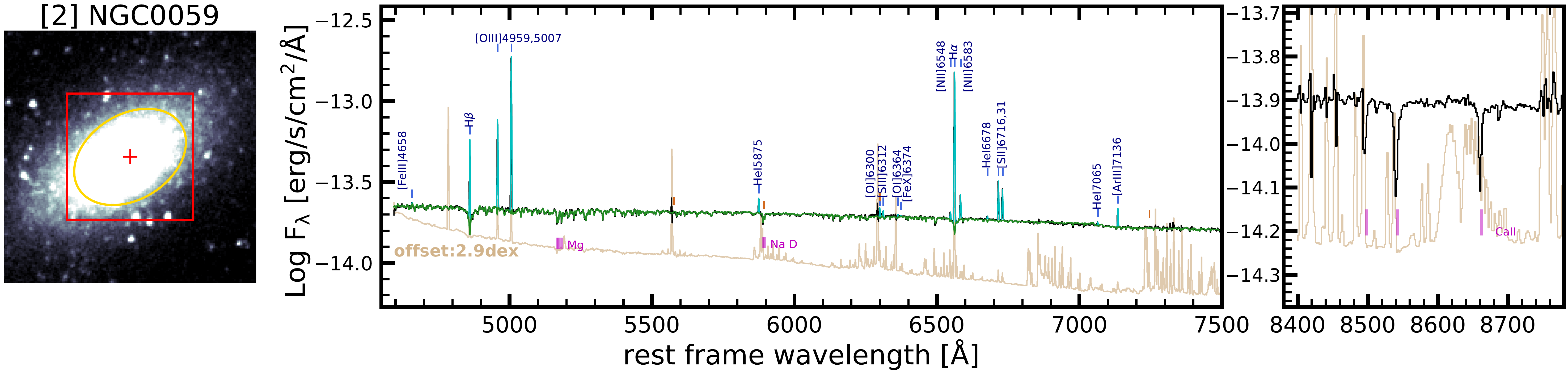}
  \newline
  \newline
  \includegraphics[width=\textwidth]{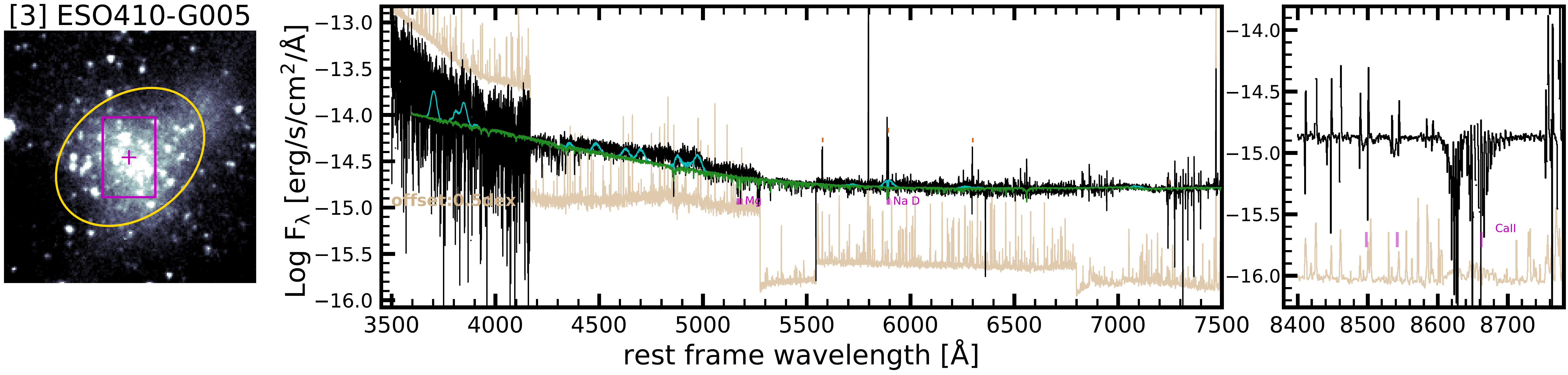}
  \includegraphics[width=\textwidth]{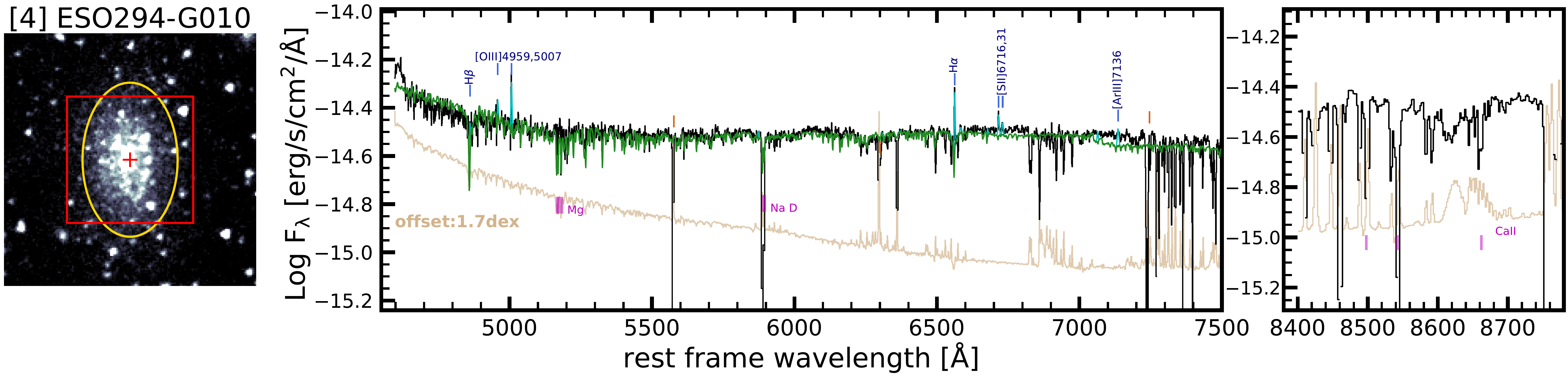}
  \includegraphics[width=\textwidth]{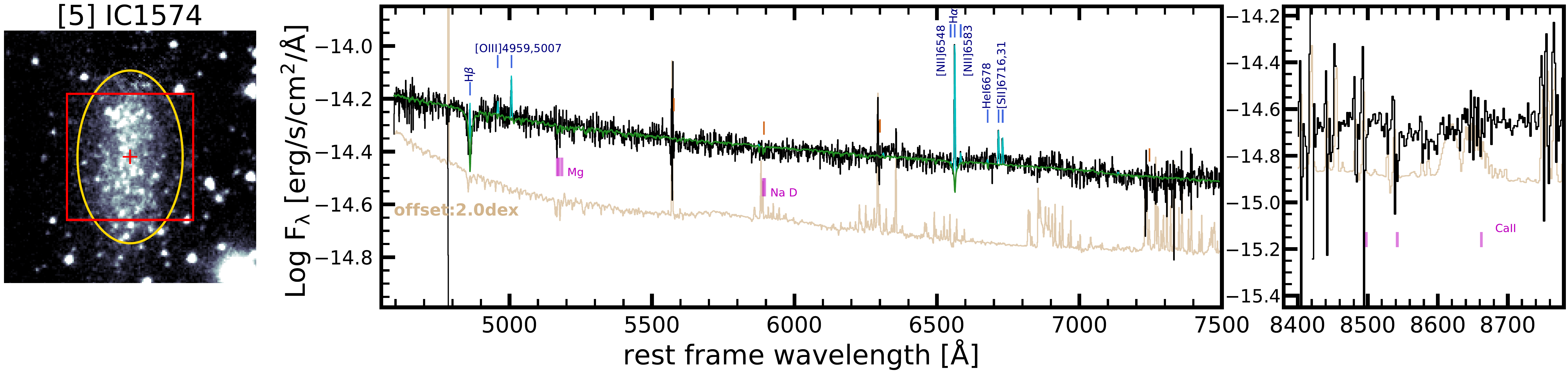}
\caption{Panoramas and global spectra of galaxies in DGIS observed by VLT/MUSE (red) and ANU-2.3m/WiFeS (magenta). {\rm left:} IRAC 3.6 $\mu$m background overlap observing FOVs (rectangles) with pointings (crosses) and R$_{\rm e}$ outlines (yellow ellipses). {\bf Middle:} global spectra (black) and errors (faint gold, adding offsets to save space) up to 7500 \AA, at rest frame. The green lines are \texttt{pPXF} fitted continuum, the sky-blue lines are \texttt{pPXF} fitted gas emitting spectra. The \texttt{pPXF} fitted gas emission lines with S/N $>10$ are labeled as short blue vertical lines and texts, the absorption lines are labeled as magenta vertical lines and texts, and the skylines are labeled as short orange vertical lines. {\bf Right:} the global spectra and errors (with the same offset) at Ca{\sc iii} triplet wavelength windows, at rest frame. The position of Ca {\sc iii} triplet is labeled as short magenta vertical lines.}
\end{figure*}

\begin{figure*}
  \includegraphics[width=0.17\textwidth]{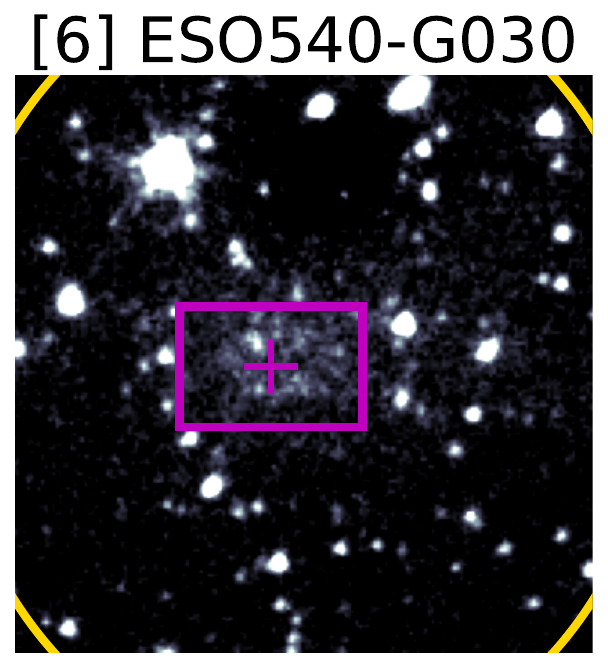}
  \newline
  \newline  
  \includegraphics[width=\textwidth]{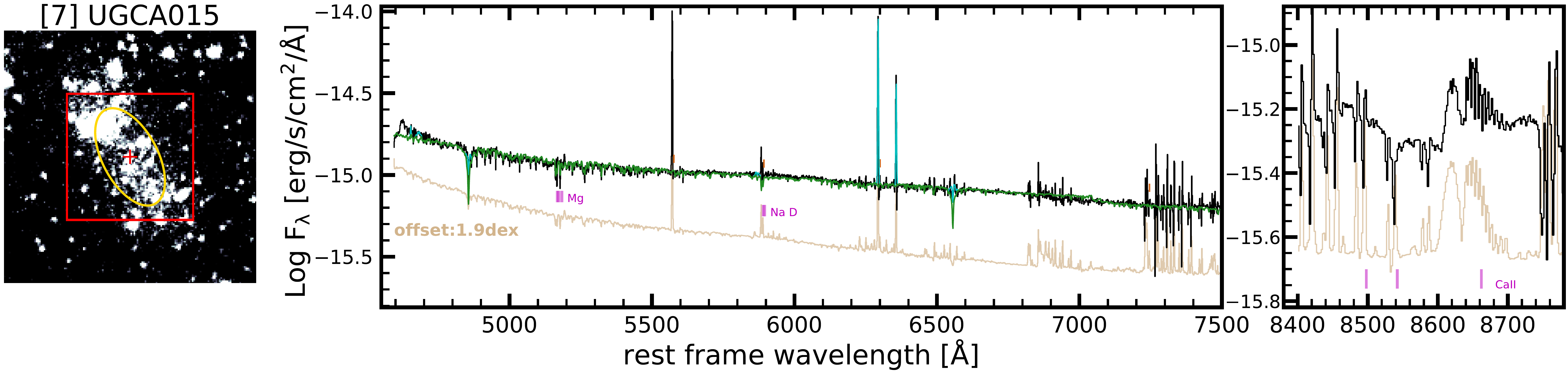}
  \includegraphics[width=\textwidth]{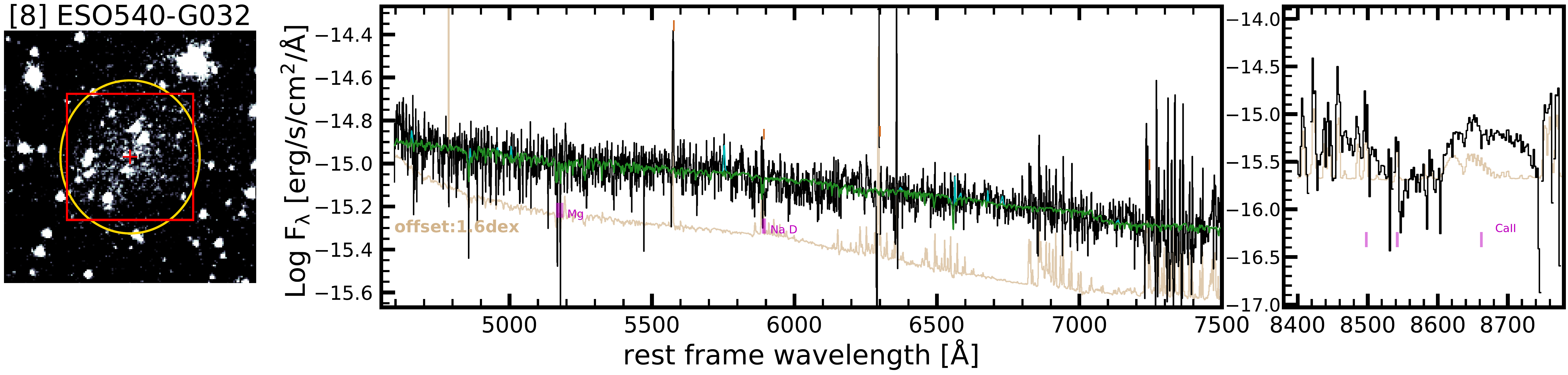}
  \includegraphics[width=\textwidth]{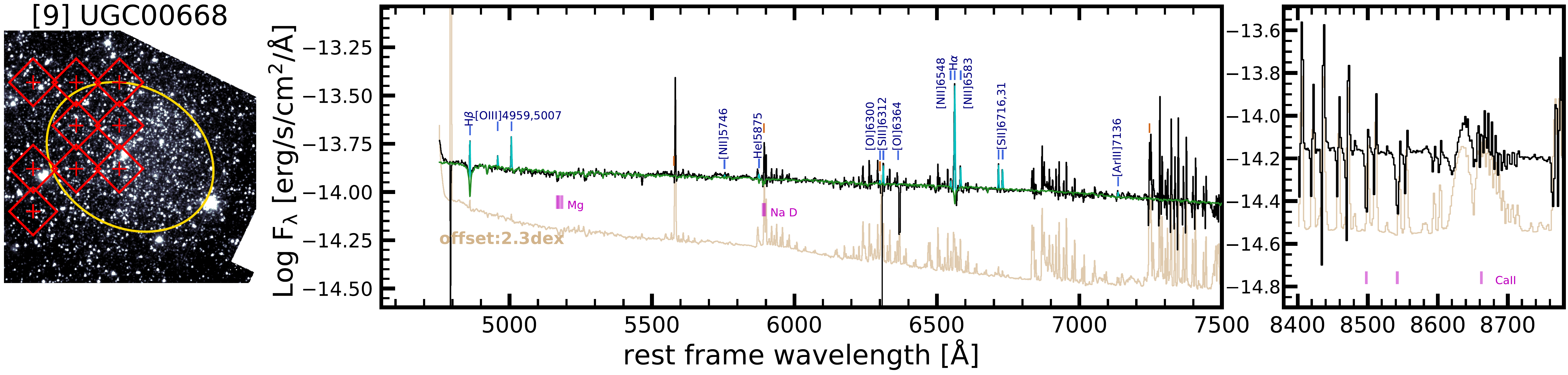}
  \includegraphics[width=\textwidth]{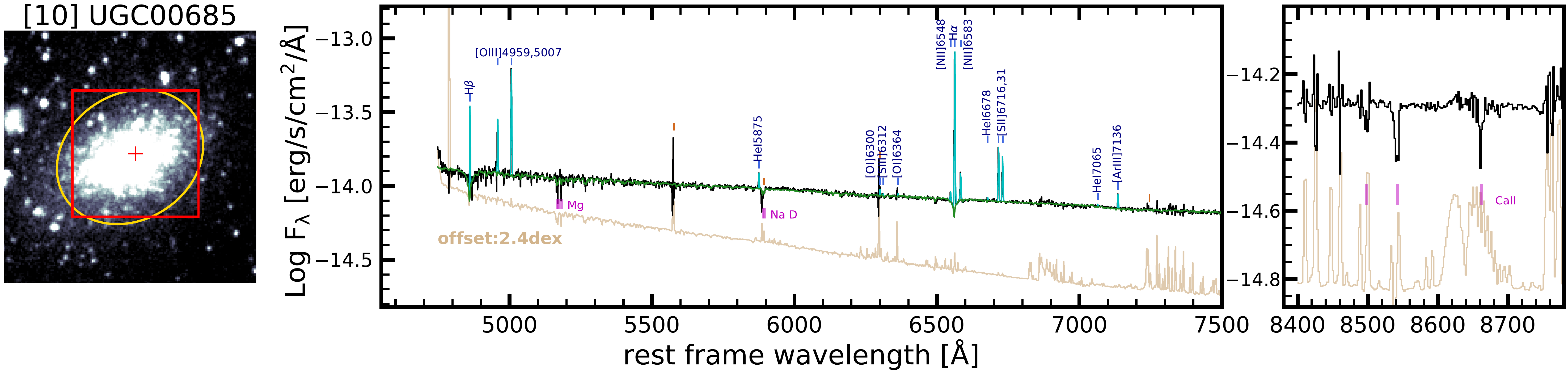}
\caption{continnue}
\end{figure*}

\begin{figure*}
  \includegraphics[width=\textwidth]{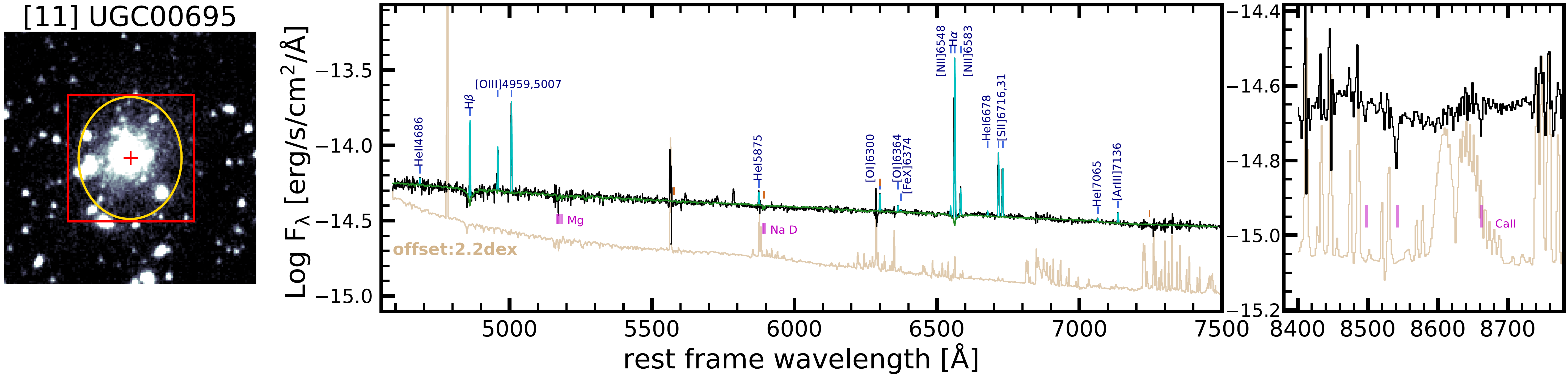}
  \includegraphics[width=\textwidth]{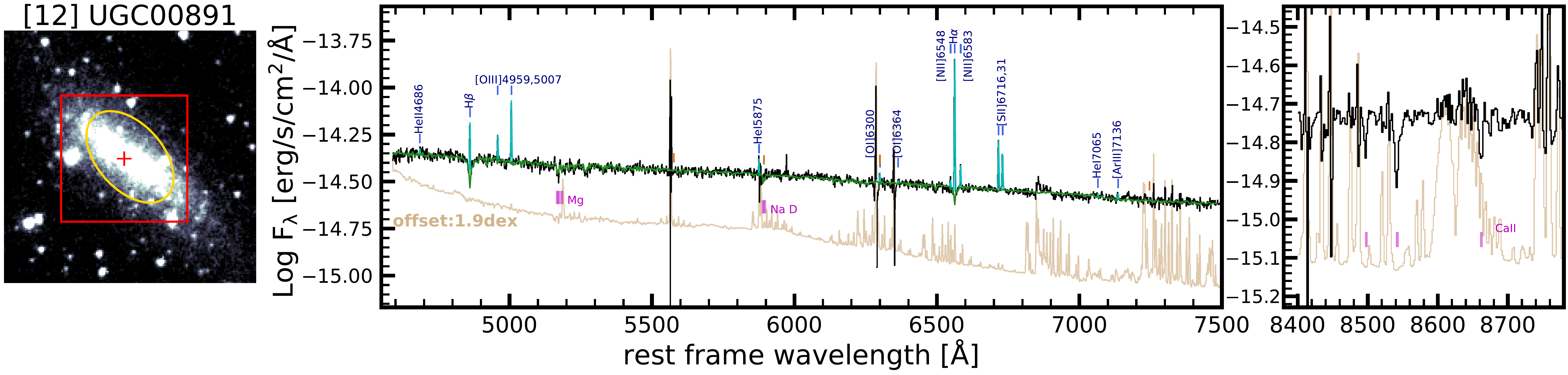}
  \includegraphics[width=\textwidth]{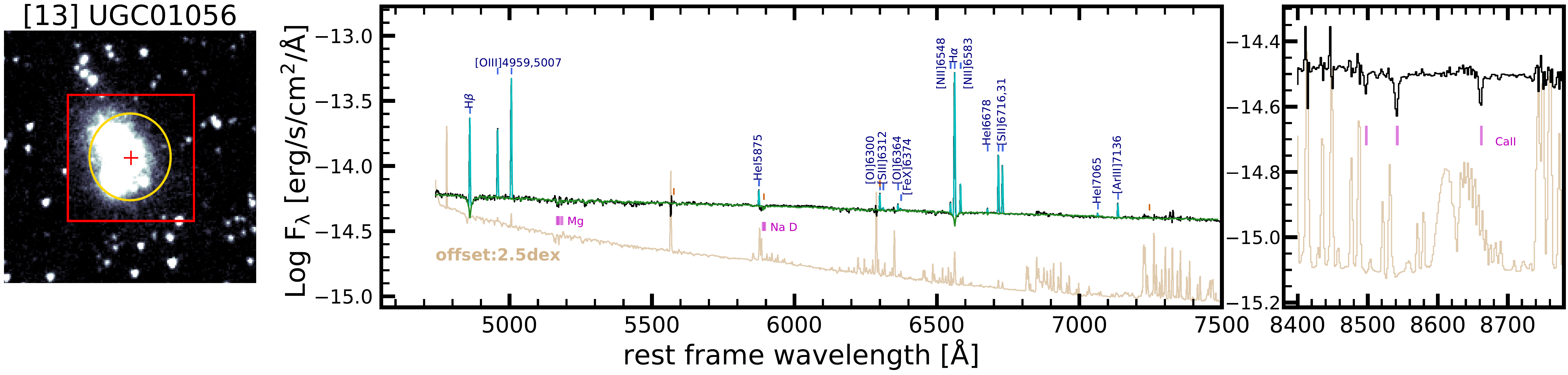}
  \includegraphics[width=0.17\textwidth]{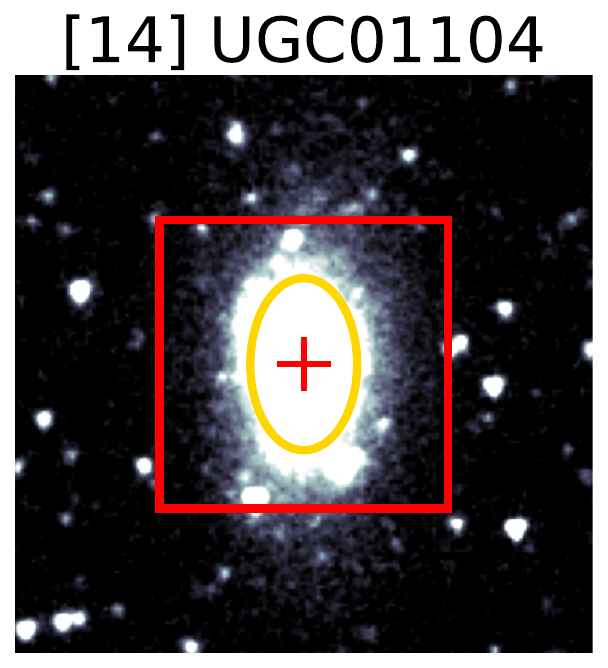}\\
  \newline
  \includegraphics[width=\textwidth]{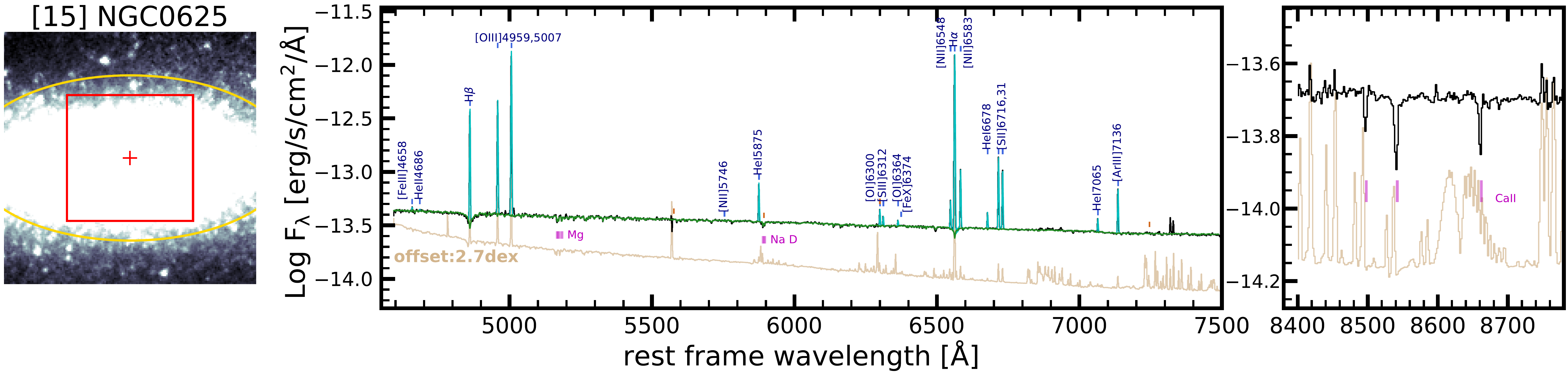}
\caption{continnue}
\end{figure*}

\begin{figure*}
  \includegraphics[width=0.17\textwidth]{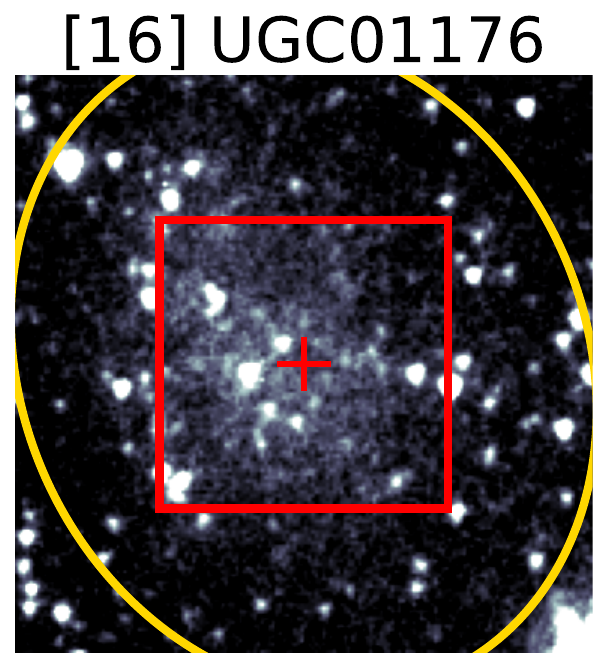}
  \newline
  \newline
  \includegraphics[width=\textwidth]{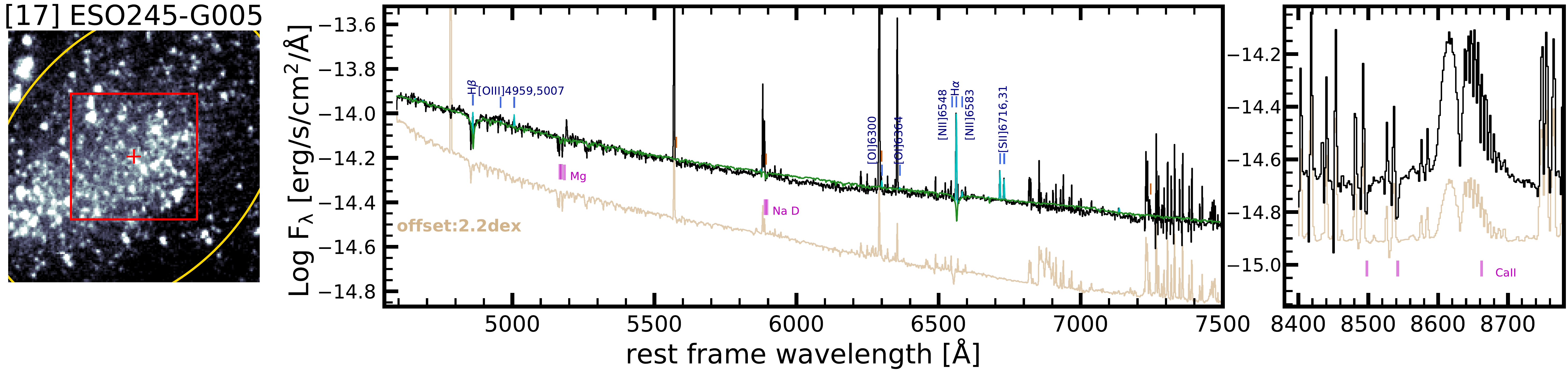}
  \includegraphics[width=0.17\textwidth]{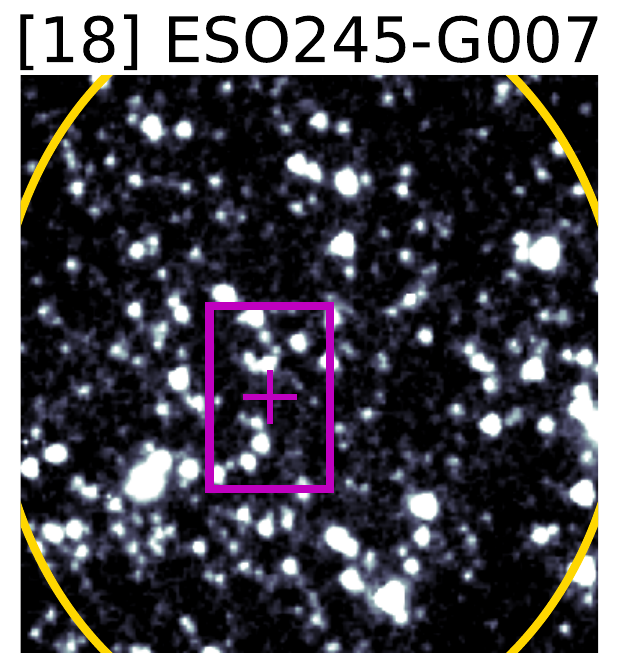}
  \newline
  \newline
  \includegraphics[width=\textwidth]{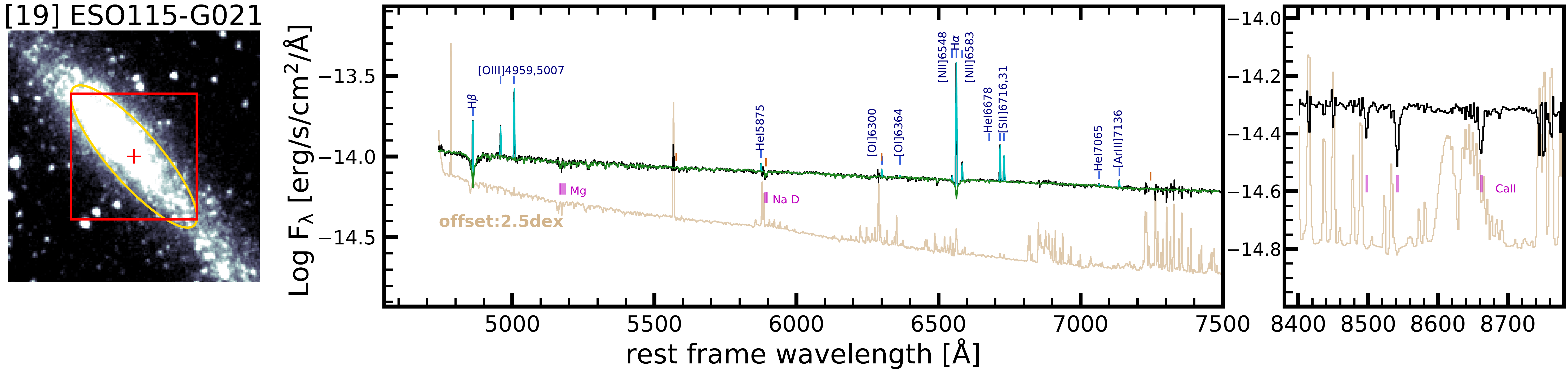}
  \includegraphics[width=0.17\textwidth]{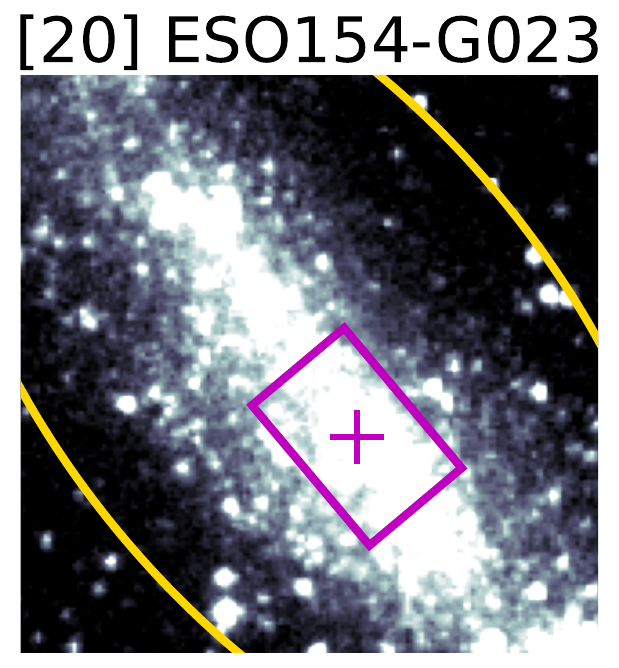}
  \newline
  \newline
\caption{continnue}
\end{figure*}

\begin{figure*}
  \includegraphics[width=\textwidth]{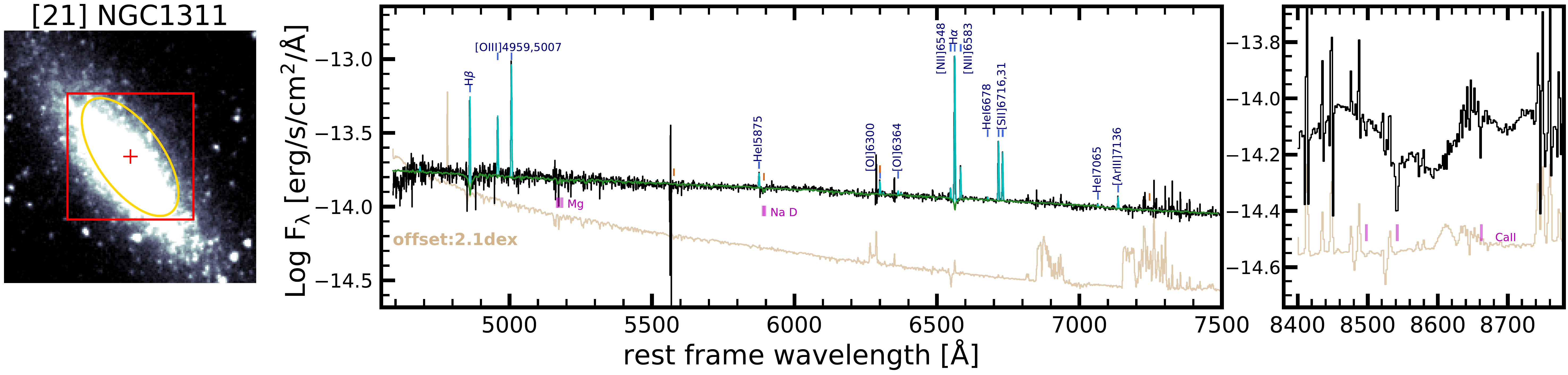}
  \includegraphics[width=\textwidth]{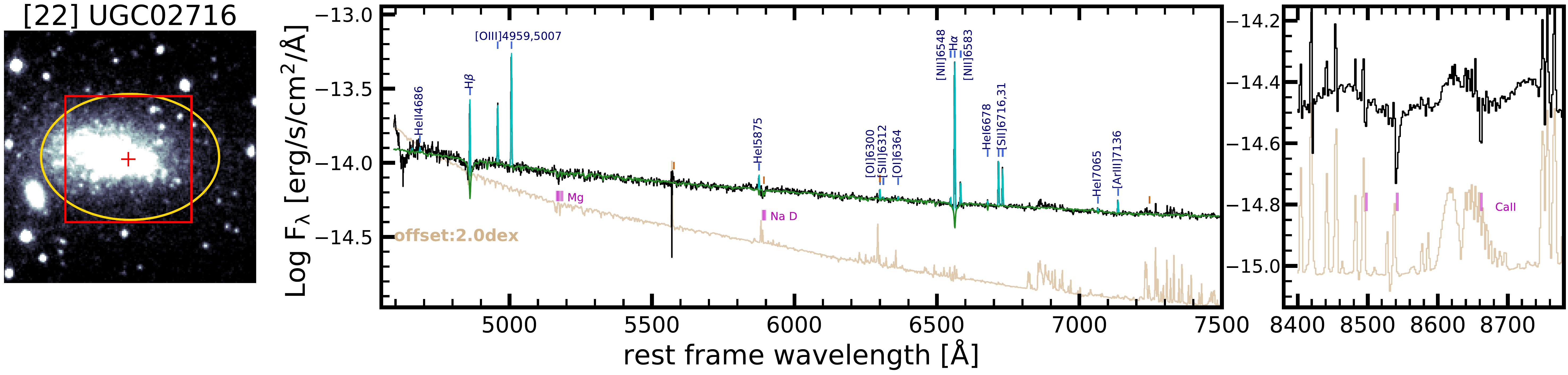}
  \includegraphics[width=\textwidth]{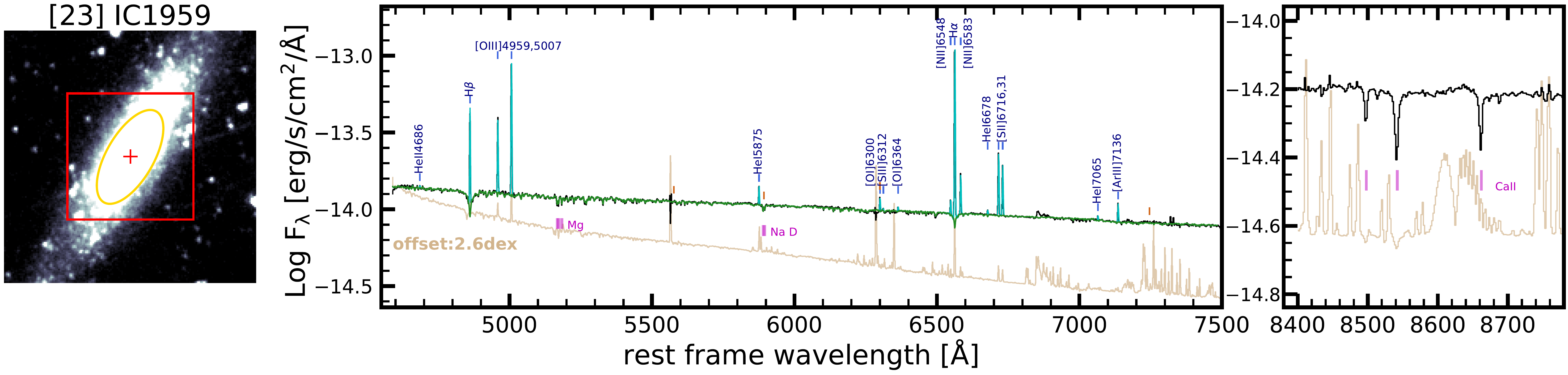}
  \includegraphics[width=\textwidth]{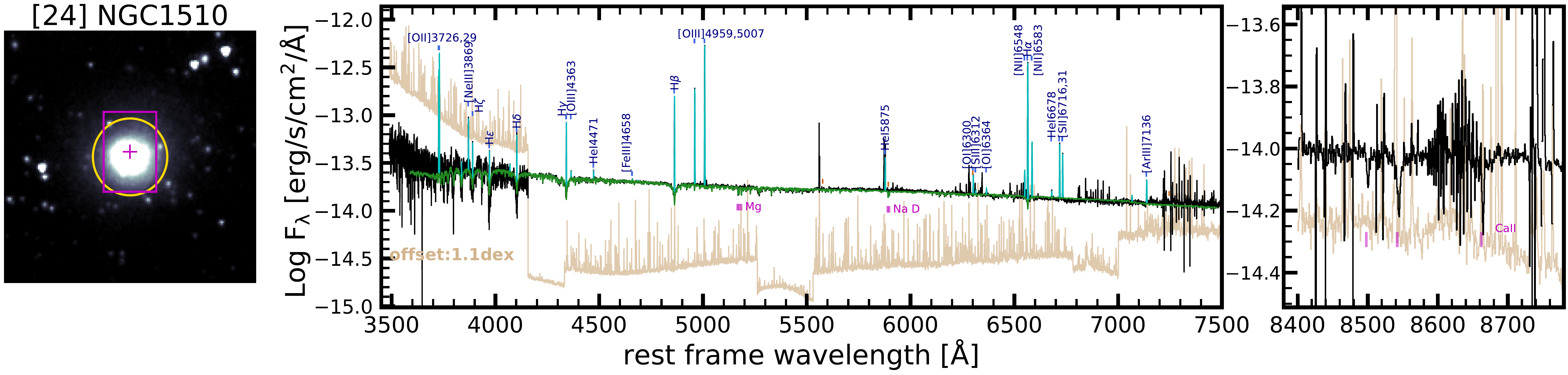}
  \includegraphics[width=\textwidth]{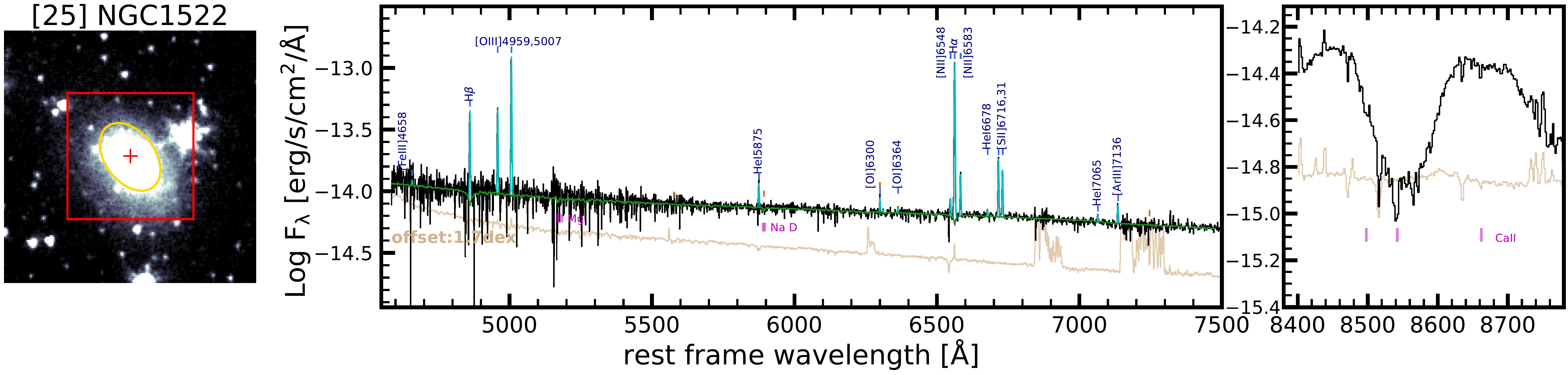}
\caption{continnue}
\end{figure*}

\begin{figure*}
  \includegraphics[width=\textwidth]{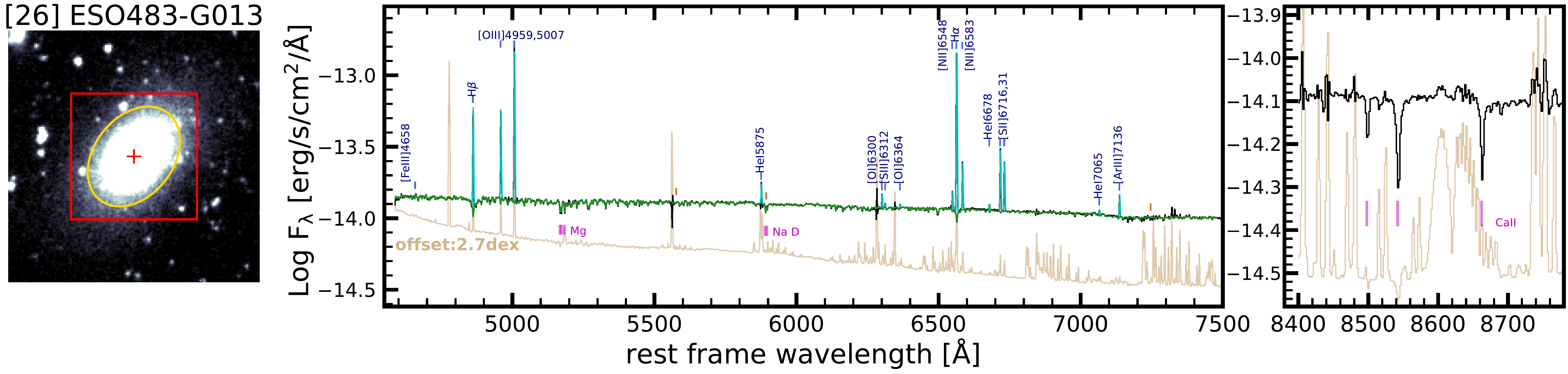}
  \includegraphics[width=\textwidth]{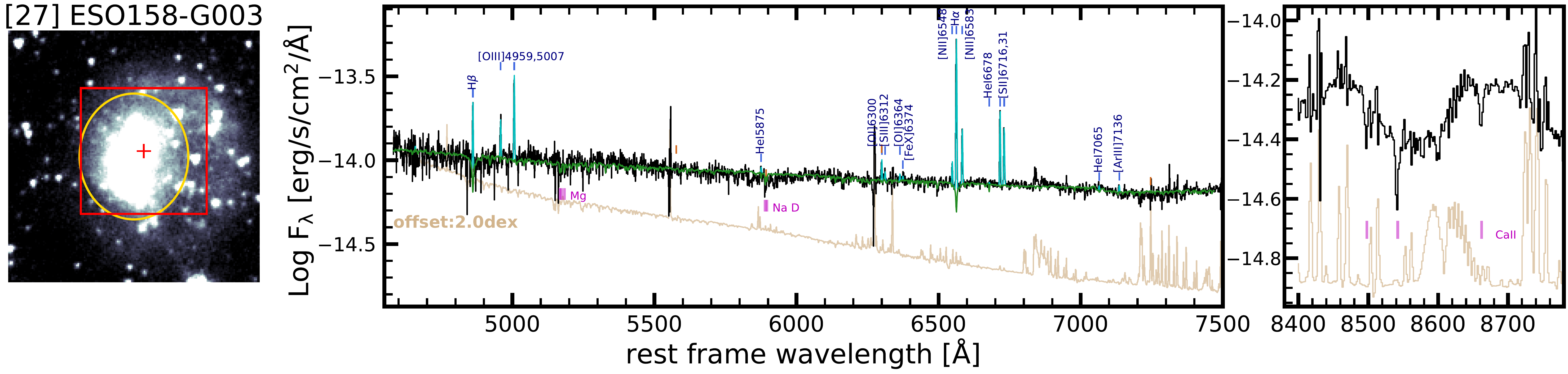}
  \includegraphics[width=\textwidth]{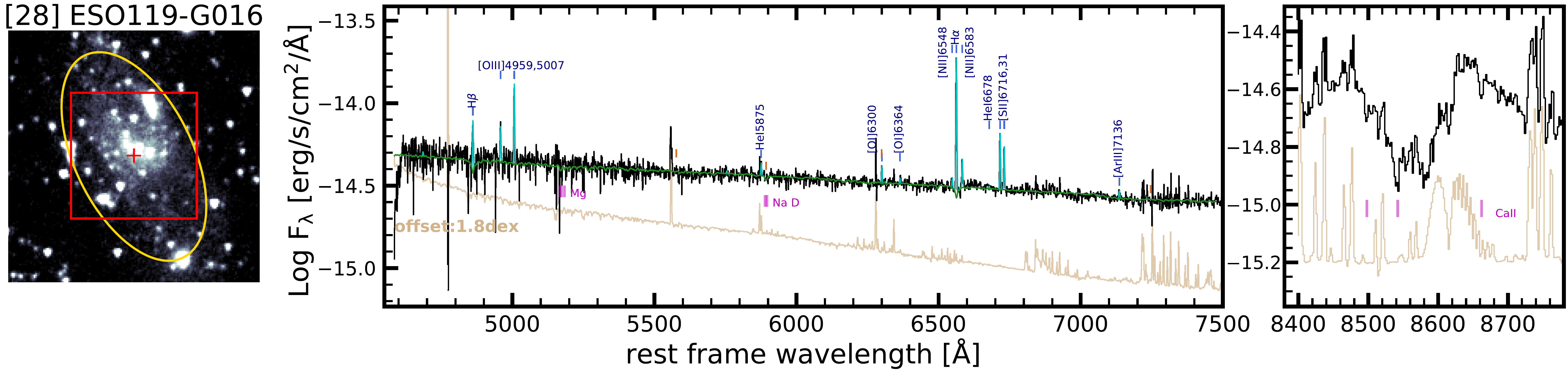}
  \includegraphics[width=0.17\textwidth]{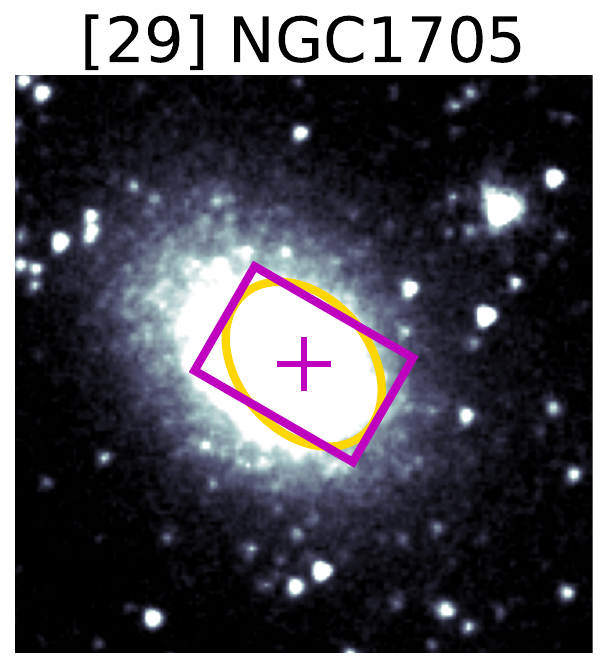}
  \newline
  \newline
  \includegraphics[width=\textwidth]{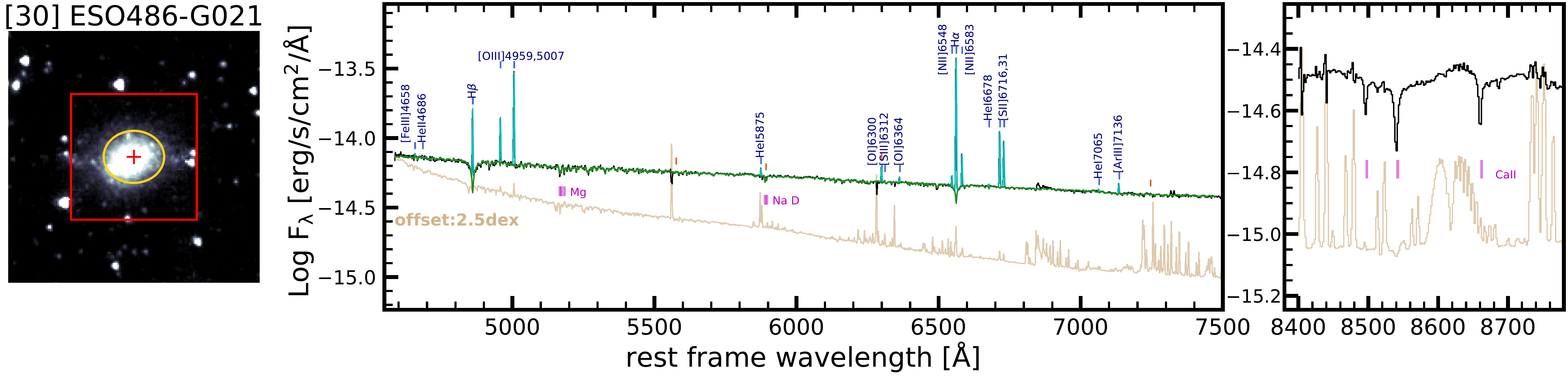}
\caption{continnue}
\end{figure*}

\begin{figure*}
  \includegraphics[width=\textwidth]{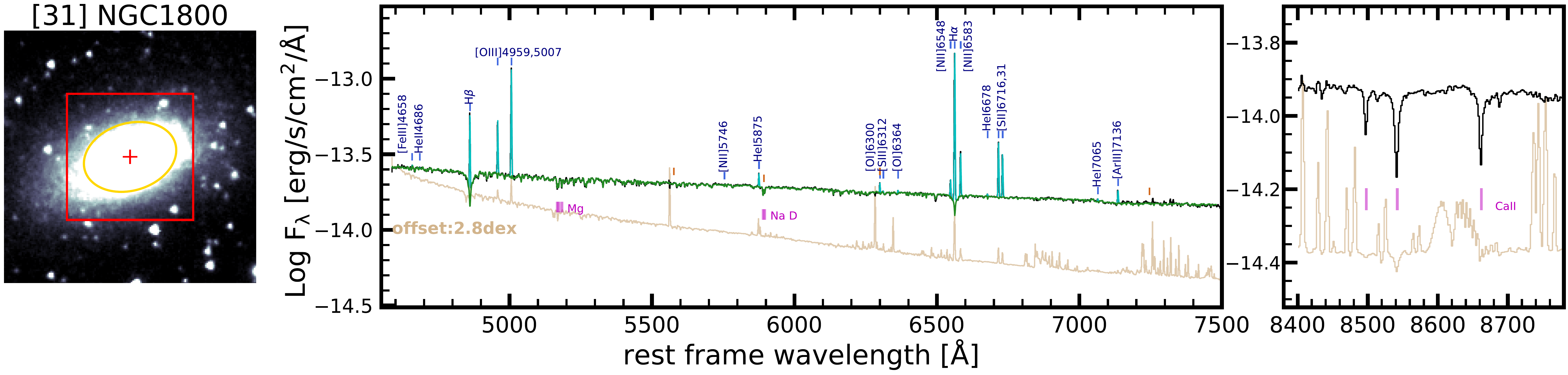}
  \includegraphics[width=0.17\textwidth]{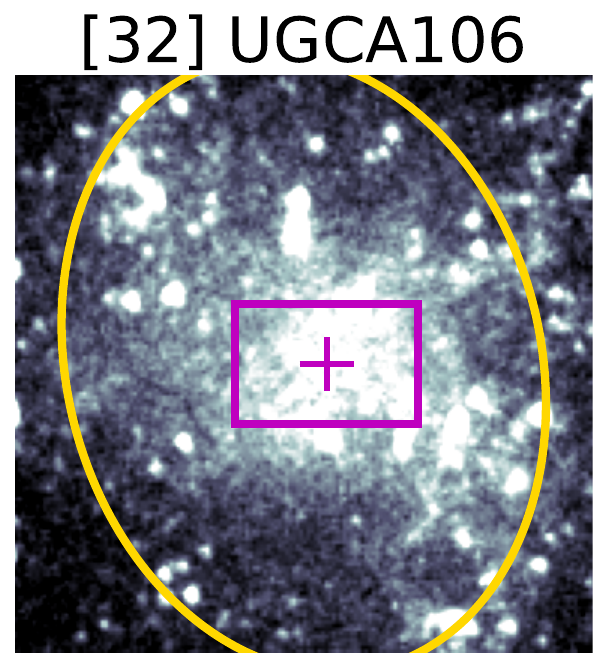}
  \newline
  \newline
  \includegraphics[width=\textwidth]{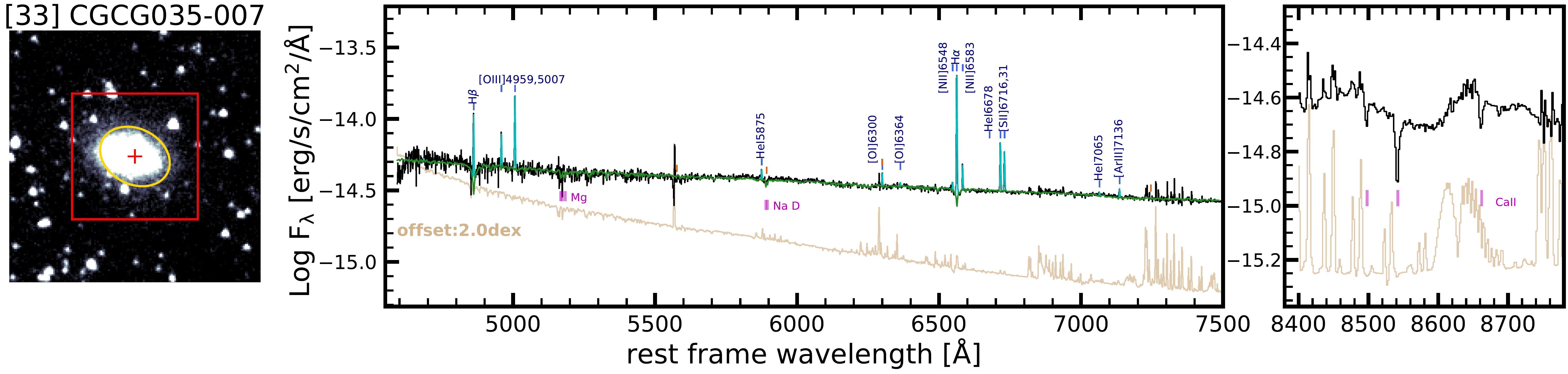}
  \includegraphics[width=0.17\textwidth]{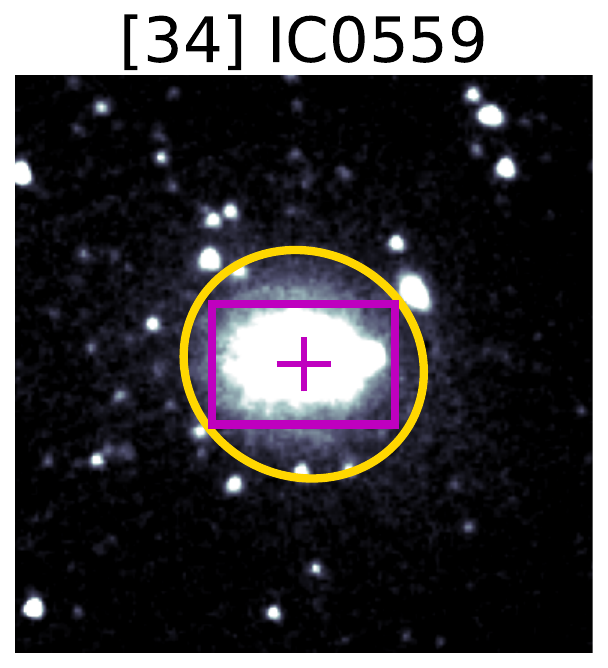}
  \newline
  \newline
  \includegraphics[width=\textwidth]{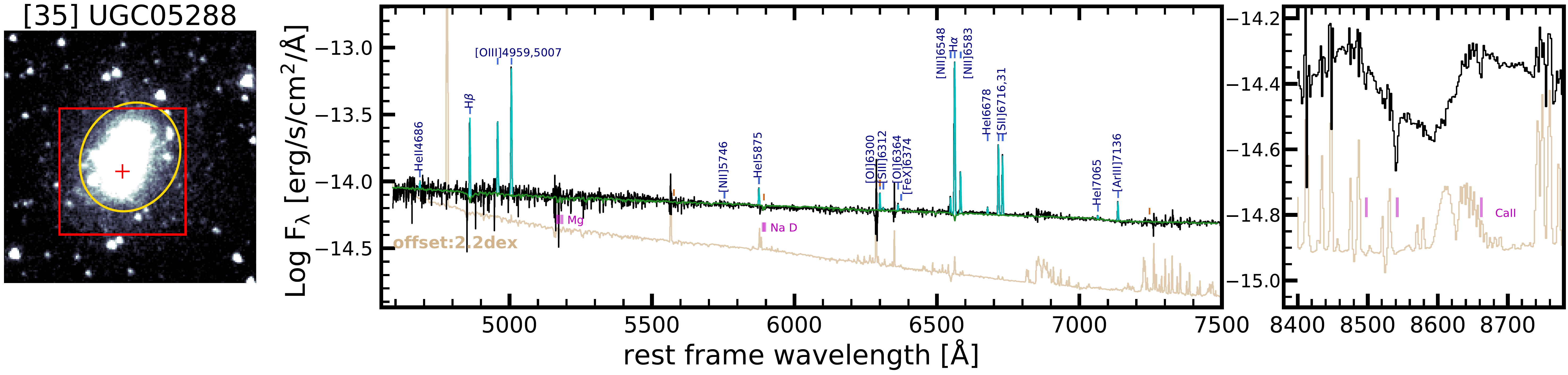}
\caption{continnue}
\end{figure*}

\begin{figure*}
  \includegraphics[width=\textwidth]{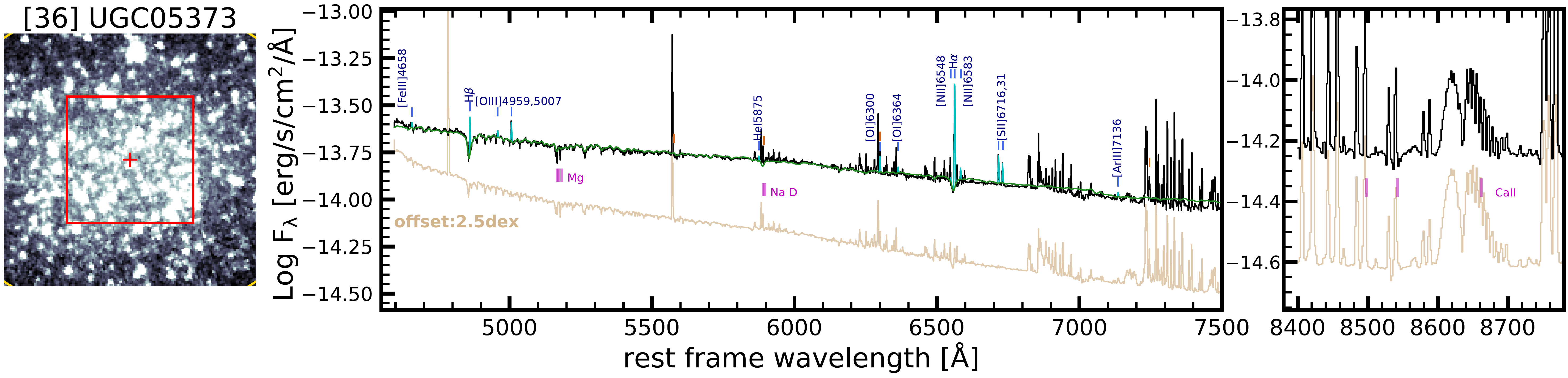}
  \includegraphics[width=\textwidth]{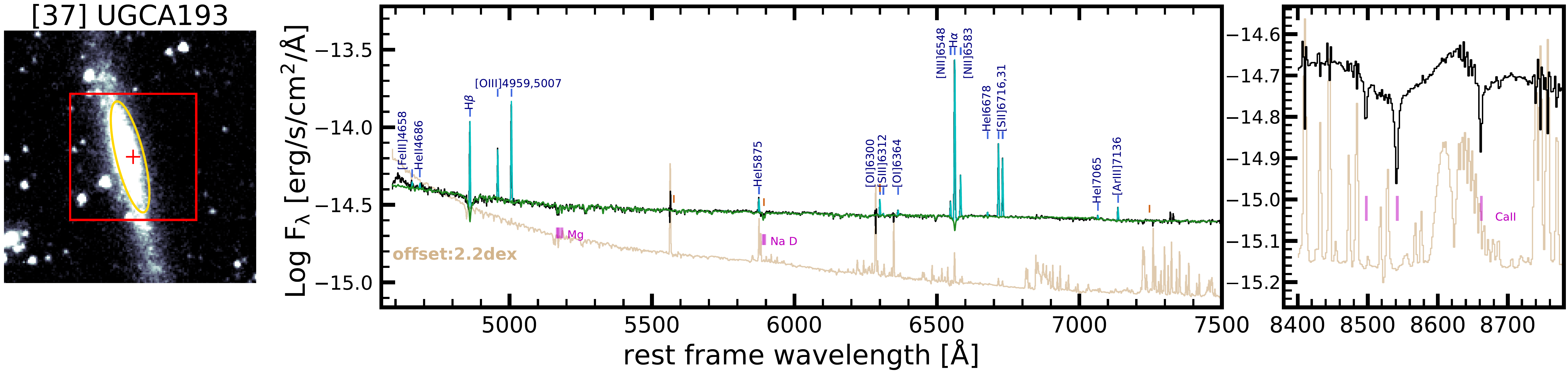}
  \includegraphics[width=0.17\textwidth]{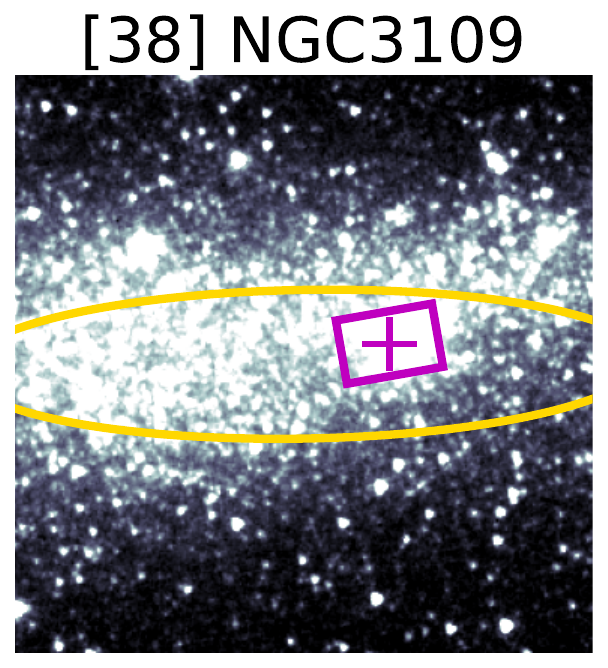}
  \newline
  \newline
  \includegraphics[width=\textwidth]{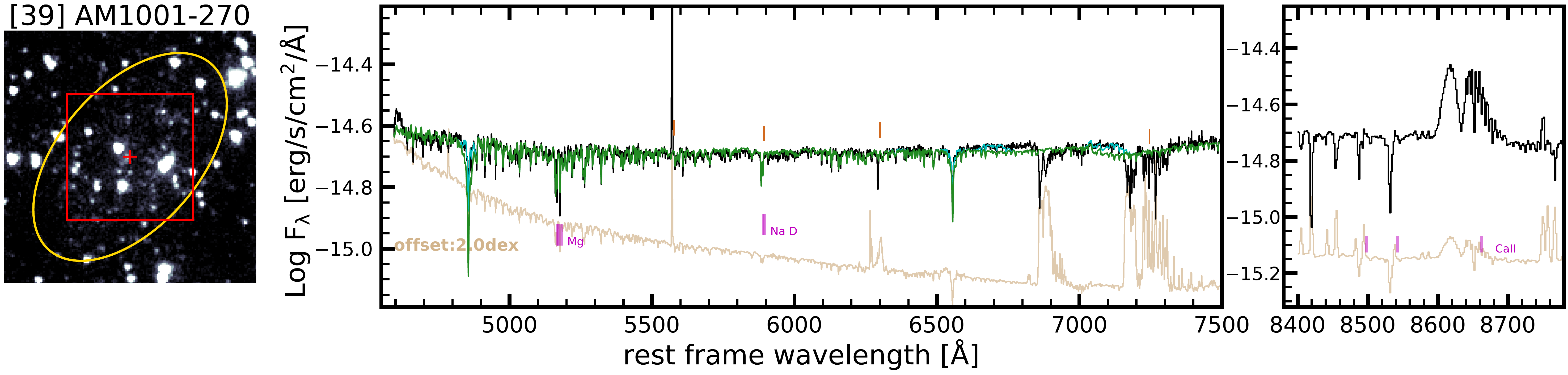}
  \includegraphics[width=0.17\textwidth]{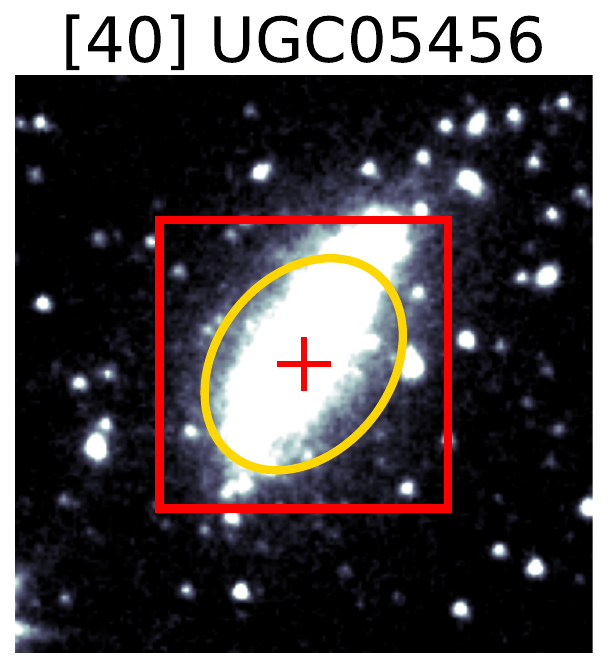}
  \newline
  \newline
\caption{continnue}
\end{figure*}

\begin{figure*}
  \includegraphics[width=0.17\textwidth]{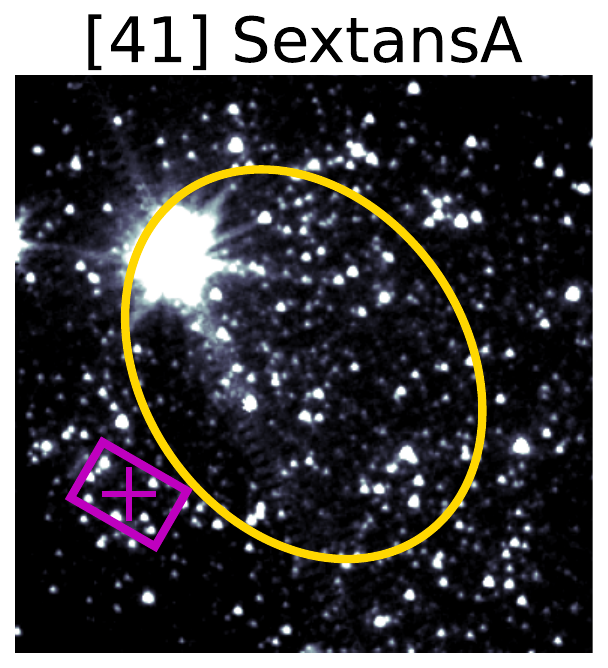}
  \newline
  \newline
  \includegraphics[width=0.17\textwidth]{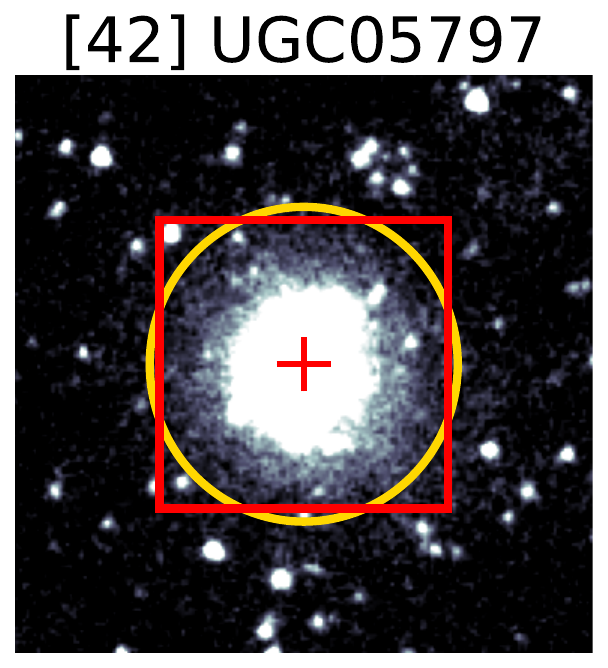}
  \newline
  \newline
  \includegraphics[width=0.17\textwidth]{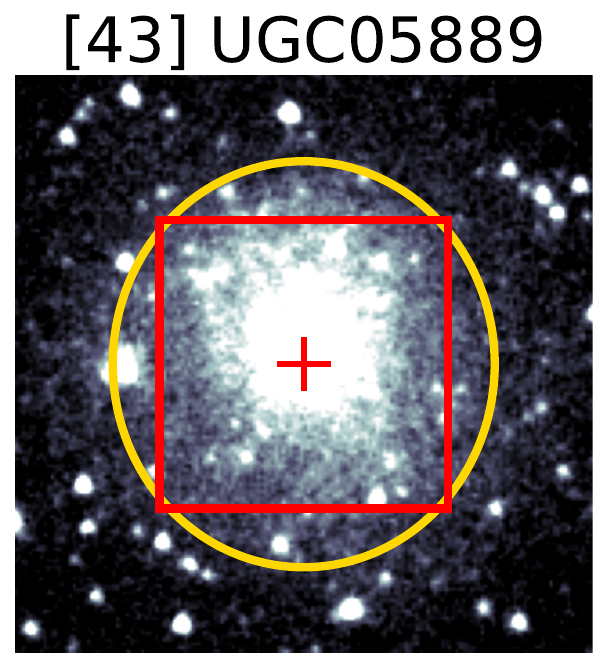}
  \newline
  \newline
  \includegraphics[width=\textwidth]{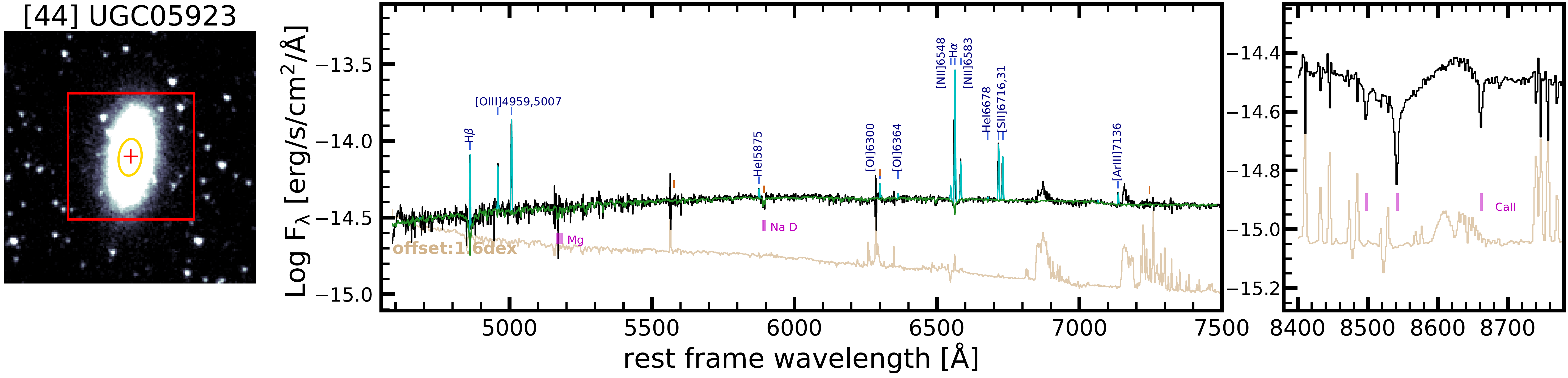}
  \includegraphics[width=0.17\textwidth]{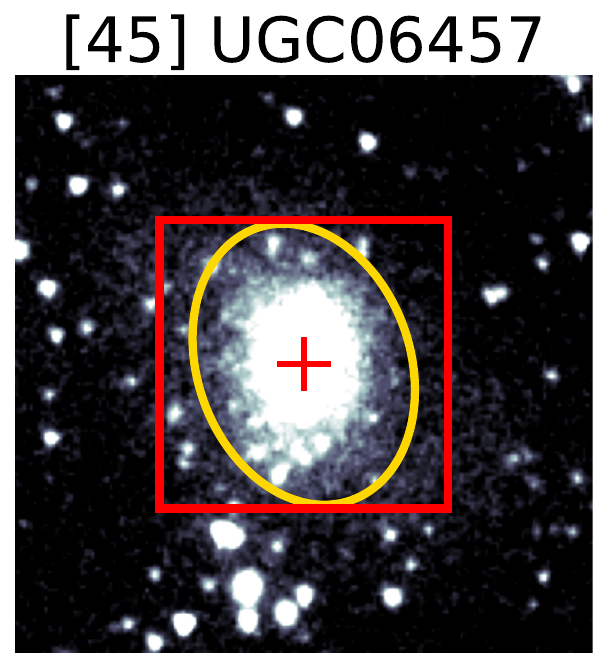}
\caption{continnue}
\end{figure*}

\begin{figure*}
  \includegraphics[width=\textwidth]{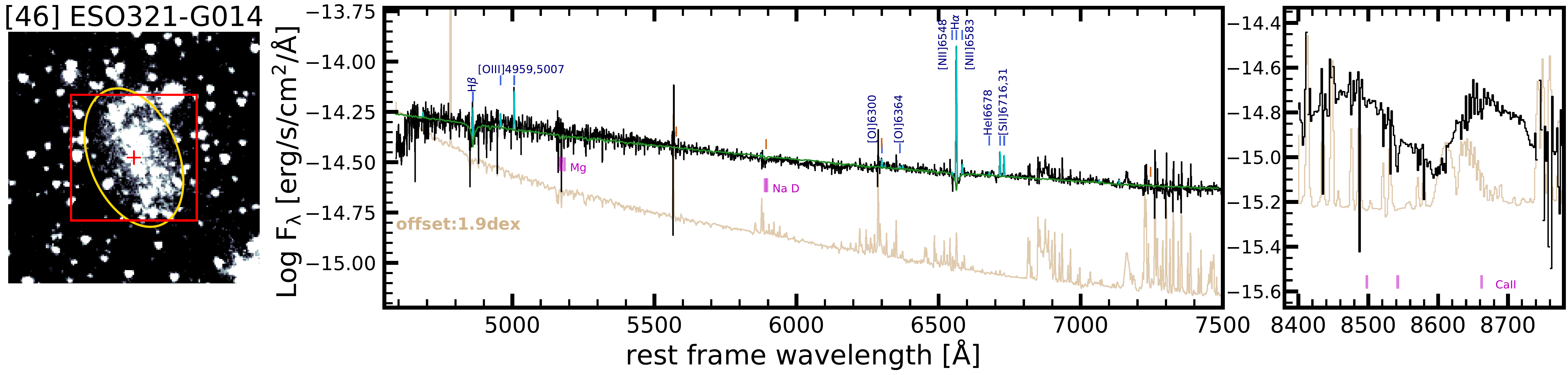}
  \includegraphics[width=\textwidth]{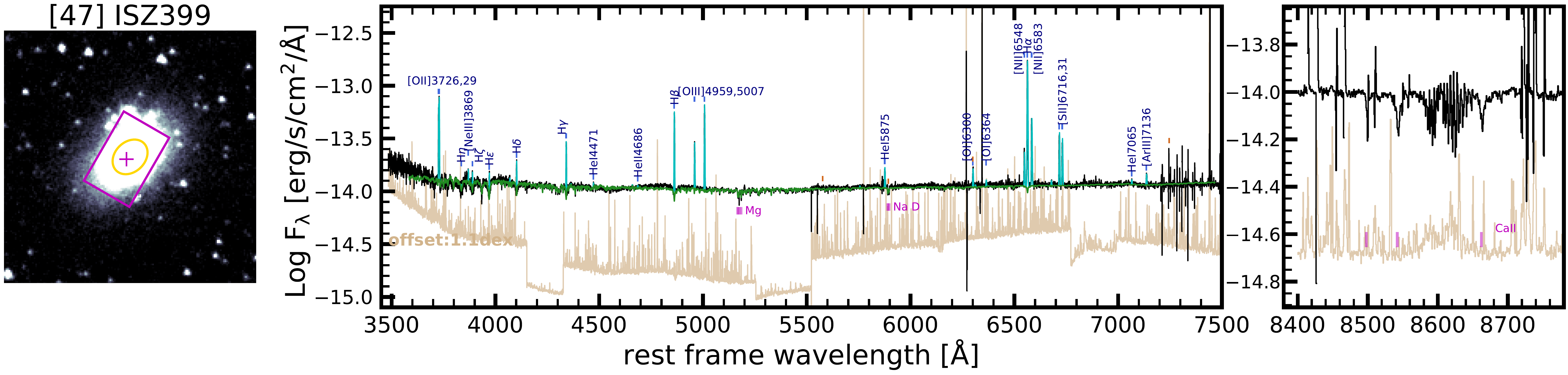}
  \includegraphics[width=\textwidth]{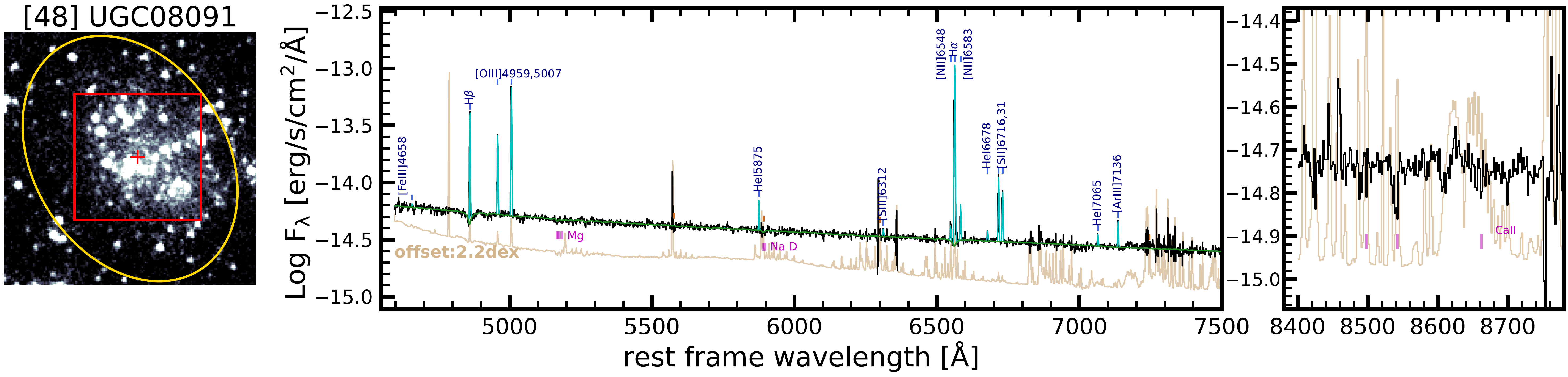}
  \includegraphics[width=0.17\textwidth]{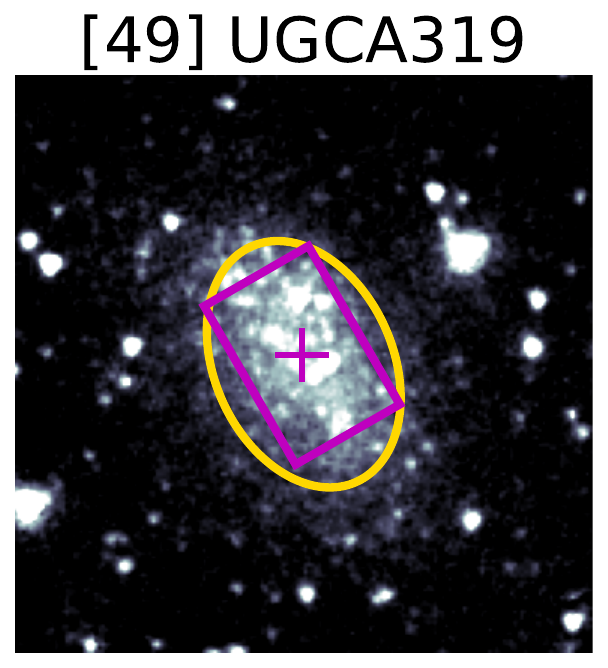}
    \newline
  \newline
  \includegraphics[width=\textwidth]{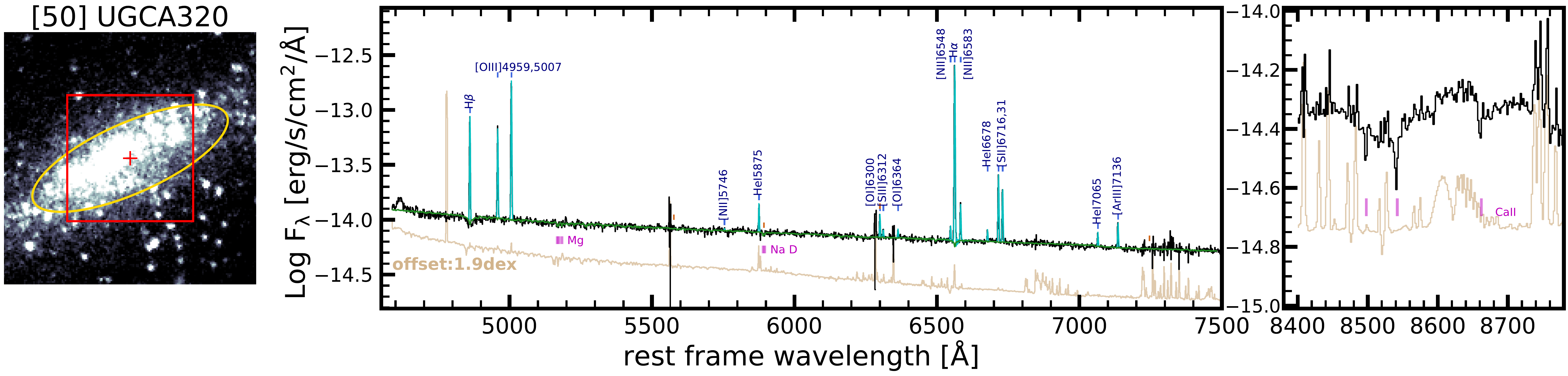}
\caption{continnue}
\end{figure*}

\begin{figure*}
\centering
  \includegraphics[width=\textwidth]{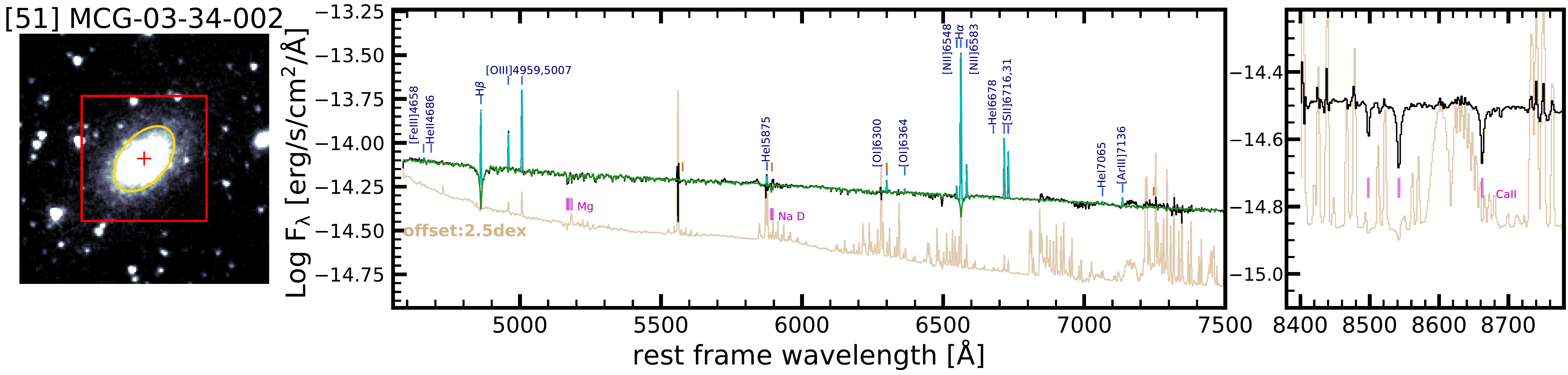}
  \includegraphics[width=\textwidth]{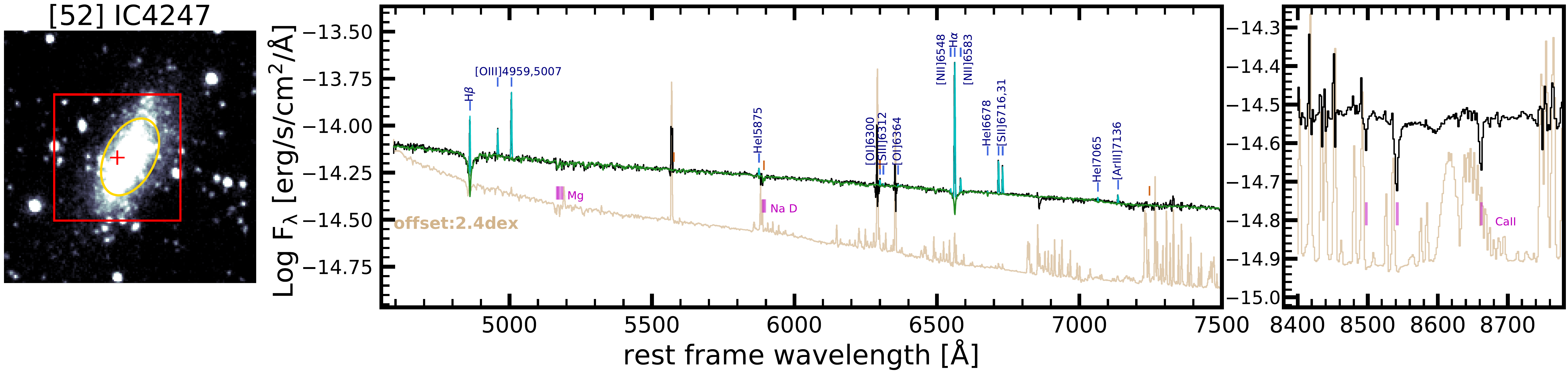}
  \includegraphics[width=\textwidth]{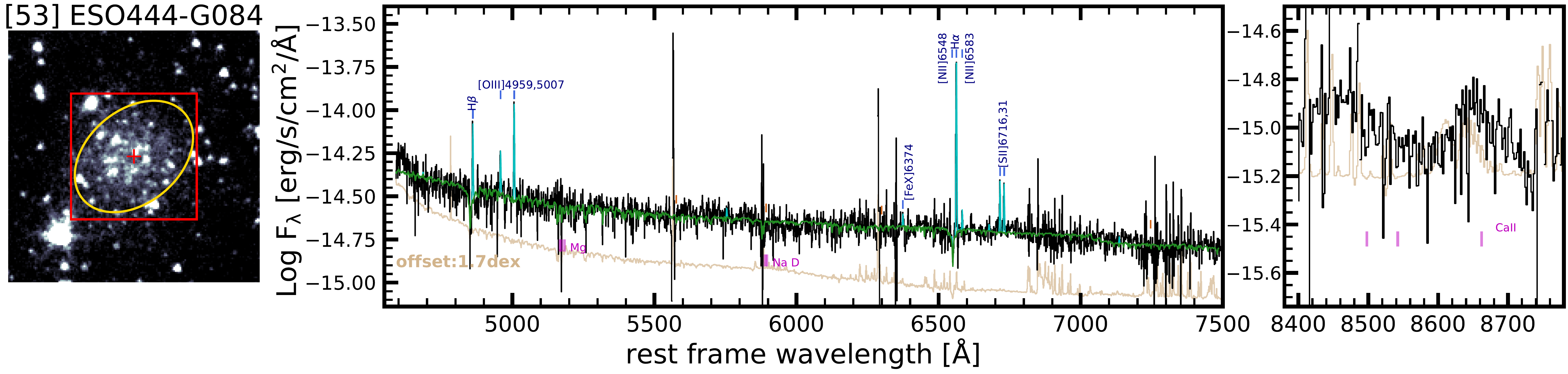}
  \includegraphics[width=\textwidth]{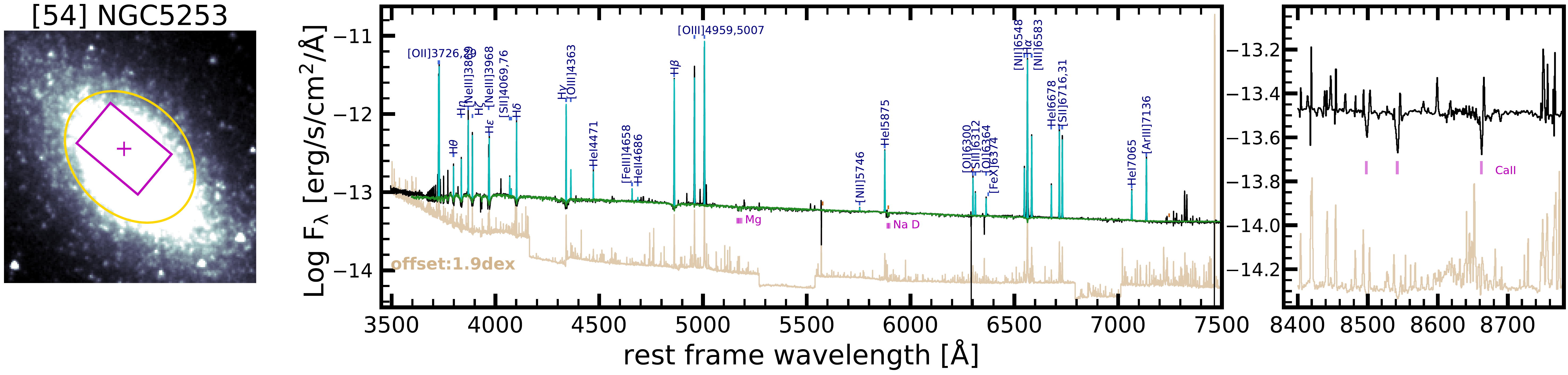}
  \includegraphics[width=\textwidth]{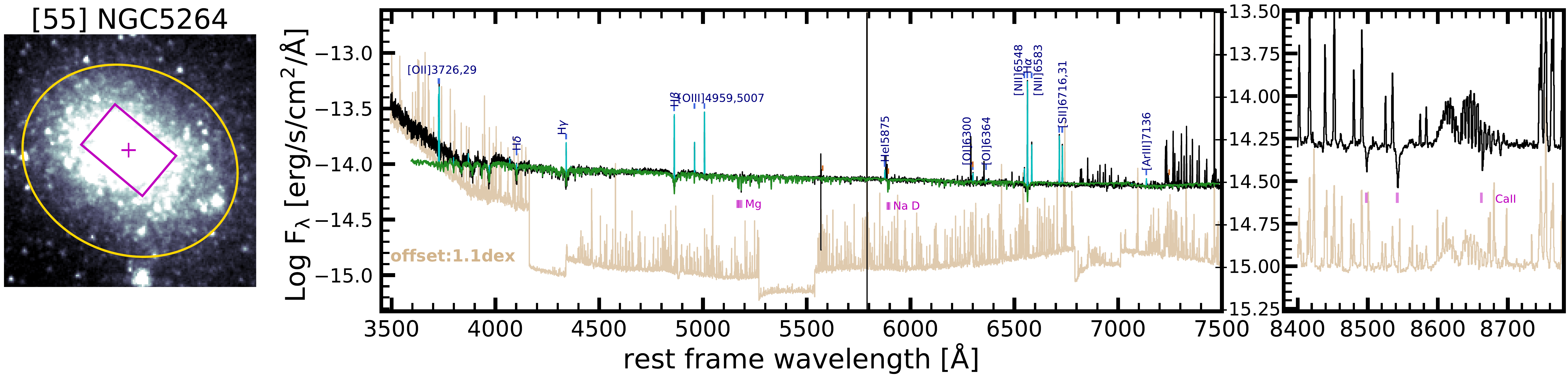}
\caption{continnue}
\end{figure*}

\begin{figure*}
  \includegraphics[width=\textwidth]{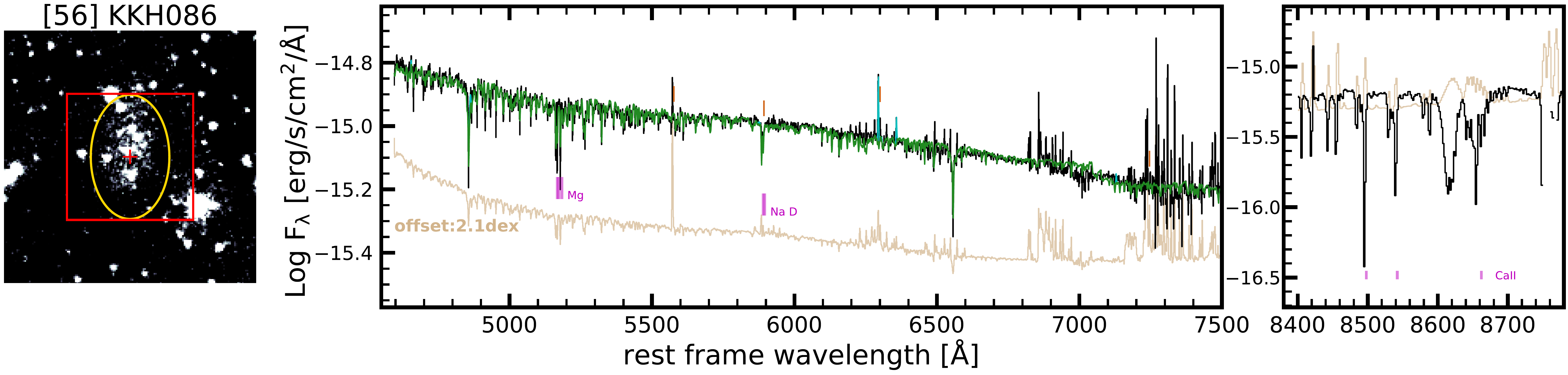}
  \includegraphics[width=\textwidth]{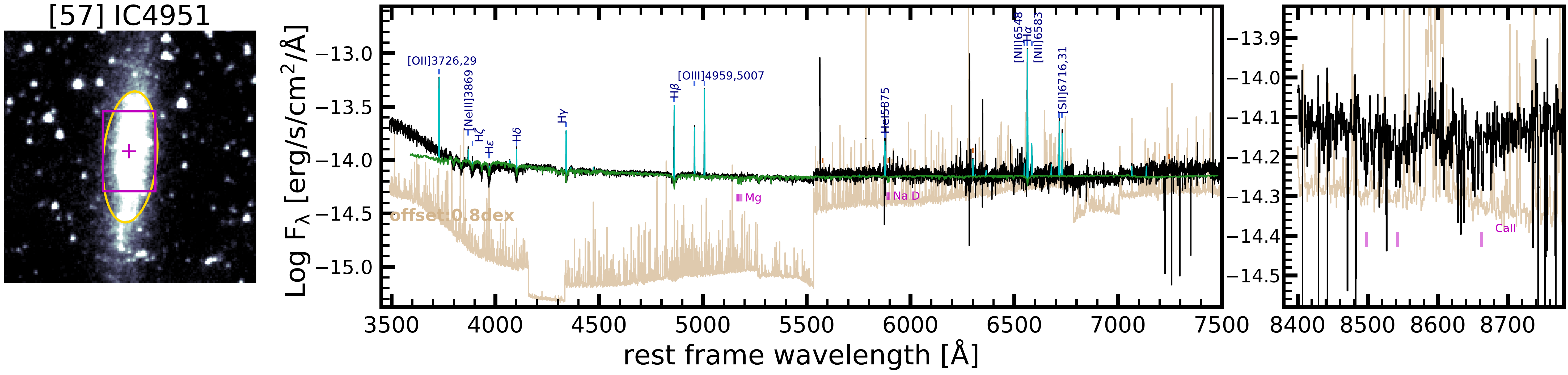}
  \includegraphics[width=\textwidth]{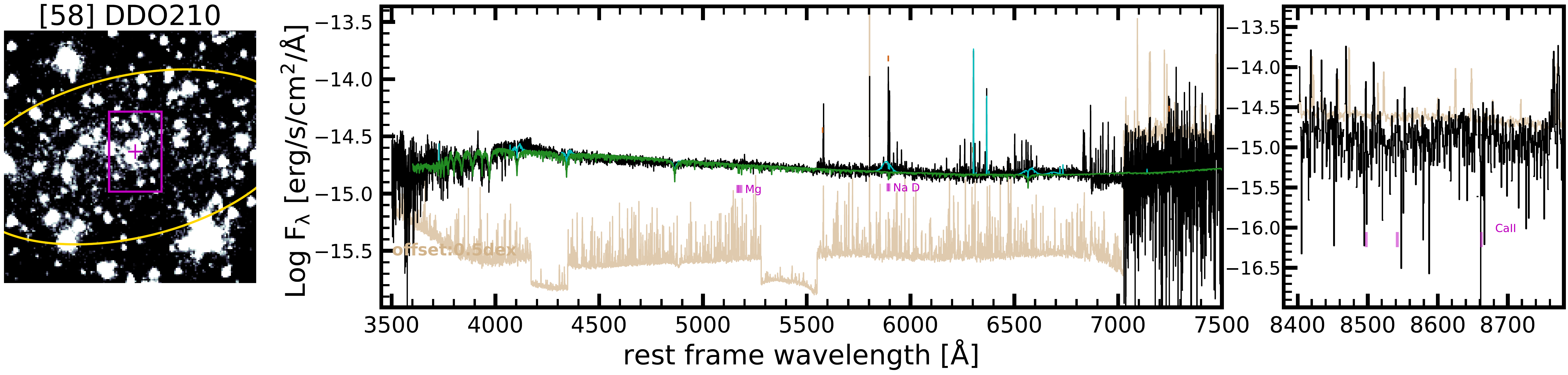}
  \includegraphics[width=\textwidth]{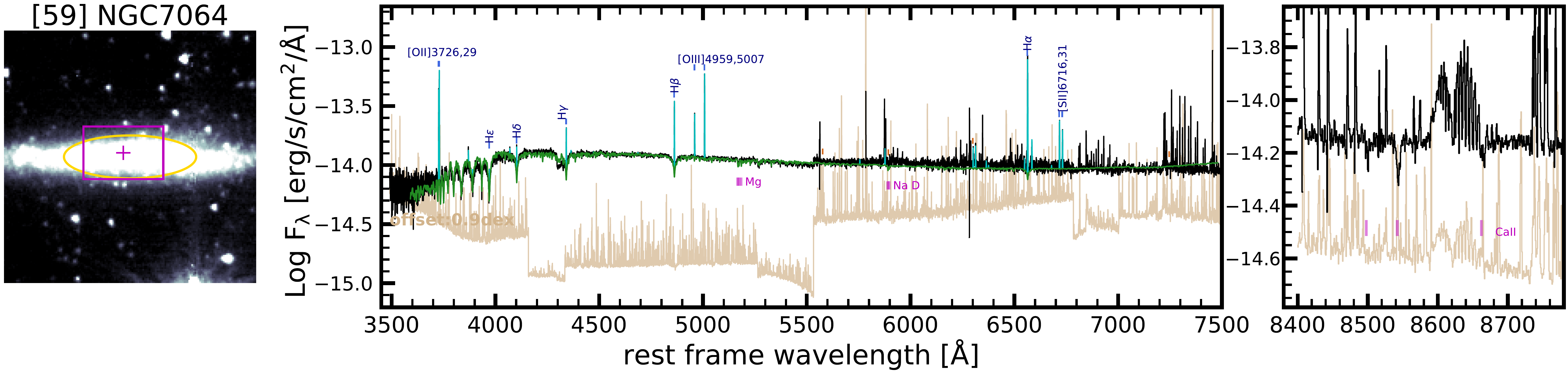}
  \includegraphics[width=\textwidth]{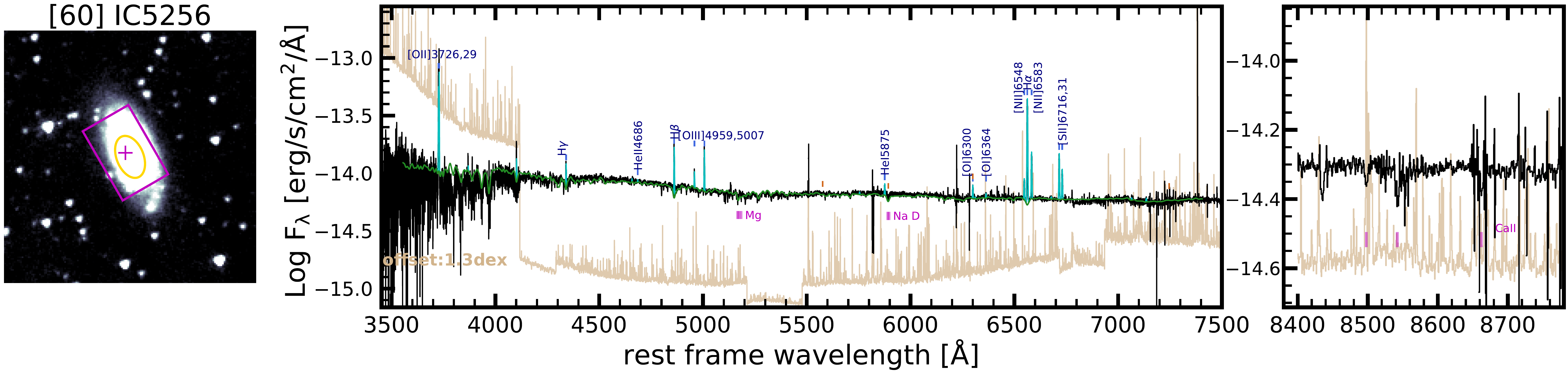}
\caption{continnue}
\end{figure*}

\begin{figure*}
  \includegraphics[width=\textwidth]{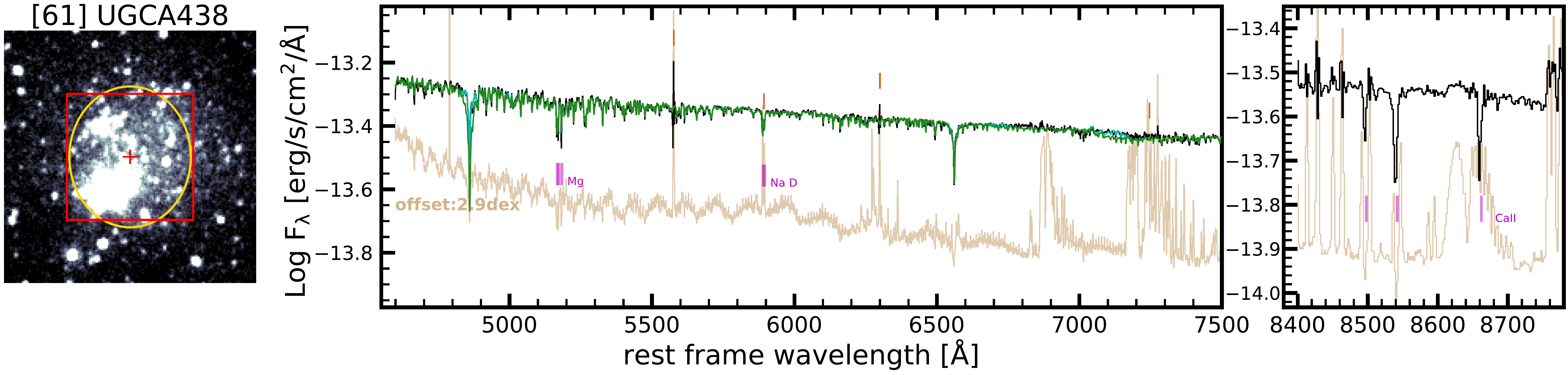}
  \includegraphics[width=0.17\textwidth]{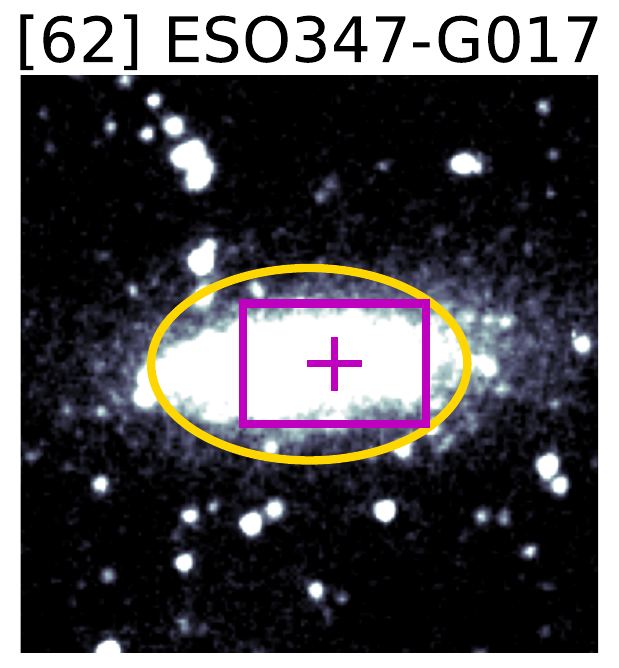}
  \newline
  \newline
  \includegraphics[width=\textwidth]{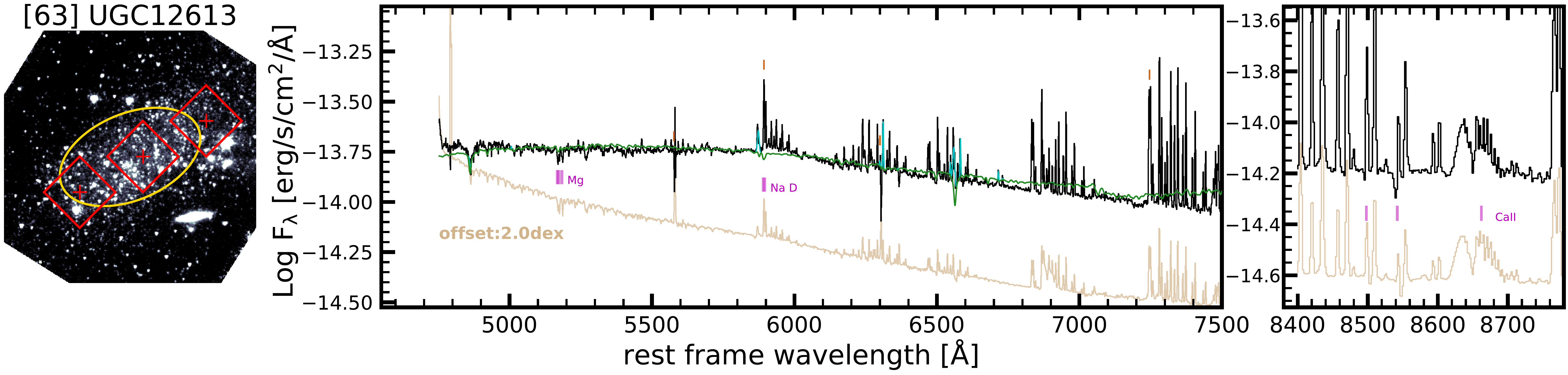}
  \includegraphics[width=\textwidth]{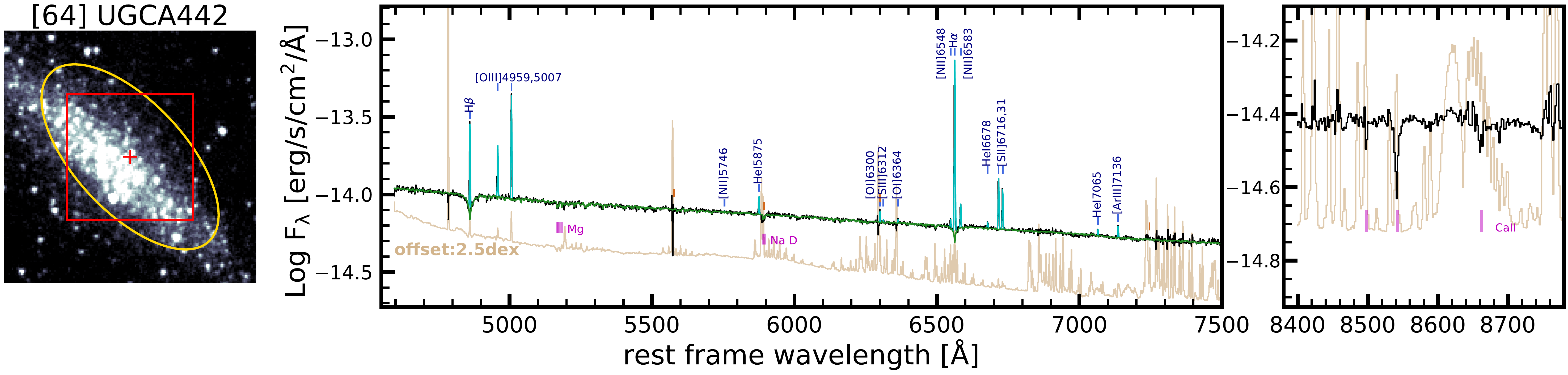}
  \includegraphics[width=0.17\textwidth]{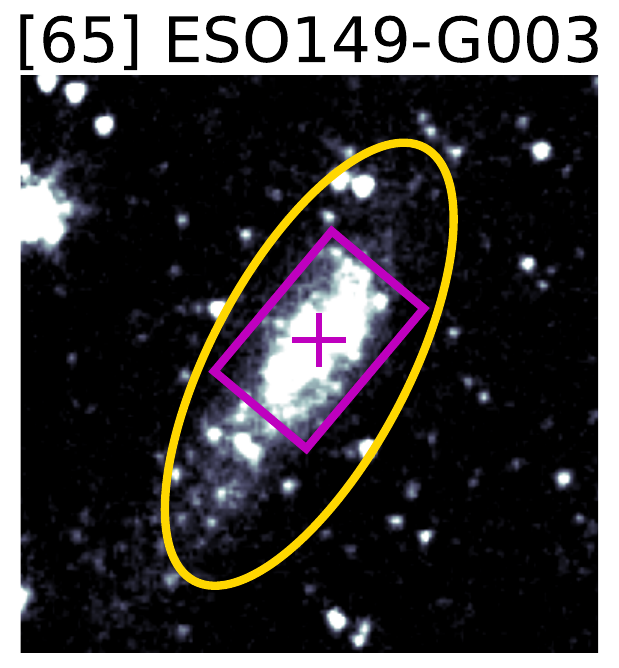}
\caption{continnue}
\end{figure*}

\clearpage
\section{Acknowledgements}

XL and YS acknowledge the support from the National Key R\&D Programe of China No. 2022YFF0503401, the National Natural Science Foundation of China (NSFC grants 12141301, 12121003, 12333002, 11825302).

X.L.Y. acknowledges the grant from the National Natural Science Foundation of China (NSFC grants 12303012), Yunnan Fundamental Research Projects (No.202301AT070242).

Based on observations collected at the European Southern Observatory under ESO programmes: 104.C-0181(C), 105.20GY.001, 105.20GY.002, 106.210Z.003, 108.21ZY.001, 109.238W.001, 110.23WR.008, 111.24UJ.002.

Based on data acquired at the ANU 2.3-metre telescope, under programes: 3190010, 1200017, 2200057, 3200031, 4200048, 1210055, 2210039, 3210057, 4210020, 3220028, 2380131, 2414211, 2425011. The automation of the telescope was made possible through an initial grant provided by the Centre of Gravitational Astrophysics and the Research School of Astronomy and Astrophysics at the Australian National University and through a grant provided by the Australian Research Council through LE230100063. The Lens proposal system is maintained by the AAO Research Data \& Software team as part of the Data Central Science Platform. We acknowledge the traditional custodians of the land on which the telescope stands, the Gamilaraay people, and pay our respects to elders past and present.

The national facility capability for SkyMapper has been funded through ARC LIEF grant LE130100104 from the Australian Research Council, awarded to the University of Sydney, the Australian National University, Swinburne University of Technology, the University of Queensland, the University of Western Australia, the University of Melbourne, Curtin University of Technology, Monash University and the Australian Astronomical Observatory. SkyMapper is owned and operated by The Australian National University's Research School of Astronomy and Astrophysics. The survey data were processed and provided by the SkyMapper Team at ANU. The SkyMapper node of the All-Sky Virtual Observatory (ASVO) is hosted at the National Computational Infrastructure (NCI). Development and support of the SkyMapper node of the ASVO has been funded in part by Astronomy Australia Limited (AAL) and the Australian Government through the Commonwealth's Education Investment Fund (EIF) and National Collaborative Research Infrastructure Strategy (NCRIS), particularly the National eResearch Collaboration Tools and Resources (NeCTAR) and the Australian National Data Service Projects (ANDS).

The Legacy Surveys consist of three individual and complementary projects: the Dark Energy Camera Legacy Survey (DECaLS; Proposal ID \#2014B-0404; PIs: David Schlegel and Arjun Dey), the Beijing-Arizona Sky Survey (BASS; NOAO Prop. ID \#2015A-0801; PIs: Zhou Xu and Xiaohui Fan), and the Mayall z-band Legacy Survey (MzLS; Prop. ID \#2016A-0453; PI: Arjun Dey). DECaLS, BASS and MzLS together include data obtained, respectively, at the Blanco telescope, Cerro Tololo Inter-American Observatory, NSF’s NOIRLab; the Bok telescope, Steward Observatory, University of Arizona; and the Mayall telescope, Kitt Peak National Observatory, NOIRLab. Pipeline processing and analyses of the data were supported by NOIRLab and the Lawrence Berkeley National Laboratory (LBNL). The Legacy Surveys project is honored to be permitted to conduct astronomical research on Iolkam Du’ag (Kitt Peak), a mountain with particular significance to the Tohono O’odham Nation.

NOIRLab is operated by the Association of Universities for Research in Astronomy (AURA) under a cooperative agreement with the National Science Foundation. LBNL is managed by the Regents of the University of California under contract to the U.S. Department of Energy.

This project used data obtained with the Dark Energy Camera (DECam), which was constructed by the Dark Energy Survey (DES) collaboration. Funding for the DES Projects has been provided by the U.S. Department of Energy, the U.S. National Science Foundation, the Ministry of Science and Education of Spain, the Science and Technology Facilities Council of the United Kingdom, the Higher Education Funding Council for England, the National Center for Supercomputing Applications at the University of Illinois at Urbana-Champaign, the Kavli Institute of Cosmological Physics at the University of Chicago, Center for Cosmology and Astro-Particle Physics at the Ohio State University, the Mitchell Institute for Fundamental Physics and Astronomy at Texas A\&M University, Financiadora de Estudos e Projetos, Fundacao Carlos Chagas Filho de Amparo, Financiadora de Estudos e Projetos, Fundacao Carlos Chagas Filho de Amparo a Pesquisa do Estado do Rio de Janeiro, Conselho Nacional de Desenvolvimento Cientifico e Tecnologico and the Ministerio da Ciencia, Tecnologia e Inovacao, the Deutsche Forschungsgemeinschaft and the Collaborating Institutions in the Dark Energy Survey. The Collaborating Institutions are Argonne National Laboratory, the University of California at Santa Cruz, the University of Cambridge, Centro de Investigaciones Energeticas, Medioambientales y Tecnologicas-Madrid, the University of Chicago, University College London, the DES-Brazil Consortium, the University of Edinburgh, the Eidgenossische Technische Hochschule (ETH) Zurich, Fermi National Accelerator Laboratory, the University of Illinois at Urbana-Champaign, the Institut de Ciencies de l’Espai (IEEC/CSIC), the Institut de Fisica d’Altes Energies, Lawrence Berkeley National Laboratory, the Ludwig Maximilians Universitat Munchen and the associated Excellence Cluster Universe, the University of Michigan, NSF’s NOIRLab, the University of Nottingham, the Ohio State University, the University of Pennsylvania, the University of Portsmouth, SLAC National Accelerator Laboratory, Stanford University, the University of Sussex, and Texas A\&M University.

BASS is a key project of the Telescope Access Program (TAP), which has been funded by the National Astronomical Observatories of China, the Chinese Academy of Sciences (the Strategic Priority Research Program “The Emergence of Cosmological Structures” Grant \# XDB09000000), and the Special Fund for Astronomy from the Ministry of Finance. The BASS is also supported by the External Cooperation Program of Chinese Academy of Sciences (Grant \# 114A11KYSB20160057), and Chinese National Natural Science Foundation (Grant \# 12120101003, \# 11433005).

The Legacy Survey team makes use of data products from the Near-Earth Object Wide-field Infrared Survey Explorer (NEOWISE), which is a project of the Jet Propulsion Laboratory/California Institute of Technology. NEOWISE is funded by the National Aeronautics and Space Administration.

The Legacy Surveys imaging of the DESI footprint is supported by the Director, Office of Science, Office of High Energy Physics of the U.S. Department of Energy under Contract No. DE-AC02-05CH1123, by the National Energy Research Scientific Computing Center, a DOE Office of Science User Facility under the same contract; and by the U.S. National Science Foundation, Division of Astronomical Sciences under Contract No. AST-0950945 to NOAO.

The Pan-STARRS1 Surveys (PS1) and the PS1 public science archive have been made possible through contributions by the Institute for Astronomy, the University of Hawaii, the Pan-STARRS Project Office, the Max-Planck Society and its participating institutes, the Max Planck Institute for Astronomy, Heidelberg and the Max Planck Institute for Extraterrestrial Physics, Garching, The Johns Hopkins University, Durham University, the University of Edinburgh, the Queen's University Belfast, the Harvard-Smithsonian Center for Astrophysics, the Las Cumbres Observatory Global Telescope Network Incorporated, the National Central University of Taiwan, the Space Telescope Science Institute, the National Aeronautics and Space Administration under Grant No. NNX08AR22G issued through the Planetary Science Division of the NASA Science Mission Directorate, the National Science Foundation Grant No. AST–1238877, the University of Maryland, Eotvos Lorand University (ELTE), the Los Alamos National Laboratory, and the Gordon and Betty Moore Foundation.

\section*{Data Availability}

 
The data will be published online.

\bibliography{ms}{}
\bibliographystyle{aasjournal}



\end{document}